\documentclass{ws-ijmpb}

\begin{document}

\markboth{V. E. Korepin \& Y. Xu}
{Entanglement in Valence-Bond-Solid States}

%
\catchline{}{}{}{}{}
%

\title{ENTANGLEMENT IN VALENCE-BOND-SOLID STATES}

\author{VLADIMIR E. KOREPIN}

\address{C.N. Yang Institute for Theoretical Physics\\
State University of New York at Stony Brook\\
Stony Brook, NY 11794-3840, U.S.\\
korepin@insti.physics.sunysb.edu}

\author{YING XU}

\address{Institut f\"ur Theoretische Physik, Universit\"at Innsbruck\\
Institut f\"ur Quantenoptik und Quanteninformation\\
der \"Osterreichischen Akademie der Wissenschaften\\
Technikerstra\ss e 21a, 6020 Innsbruck, Austria\\
ying.xu@uibk.ac.at}

\maketitle

\begin{history}
\received{Day Month Year}
\revised{Day Month Year}
\end{history}

\begin{abstract}

This article reviews the quantum entanglement in Valence-Bond-Solid (VBS) states defined on a lattice or a graph. The subject is presented in a self-contained and pedagogical way. The VBS state was first introduced in the celebrated paper by I. Affleck, T. Kennedy, E. H. Lieb and H. Tasaki (abbreviation AKLT is widely used). It became essential in condensed matter physics and quantum information (measurement-based quantum computation). Many publications have been devoted to the subject. Recently entanglement was studied in the VBS state. In this review we start with the definition of a general AKLT spin chain and the construction of VBS ground state. In order to study entanglement, a block subsystem is introduced and described by the density matrix. Density matrices of $1$-dimensional models are diagonalized and the entanglement entropies (the von Neumann entropy and R\'enyi entropy) are calculated. In the large block limit, the entropies also approach finite limits. Study of the spectrum of the density matrix led to the discovery that the density matrix is proportional to a projector.\end{abstract}


\section{Introduction}
\label{sec:intro1}

The fields of statistical mechanics, condensed matter physics and quantum information theory share a common interest in the study of interacting quantum many body systems. The concept of \textbf{entanglement} in quantum mechanics has significant importance in all these areas, it was introduced in  \cite{S}. Roughly speaking, entanglement \cite {NC} is a phenomenon of quantum mechanical nature in which quantum states of physical systems are linked together so that one system can not be adequately described without full mention of its counterpart, even when the individual systems may be spatially separated. Entanglement becomes particularly interesting in a many body interacting system where a subsystem strongly correlates with its environment (other parts of the system). The correlations may reject the principle of local realism, which states that information about the state of a system can only be mediated by interactions in its immediate surroundings (neighbors). The characteristic length of entanglement may be diverging while the usual correlation length remains finite \cite{VMC}. Quantum entanglement is a fundamental measure of `quantumness' of a system: how much quantum effects we can observe and use to control one quantum system by another. It is the primary resource in emerging technologies of quantum computation and quantum information processing \cite{BD,L}. Entanglement properties play an important role in condensed matter physics, such as phase transitions \cite{OAFF,ON} and macroscopic properties of solids \cite{GRAC}.  There is an excellent review \cite{GT} on entanglement detection, which covers many topics (such as multi-partite entanglement, concurrence, Bell inequalities, entanglement witnesses and experiments) beyond the scope of our present discussion.

Much of current research seeks to elucidate quantum entanglement in a variety of interacting systems. Extensive research has been undertaken to investigate quantum entanglement in strongly correlated states such as spin chains, correlated electrons, interacting bosons as well as other models. (See \cite{AFOV,ABV,BH,CBO,GDLL,EC,FIJK,FIK,HZS,HJPW,HLW,IJK,JK,JKLPPSW,KM,K,LORV,LRV,OL,P,PP,PS,VMC,VLRK,ZR} for reviews and references.) A general approach studies the density matrix of a certain subsystem of a strongly entangled state. The concept of the reduced density matrix was first introduced by P. A. M. Dirac in \cite{D}. Following C. H. Bennett, H. J. Bernstein, S. Popescu and B. Schumacher \cite{BBPS}, the entropy (derived from the spectrum of the density matrix) of a subsystem serves as the measure of entanglement for a \textbf{pure state}. The von Neumann entropy and its generalization (the R\'enyi entropy) are typical quantifications of entanglement. These characteristic functions may depend on the physical parameters (size, coupling constants, external fields, \textit{etc}.) in various different ways. An area law for the von Neumann entropy in harmonic lattice systems has been extensively studied \cite{CEP,CEPD,Has}. The area law states that the entropy scales proportional to the size (area) of the boundary of the subsystem (this statement was strictly proved for gapped models only). The entropy of the whole system vanishes if the system is in a pure state (usually the unique ground state), but it can be positive for a subsystem (this means entanglement). This is in contrast with a classical system in which if the entropy is equal to zero for the whole system, then it also vanishes for any subsystem. In quantum mechanics, when the whole system is in a pure state, a subsystem can be in a mixed state. \textit{e.g.} The Bell pair (or the Einstein-Podolsky-Rosen state) $(|\uparrow\downarrow\rangle-|\downarrow\uparrow\rangle)/\sqrt{2}$ is a pure state. A single spin subsystem is in a completely mixed state. The density matrix of one spin in an EPR state is proportional to identity matrix: $\boldsymbol{\rho}=(|\uparrow\rangle\langle\uparrow|+|\downarrow\rangle\langle\downarrow|)/2$,  which has  entropy $\ln 2$. In a more general case, we shall consider a pure system consisting of two subsystems $B$ (Block) and $E$ (Environment) with two different Hilbert spaces $\mathbf{H}_{B}$ and $\mathbf{H}_{E}$ of dimensions $dim_{B}$ and $dim_{E}$, respectively (assuming that $dim_{B}\leq dim_{E}$). So the entropy of the whole system $B\cup E$ is zero. If the entropy of a subsystem ($B$) is positive then the wave function of the whole system does not factorize:
\begin{eqnarray}
|B\cup E\rangle\neq|B\rangle\otimes|E\rangle.
\end{eqnarray}
Let $\{|b_{j}\rangle\}$ and $\{|e_{k}\rangle\}$ be any fixed orthonormal bases for subsystems $B$ and $E$, respectively. Then the wave function of the whole system can be written as
\begin{eqnarray}
	|B\cup E\rangle=\sum_{jk}A_{jk}|b_{j}\rangle\otimes|e_{k}\rangle,
\end{eqnarray}
for some matrix $A$ of complex elements $A_{jk}$. Then by the singular value decomposition (see \S$\,2.1.10\,$ of \cite{NC}), $A=U\mathcal{D}V$, where $\mathcal{D}$ is a diagonal matrix with non-negative elements, and $U$ and $V$ are unitary matrices ($A$ is a rectangular matrix if $dim_{B}\neq dim_{E}$, in that case $U$ and $V$ have different dimensions). Define $|B_{i}\rangle\equiv\sum_{j}U_{ji}|b_{j}\rangle$, $|E_{i}\rangle\equiv\sum_{k}V_{ik}|e_{k}\rangle$, and $\sqrt{\lambda_{i}}\equiv \mathcal{D}_{ii}$. With such a choice of bases, we find that the wave function can be put in the following form:
\begin{eqnarray}
	|B\cup E\rangle=\sum^{D}_{j=1}\sqrt{\lambda_{j}}\,|B_{j}\rangle\otimes|E_{j}\rangle
\end{eqnarray}
where $\{|B_{j}\rangle\}$, as well as $\{|E_{j}\rangle\}$, are orthonormal bases. The coefficients $0<\sqrt{\lambda_{j}}<1$ satisfy
\begin{eqnarray}
	\sum^{D}_{j=1}\lambda_{j}=1.
\end{eqnarray}
Once again, if the entropy of the subsystem $S[B]>0$ then $D>1$ ($D\leq dim_{B}$). In this case we can control one subsystem by the other. Indeed if we measure subsystem $B$ (in $|B_{j}\rangle$ basis), then the wave function of the whole system will change (collapse) into
\begin{eqnarray}
	|B_{j_{0}}\rangle\otimes|E_{j_{0}}\rangle
\end{eqnarray}
with probability $\lambda_{j_{0}}$ for some $j_{0}$. So that a measurement of $B$ changes the state of $E$ and this is quantum control. This is an important resource for quantum device building (and maybe even quantum computation). Much insight in understanding entanglement of quantum systems has been obtained by studying exactly solvable models in statistical mechanics, in which it is possible to solve the subsystem density matrix and calculate the entropy exactly. This review is devoted to one particular set of models -- the AKLT models.

In 1987, I. Affleck, T. Kennedy, E. H. Lieb and H. Tasaki proposed a spin interacting model known as the AKLT model \cite{AKLT1,AKLT2}. The model consists of spins on lattice sites and the Hamiltonian describes interactions between nearest neighbors. The Hamiltonian density is a linear combination of projectors. Each projector is written as a polynomial of the inner product of a pair of interacting spin vectors. The authors (AKLT) of \cite{AKLT1,AKLT2} found the exact ground state, which has an exponentially-decaying correlation function and a finite energy gap. In their early works, authors discussed the $1$-dimensional AKLT lattices with open and periodic boundary conditions and $2$-dimensional models such as the hexagonal lattice. D. A. Arovas, A. Auerbach and F. D. M. Haldane studied the model \cite{AAH} in the Schwinger boson representation (see \S$\,\ref{sec:vbssch}\,$ below) and calculated the correlation functions using the coherent state basis (see \S$\,\ref{sec:cohe}\,$ below). In their work \cite{AAH}, it was shown that the quantum spins in the AKLT model is equivalent to classical spins. This model has been attracting enormous research interests since then \cite{AAH,FM,KHH,KLT,KK,KSZ,RB}. It can be defined and solved in higher dimensional lattices \cite{AKLT2,FKR2,KLT,RB}. It is generalizable to the inhomogeneous (non-translational invariant) case (spins at different lattice sites may take different values) and an arbitrary graph \cite{AKLT1,AKLT2,KK,XK}. The AKLT model has also been generalized to the $SU(n)$ version \cite{GR,GRS,KHK,RSSTG,RTFSG}. Given certain conditions (see \S$\,\ref{sec:unique}\,$), the ground state has proven to be unique \cite{AKLT1,AKLT2,KLT,KK}. It is known as the Valence-Bond-Solid (VBS) state. The VBS state is interesting to different research fields. The Schwinger boson representation of the VBS state (see (\ref{generalizedvbs})) relates to the Laughlin ansatz of the fractional quantum Hall effect \cite{AAH,HZS,ILO}. The Laughlin wave function of the fractional quantum Hall effect is the VBS state on the complete graph \cite{HR0}. The VBS state illustrates ground state properties of anti-ferromagnetic integer-spin chains with a Haldane gap \cite{H}. In $1$-dimension, the VBS state is related to the matrix product state discovered by A. Kl\"umper, A. Schadschneider and J. Zittartzand, see \cite{KSZ1,KSZ} where a `q--deformed' VBS-model (a generalized anisotropic version) was studied. 
The geometric entanglement for a $1$-dimensional spin-$1$ VBS state was studied in \cite{O}. The VBS state can be used as a resource state in measurement-based quantum computation invented by R. Raussendorf and H. Briegel \cite{RB1}. VBS state allows universal quantum computation \cite{VC} . Many people developed these ideas \cite{BM,GESG}.  AKLT Hamiltonian can be  implemented in optical lattices \cite{GMC}. \footnote{Many more people work on and contribute to the subject and our list of references is far from complete.}

This review introduces some of the main results on quantum entanglement in VBS states defined on a lattice or a graph. We take a pedagogical approach, starting with the basics of the AKLT model, the construction of VBS states and measures of entanglement (special attention being paid to the uniqueness of the ground state for a finite lattice or graph). We shall consider a part (subsystem) of the ground state, \textit{i.e.} a block of spins. It is described completely by the reduced density matrix of the block, which we call \textit{the density matrix} later for short. The density matrix has been studied extensively. It contains information of all correlation functions \cite{AAH,JK,KHH,XKHK}. The entanglement properties of the VBS states has been studied by means of the density matrix as in \cite{FK,FKR,FKR2,FM,KHH,VMC,XKHK,XKHK2,XK}. Note that throughout the review we are considering \textbf{zero temperature} so that the AKLT spin system lies in the VBS ground state. 

The review is divided into eight sections including a complete treatment of $1$-dimensional models with open boundary conditions:
\begin{enumerate}
	\item A brief introduction to the topic. (\S$\,\ref{sec:intro1}\,$)
	\item The construction of the general AKLT Hamiltonian. Introduction of the VBS ground state. Definition of different versions of the AKLT model: 1) The \textit{basic} model; 2) The \textit{generalized} (including the inhomogeneous) model. Proof of the uniqueness of the VBS ground state on a finite graph. In this and the next sections, we consider the AKLT model on an arbitrary connected finite graph. This includes all lattices in any dimension. (\S$\,\ref{sec:general}\,$)
	\item In order to study entanglement, the graph (or lattice) is divided into two subsystems (the \textit{block} and the \textit{environment}). Different measures of entanglement, namely, the \textit{von Neumann Entropy} and the \textit{R\'enyi entropy} are defined. We define anther Hamiltonian called the \textit{block Hamiltonian}. The block Hamiltonian is the AKLT Hamiltonian for the block, but the uniqueness condition is violated. The block Hamiltonian is used to describe general properties of the block density matrix. We discuss the relation between the spectrum of the density matrix and the degenerate ground states of the block Hamiltonian. (\S$\,\ref{sec:subandmeas}\,$)
	\item The simplest $1$-dimensional AKLT model with spin-$1$. Calculation and diagonalization of the density matrix (in an algebraic approach). Calculation of the \textbf{entanglement entropies}. (\S$\,\ref{sec:1dspin1}\,$)
	\item The $1$-dimensional AKLT model with spin-$S$. Calculation and diagonalization of the density matrix using the Schwinger boson representation. Discussion of the entanglement entropies. Derivation of the relation between the density matrix and correlation functions. (\S$\,\ref{sec:1dssh}\,$)
	\item The $1$-dimensional inhomogeneous model (spins at different lattice sites can be different). Discussion of entanglement entropies. (\S$\,\ref{sec:1dinhom}\,$)
	\item The $1$-dimensional $SU(n)$ model (in the adjoint representation). Calculation of entanglement entropies. (\S$\,\ref{sec:su(n)}\,$)   
	\item A summary including open problems and conjectures. (\S$\,\ref{sec:sum1}\,$)
\end{enumerate}

\section{The General AKLT Model}
\label{sec:general}

In the following we give the most general AKLT Hamiltonian and VBS states. The definition applies to both graphs and arbitrary lattices.
\subsection{The Hamiltonian}
\label{sec:hamil}

The original AKLT Hamiltonian describes a spin interacting system, in which spins sitting at lattice sites interact with nearest neighbors. One of the most simple versions is an ($1$-dimensional) open chain of $N$ sites with spin-$1$ at each site, and the Hamiltonian is given by \cite{AKLT1}
\begin{eqnarray}
H=\frac{1}{2}\sum^{N-1}_{j=0}\left(\boldsymbol{S}_{j}\cdot\boldsymbol{S}_{j+1}+\frac{1}{3}(\boldsymbol{S}_{j}\cdot\boldsymbol{S}_{j+1})^{2}+\frac{2}{3}\right), \label{simphami}
\end{eqnarray}
in which $\boldsymbol{S}_{j}$ denotes the quantum operator for a spin vector $\boldsymbol{S}_{j}=\left(S^{x}_{j}, S^{y}_{j}, S^{z}_{j}\right)$ at site $j$. This Hamiltonian (\ref{simphami}) looks like the Heisenberg Hamiltonian with an extra quadratic term (the proportionality factor $1/2$ and the additive constant $2/3$ are sometimes neglected which only shifts and scales the energy spectrum as a whole), but the physical system behaves quite differently. It was later generalized: 1) The spin $\boldsymbol{S}_{j}$ at each site can take higher values; 2) Different lattice sites can have different spins; 3) Different boundary conditions (\textit{e.g.} an open boundary condition) can be applied. An arbitrary boundary condition or distribution of spin values over sites may not yield a unique ground state (\textit{e.g.} Hamiltonian (\ref{simphami}) has $4$-fold degenerate ground states). We could find the condition for the uniqueness of the ground state (see \S$\,\ref{sec:unique}\,$ and \cite{KK}). The Hamiltonian can be defined on higher dimensional lattices (\textit{e.g.} $2$-dimensional square or hexagonal lattice \cite{AKLT1,AKLT2}) or an arbitrary graph (A graph consists of two types of elements, namely \textit{vertices} and \textit{edges}. Every edge connects two vertices. Spins locate on vertices and two vertices connected by an edge are nearest neighbors. An edge is also called a \textit{bond} physically. In our considerations only nearest neighbors are present in the Hamiltonian. More formal and detailed explanation of a graph is given in \S$\,\ref{sec:basichamil}\,$). These different versions of the AKLT Hamiltonian under consideration share two common features:
\begin{enumerate}
	\item The Hamiltonian is a sum of terms with only nearest neighbor interactions. \textit{i.e.}
		\begin{eqnarray}
			H=\sum_{\langle kl\rangle}H(k,l), \qquad \langle kl\rangle\in\{\mathrm{edges}\}. \label{hamisum}
		\end{eqnarray}
		Here the Hamiltonian density $H(k,l)$ describes the interaction between two spins at vertices $k$ and $l$ for a connected graph. The construction for a lattice is similar. Only spins at nearest neighbor pairs $\langle kl \rangle$'s called \textit{bonds} (edges of a graph) interact.
	\item The Hamiltonian density $H(k,l)$ is a sum of several terms. Each term is proportional to a projector. The proportionality coefficients are all positive numbers. \textit{i.e.}
		\begin{eqnarray}
			H(k,l)=\sum_{J}C_{J}(k,l)\,\pi_{J}(k,l). \label{hamiden}
		\end{eqnarray}
		Here $\pi_{J}(k,l)$'s are projectors and $C_{J}(k,l)$'s are positive coefficients. Note that the coefficients may depend on the bond $\langle kl\rangle$ and the projector $\pi_{J}(k,l)$. The projector $\pi_{J}(k,l)$ projects $\boldsymbol{S}_{k}$ and $\boldsymbol{S}_{l}$ on the joint value $J$ of the bond spin $\boldsymbol{J}_{kl}=\boldsymbol{S}_{k}+\boldsymbol{S}_{l}$. (We also call the joint spin value $J$ the \textit{bond spin} value). The meaning is this: The spin $\boldsymbol{S}_{k}$ with spin value $S_{k}$ at site (or vertex) $k$ is a $(2S_{k}+1)$-dimensional representation of the $SU(2)$ Lie algebra, while $\boldsymbol{S}_{l}$ is a $(2S_{l}+1)$-dimensional representation. The direct product of these two representations is reducible to a direct sum of irreducible representations with dimensions $2J+1$ and $J$ runs from $|S_{k}-S_{l}|$ to $S_{k}+S_{l}$. The Hilbert space `splits' into these invariant subspaces labeled by $J$ which is called the \textit{bond spin} value of $\boldsymbol{S}_{k}$ and $\boldsymbol{S}_{l}$. (The eigenvalues of the Casimir operator -- the square of the bond spin $(\boldsymbol{S}_{k}+\boldsymbol{S}_{l})^{2}$ is $J(J+1)$.) The projector $\pi_{J}(k,l)$ projects on the invariant subspace with bond spin $J$. If we choose an orthonormal basis $\{|J,m\rangle\ |\ m=-J,\ldots,J\}$ for the subspace, such that $(\boldsymbol{S}_{k}+\boldsymbol{S}_{l})^{2}|J,m\rangle=J(J+1)|J,m\rangle$ and $(S^{z}_{k}+S^{z}_{l})|J,m\rangle=m|J,m\rangle$, then the projector could be written as
		\begin{eqnarray}
			\pi_{J}(k,l)=\sum^{J}_{m=-J}|J,m\rangle\langle J,m|. \label{projstates}
		\end{eqnarray}
		This form (\ref{projstates}) is cumbersome in practical use and it is preferred to express the projector $\pi_{J}(k,l)$ explicitly in terms of spin operators $\boldsymbol{S}_{k}$ and $\boldsymbol{S}_{l}$. We shall do that in the next section (\S$\,\ref{sec:proj}\,$).
\end{enumerate}
Even without an explicit form of the projectors, an immediate consequence of these two properties is that the Hamiltonian is \textit{positive semi-definite}. \footnote{The Hamiltonian is essentially a sum of projectors with positive coefficients. A projector $\pi$ satisfies $\pi^2=\pi$. So that for an arbitrary state $|\psi\rangle$, we have $\langle\psi|\pi|\psi\rangle=\langle\psi|\pi^{2}|\psi\rangle\geq0$ because it is an inner product of $\pi|\psi\rangle$ with itself.} Furthermore, because of this, if we could construct a state $|\psi\rangle$ which has no projection on any of the specified bond spin-$J$ states appearing in (\ref{hamiden}) for each bond, \textit{i.e.} $\pi_{J}(k,l)|\psi\rangle=0$, $\forall\ \langle kl\rangle$, then it has to be a ground state (with energy equal to zero) regardless of the specific values of the coefficients. \footnote{Some authors add or omit additive constants in the expression of projectors. \textit{e.g.} in (\ref{simphami}) the constant $2/3$ can be dropped. This may shift the ground state energy but does not affect the form of the ground state because the ground state is constructed to have no projection on the specified subspaces for every bond.} The uniqueness condition of the ground state will be discussed later in \S$\,\ref{sec:unique}\,$.

\subsection{The Projector}
\label{sec:proj}

In order to complete the definition of the general AKLT Hamiltonian (\ref{hamisum}) and Hamiltonian density (\ref{hamiden}), we have to give an explicit expression of the projector $\pi_{J}(k,l)$ in terms of spin operators $\boldsymbol{S}_{k}$ and $\boldsymbol{S}_{l}$. We derive the expression in two steps. The forms of $\pi_{J}(k,l)$ for a specific model such as the expression (\ref{simphami}) or those for $1$-dimensional models in \S$\,\ref{sec:1dspin1}\,$, \S$\,\ref{sec:1dssh}\,$ and \S$\,\ref{sec:1dinhom}\,$ can all be obtained through this approach as follows. (An explicit construction of the projector was given in \cite{KK}.)

\begin{enumerate}
	\item Consider the following two sets of operators: the projectors
	\begin{eqnarray}
	\{\pi_{J}(k,l)\ |\ J=|S_{k}-S_{l}|,\ldots,S_{k}+S_{l}\}
	\end{eqnarray}
	 and the powers of the inner product $(\boldsymbol{S}_{k}\cdot\boldsymbol{S}_{l})$
	\begin{eqnarray}
	\{(\boldsymbol{S}_{k}\cdot\boldsymbol{S}_{l})^{n}\ |\ n=0,\ldots,2S_{<}\}, \qquad S_{<}\equiv \mathrm{min}\{S_{k}, S_{l}\}.
	\end{eqnarray}
One set is expressible in terms of the other. They are related by a linear transformation:
		\begin{eqnarray}
			&&(\boldsymbol{S}_{k}\cdot\boldsymbol{S}_{l})^{n} \nonumber\\
			&=&\left(\frac{1}{2}\right)^{n}\left[(\boldsymbol{S}_{k}+\boldsymbol{S}_{l})^{2}-S_{k}(S_{k}+1)-S_{l}(S_{l}+1)\right]^{n}\sum^{S_{k}+S_{l}}_{J=|S_{k}-S_{l}|}\pi_{J}(k,l) \nonumber\\
			&=&\sum^{S_{k}+S_{l}}_{J=|S_{k}-S_{l}|}\left(\frac{1}{2}\right)^{n}\left[J(J+1)-S_{k}(S_{k}+1)-S_{l}(S_{l}+1)\right]^{n}\pi_{J}(k,l) \nonumber\\
		\label{lineartr}
		\end{eqnarray}
	for $n=0,\ldots,2S_{<}$. In (\ref{lineartr}) we have used
 	\begin{eqnarray}
	\sum^{S_{k}+S_{l}}_{J=|S_{k}-S_{l}|}\pi_{J}(k,l)=I
	\end{eqnarray}
 	being the identity. This set of $2S_{<}+1$ linear equations (\ref{lineartr}) can be inverted, which express the projector $\pi_{J}(k,l)$ as a polynomial of the inner product $(\boldsymbol{S}_{k}\cdot\boldsymbol{S}_{l})$. We shall not pursue with the inversion of (\ref{lineartr}) but to construct the projector in the next step.
	\item The next step is to realize that if an operator $\mathcal{P}(k,l)$ satisfies the following conditions
		\begin{eqnarray}
			\mathcal{P}(k,l)\pi_{J'}(k,l)=\delta_{JJ'}\pi_{J}(k,l), \qquad \forall \ J' \label{cond}
		\end{eqnarray}
		then the operator $\mathcal{P}(k,l)$ is identified with $\pi_{J}(k,l)$ because
		\begin{eqnarray}
			\mathcal{P}(k,l)&=&\mathcal{P}(k,l)\sum^{S_{k}+S_{l}}_{J'=|S_{k}-S_{l}|}\pi_{J'}(k,l) \nonumber\\
			&=&\left(\sum^{S_{k}+S_{l}}_{J'=|S_{k}-S_{l}|}\delta_{JJ'}\right)\pi_{J}(k,l)=\pi_{J}(k,l).
		\end{eqnarray}
		Therefore we could construct an operator satisfying the condition (\ref{cond}):
		\begin{eqnarray}
			\mathcal{P}(k,l)=\prod^{j\neq J}_{|S_{k}-S_{l}|\leq j\leq S_{k}+S_{l}}\frac{(\boldsymbol{S}_{k}+\boldsymbol{S}_{l})^{2}-j(j+1)}{J(J+1)-j(j+1)}. \label{piproj}
		\end{eqnarray}
		When $\mathcal{P}(k,l)$ acting on $\pi_{J'}(k,l)$, we have:
		\begin{enumerate}
			\item If $J'\neq J$, then the numerator of one factor in the product vanishes, so that $\mathcal{P}$ vanishes. \textit{i.e.} $\mathcal{P}(k,l)\pi_{J'}(k,l)=0$, if $J'\neq J$;
			\item If $J'=J$, all factors in the product become equal to $1$, so as the expression $\mathcal{P}$. \textit{i.e.} $\mathcal{P}(k,l)\pi_{J}(k,l)=\pi_{J}(k,l)$.
		\end{enumerate}
		So that (\ref{piproj}) is the projector $\pi_{J}(k,l)$, \textit{i.e.} $\pi_{J}(k,l)=\mathcal{P}(k,l)$. Operator 
\begin{eqnarray}
	\pi_{J}(k,l)=\prod^{j\neq J}_{|S_{k}-S_{l}|\leq j\leq S_{k}+S_{l}}\frac{(\boldsymbol{S}_{k}+\boldsymbol{S}_{l})^{2}-j(j+1)}{J(J+1)-j(j+1)} \label{pi}
\end{eqnarray}
projects the bond spin $\boldsymbol{J}_{kl}\equiv\boldsymbol{S}_{k}+\boldsymbol{S}_{l}$ on the subspace with fixed total spin value $J$ and $|S_{k}-S_{l}|\leq J\leq S_{k}+S_{l}$. Note that we could expand $(\boldsymbol{S}_{k}+\boldsymbol{S}_{l})^{2}=2\boldsymbol{S}_{k}\cdot\boldsymbol{S}_{l}+S_{k}(S_{k}+1)+S_{l}(S_{l}+1)$. Therefore projector $\pi_{J}(k,l)$ in (\ref{pi}) is a polynomial in the scalar product $(\boldsymbol{S}_{k}\cdot\boldsymbol{S}_{l})$ of degree $2S_{<}$, where $S_{<}\equiv \mathrm{min}\{S_{k}, S_{l}\}$ is the minimum of the two spin values of the same bond $\langle kl\rangle$. For example with $S_{k}=S_{l}=1$, we may have a quadratic polynomial as in (\ref{simphami}):
\begin{eqnarray}
	\pi_{2}(k,l)=\frac{1}{6}(\boldsymbol{S}_{k}\cdot\boldsymbol{S}_{l})^{2}+\frac{1}{2}(\boldsymbol{S}_{k}\cdot\boldsymbol{S}_{l})+\frac{1}{3}. \label{examplepi2}
\end{eqnarray}
\end{enumerate}


\subsection{The Basic AKLT Model}
\label{sec:basic}

\subsubsection{The Hamiltonian}
\label{sec:basichamil}

Once we have the building blocks for the Hamiltonian from \S$\,\ref{sec:hamil}\,$ and \S$\,\ref{sec:proj}\,$, various types of the AKLT model can be constructed. Let us start with the definition of the \textit{basic} AKLT model on a connected graph or lattice. (Any lattice is a special graph with periodic structure; our notations and definitions refer to the most general). A \textit{graph} consists of two types of elements, namely \textit{vertices} and \textit{edges}. Every edge \textit{connects} two vertices. As in Figure (\ref{figure1}), a vertex is drawn as a (large) circle and an edge is drawn as a solid line connecting two vertices. For every pair of vertices in the \textit{connected} graph, there is a \textit{walk} \footnote{A \textit{walk} is an alternating sequence of vertices and edges, beginning and ending with a vertex, in which each vertex is incident to the two edges that precede and follow it in the sequence, and the vertices that precede and follow an edge are the endvertices of that edge.} from one to the other. Vertices can also be called \textit{sites} and edges sometimes called \textit{links} or \textit{bonds}. In a graph, a pair of vertices connected by an edge is regarded nearest neighbors, \textit{i.e.} the terms edge, bond and nearest neighbor are equivalent and interchangeable. (For a lattice, vertices become sites and bonds become lattice vectors connecting nearest neighboring sites.) In the case of a disconnected graph, the Hamiltonian (\ref{hamisum}) is a direct sum with respect to connected components and the ground state is a direct product. Therefore we shall need only to study a connected graph. Also, assuming that the graph consists of more than one vertices to avoid the trivial case where there would be no interaction at all.

Let us introduce mathematical notations. By $\boldsymbol{S}_{l}$ we shall denote the spin operator located at vertex $l$ with spin value $S_{l}$. In the \textit{basic} model we require that $S_{l}=z_{l}/2$, where $z_{l}$ is the number of incident edges connected to vertex $l$, also known as the \textit{valence} or \textit{coordination number} (the number of nearest neighbors of the vertex $l$). The relation between the spin value and coordination number must be true for any vertex $l$, including boundaries. This will guarantee the uniqueness of the ground state, see \S$\,\ref{sec:unique}\,$. For a lattice, this condition would also yield that bulk spins (spins not on the boundary) take the same value $z/2$ because the number of nearest neighbors $z$ is a constant.

In the \textit{basic} model we define the Hamiltonian density $H(k,l)$ for bond (edge) $\langle kl\rangle$ as
\begin{eqnarray}
	H(k,l)=C(k,l)\,\pi_{S_{k}+S_{l}}(k,l), \qquad H(k,l)\geq 0 \label{basichd}
\end{eqnarray}
with $C(k,l)$ an arbitrary positive real coefficient (it may depend on the bond $\langle kl\rangle$). So that the Hamiltonian density for each bond (edge) is proportional to the projector on the subspace with the highest possible bond spin value $(S_{k}+S_{l})$. The physical meaning is that interacting spins do not form the highest possible bond spin (this will increase the energy) in the ground state. Then the basic AKLT Hamiltonian on an arbitrary connected graph according to (\ref{hamisum}) is
\begin{eqnarray}
	H=\sum_{\langle kl\rangle}H(k,l)=\sum_{\langle kl\rangle}C(k,l)\,\pi_{S_{k}+S_{l}}(k,l). \label{basich}
\end{eqnarray}
Here we sum over all bonds (edges) $\langle kl\rangle$. Note that for a lattice all the highest bond spin values for bulk spins are the same and equal to the coordination number $z$. For example, the basic model defined on a $2$-dimensional square lattice must have spin-$2$ in each bulk vertex and $\pi_{4}$ in the Hamiltonian; also, the Hexagonal lattice has spin-$3/2$ in each vertex and $\pi_{3}$ in the Hamiltonian.

\subsubsection{The VBS State -- Pictorial Method}
\label{sec:graphvbs}

In this section we consider the Hamiltonian (\ref{basich}) and construct a ground state which is denoted by $|\mathrm{VBS}\rangle$. Later we shall see in \S$\,\ref{sec:unique}\,$ that it is the unique ground state.
 
\begin{figure}
	\centering
		\includegraphics[width=4in]{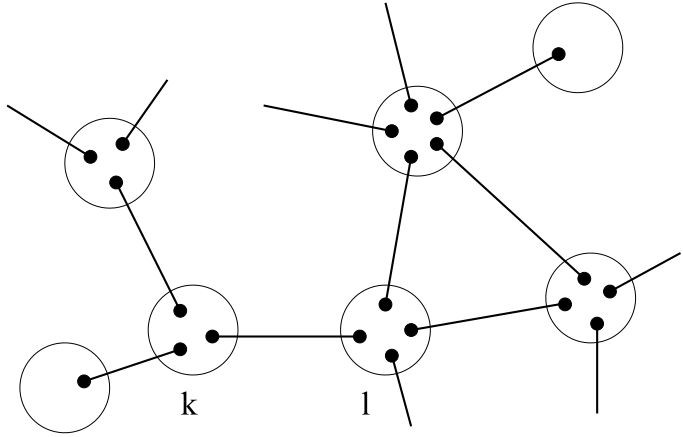}
	\caption{Example of a part of the graph including vertex $k$ with $z_{k}=3$ and vertex $l$ with $z_{l}=4$. Black dots represent spin-$1/2$ states, which are enclosed by large circles representing vertices and symmetrization of the product of spin-$1/2$'s at each vertex. Solid lines represent edges (bonds) which anti-symmetrize the pair of connected spin-$1/2$'s.}
	\label{figure1}
\end{figure}

The Hamiltonian (\ref{basich}) with condition
\begin{eqnarray}
	S_{l}=\frac{1}{2}z_{l} \label{basiccond}
\end{eqnarray}
has a unique ground state \cite{AKLT1,AKLT2,AAH,KK} known as the Valence-Bond-Solid (VBS) state. It can be constructed in a Pictorial way as follows (see Figure \ref{figure1}).
Each vertex $l$ is represented by $z_{l}$ spin-$1/2$'s. We associate each spin-$1/2$ with an incident edge. In such a way each edge has two spin-$1/2$'s at its ends. We anti-symmetrize the spin states of these two spin-$1/2$'s. So that anti-symmetrization is done along each edge. These anti-symmetrizations ensure that there is no projection on the highest bond spin states for every bond. Then we also symmetrize the product of spin-$1/2$'s at each vertex (each large circle in Figure \ref{figure1}). These symmetrizations preserve the correct spin value at each vertex. 

Let us write down the VBS ground state algebraically following this approach. We label the particular dot from vertex $l$ connected with some dot from vertex $k$ by $l_{k}$ (correspondingly, that dot from vertex $k$ is labeled by $k_{l}$). In this way we have specified a unique prescription of indices with dots. Then the anti-symmetrization results in the singlet state
\begin{eqnarray}
	|\Phi\rangle_{kl}=\frac{1}{\sqrt{2}}\left(|\uparrow\rangle_{l_{k}}|\downarrow\rangle_{k_{l}}-|\downarrow\rangle_{l_{k}}|\uparrow\rangle_{k_{l}}\right), \label{antisymm}
\end{eqnarray}
where $|\uparrow\rangle$ (or $\downarrow\rangle$) denotes spin up (or down) states referring to a basis. The direct product of all these $|\Phi\rangle$ singlet states corresponds to all edges (bonds) in our Figure \ref{figure1}:
\begin{eqnarray}
	\prod_{\langle kl\rangle}|\Phi\rangle_{kl}. \label{prodsing}
\end{eqnarray}
We still have to complete the symmetrization (circles) at each vertex. We denote the symmetrization operator of $z_{l}$ black dots in vertex $l$ by $\mathbf{P}(l)$. The action of $\mathbf{P}(l)$ on any product of $z_{l}$ spin-$1/2$'s is
\begin{eqnarray}
	&&\mathbf{P}(l)|\chi_{l_{k_{1}}}\rangle_{l_{k_{1}}}|\chi_{l_{k_{2}}}\rangle_{l_{k_{2}}}\cdots|\chi_{l_{k_{z_{l}}}}\rangle_{l_{k_{z_{l}}}} \label{symmtri}\\
&=&\sum^{z_{l}!\ \mathrm{terms}}_{\sigma}|\chi_{l_{k_{\sigma(1)}}}\rangle_{l_{k_{1}}}|\chi_{l_{k_{\sigma(2)}}}\rangle_{l_{k_{2}}}\cdots|\chi_{l_{k_{\sigma(z_{l})}}}\rangle_{l_{k_{z_{l}}}}, \qquad \chi=\uparrow \mathrm{or} \downarrow \nonumber
\end{eqnarray}
where $k_{1}, k_{2}, \ldots, k_{z_{l}}$ are the $z_{l}$ spin-$1/2$'s (block dots) belonging to vertex $l$ (the index also refers to the $z_{l}$ vertices connected to vertex $l$ by an edge) and $\sigma$ is a permutation of the indices:
\begin{eqnarray}
	\sigma=\left(\begin{array}{cccc} 1 & 2 & \cdots & z_{l} \\ \sigma(1) & \sigma(2) & \cdots & \sigma(z_{l})\end{array}\right).
\end{eqnarray}
All permutations are summed up in (\ref{symmtri}). Then the symmetrization at each vertex is carried out by taking the product $\prod_{l}\mathbf{P}(l)$ of all vertices. Finally, the unique VBS ground state can be written as
\begin{eqnarray}
	|\mathrm{VBS}\rangle=\prod_{l}\mathbf{P}(l)\prod_{\langle kl\rangle}|\Phi\rangle_{kl}. \label{basicvbs}
\end{eqnarray}
Here the first product runs over all vertices and the second over all edges (bonds). So that we have constructed a ground state of the AKLT Hamiltonian (\ref{basich}) such that
\begin{eqnarray}
	H|\mathrm{VBS}\rangle=0, \qquad \pi_{S_{k}+S_{l}}(k,l)|\mathrm{VBS}\rangle=0, \qquad \forall\ \langle kl\rangle. \label{hvbs=0}
\end{eqnarray}
Note that the VBS state in (\ref{basicvbs}) is not normalized in general. If the coordination number $z_{l}$ is a constant over all vertices in the graph except for boundaries (such as in the case of a lattice), then we would have the same spin value at each bulk vertex. In that case the \textit{basic} model is also referred to as the \textit{homogeneous} model.

\subsection{The Generalized AKLT Model}
\label{sec:generalized}

\subsubsection{The Hamiltonian}
\label{sec:generalhamil}

In the \textit{generalized} AKLT model, relation (\ref{basiccond}) is generalized. We associate a positive integer $M_{kl}$ ($M_{kl}\equiv M_{lk}$) to each edge $\langle kl\rangle$ of the graph (or each bond of a lattice). We shall call $M_{kl}$ \textit{multiplicity numbers}. Similar to the basic model, the AKLT Hamiltonian describes interactions between nearest neighbors (vertices connected by an edge):
\begin{eqnarray}
	H=\sum_{\langle kl\rangle}H(k,l). \label{generalizedh}
\end{eqnarray}
However, the Hamiltonian density is no longer proportional to a single projector as in (\ref{basichd}) in general. Instead, it is a linear combination of projectors
\begin{eqnarray}
	H(k,l)=\sum^{S_{k}+S_{l}}_{J=S_{k}+S_{l}-M_{kl}+1}C_{J}(k,l)\,\pi_{J}(k,l), \label{generalizedhd} \\
	1\leq M_{kl}\leq 2S_{<}, \qquad S_{<}\equiv \mathrm{min}\{S_{k}, S_{l}\}. \nonumber 
\end{eqnarray}
Projector $\pi_{J}(k,l)$ is given by (\ref{pi}), and $C_{J}(k,l)$'s are arbitrary positive coefficients. So that $H(k,l)$ projects the bond spin on the subspace with bond spin value $J$ greater than $S_{k}+S_{l}-M_{kl}$. Physically formation of any bond spin higher than $S_{k}+S_{l}-M_{kl}$ would increase the energy.

The condition of uniqueness of the ground state was introduced by Kirillov and Korepin in \cite{KK}:
\begin{eqnarray}
	2S_{l}=\sum_{k}M_{kl}, \qquad \forall \ l. \label{condition1}
\end{eqnarray}
Here $S_{l}$ is the spin value at vertex $l$ and we sum over all edges incident to vertex $l$ (connected to vertex $l$). \textit{i.e.} $k$ are nearest neighbors. The Hamiltonian (\ref{generalizedh}) has a unique ground state if (\ref{condition1}) is valid, see \S$\,\ref{sec:unique}\,$. The relation $2S_{l}=z_{l}$ for the \textit{basic} model is a special case when all $M_{kl}=1$. Also, when $M_{kl}=1$, the Hamiltonian density (\ref{generalizedhd}) reduces to that of the basic model (\ref{basichd}). The condition (\ref{condition1}) can be put into an invariant form. Let us define a column vector $\mathbf{S}$, the $l^{\mathrm{th}}$ component of which is associated with vertex $l$ of the graph and equal to $S_{l}$. The number of components is equal to the number of vertices $N$ in the whole graph. Next, we define another column vector $\mathbf{M}$ with its dimension equal to the number of edges $M$ in the graph. The $k^{\mathrm{th}}$ and $l^{\mathrm{th}}$ components of this vector are associated with edge $\langle kl\rangle$ and both equal to $M_{kl}$. The most important geometrical characteristic of the graph is the vertex-edge incidence matrix $\hat{\mathbf{I}}$ (see \cite{Har}). This is a rectangular matrix with $N$ rows and $M$ columns. Each row is associated with the vertex and each column is associated with the edge. If the vertex belongs to the edge the corresponding matrix element is equal to one, otherwise zero. Then the condition (\ref{condition1}) of uniqueness can be re-written as
\begin{eqnarray}
	2\,\mathbf{S}=\hat{\mathbf{I}}\cdot\mathbf{M}. \label{condition2} 
\end{eqnarray}
For more details we refer to \cite{KK}.

\subsubsection{The VBS State -- Schwinger Boson Representation}
\label{sec:vbssch}

Under condition (\ref{condition1}) or (\ref{condition2}), the unique ground state of Hamiltonian (\ref{generalizedh}) is referred to as the generalized VBS state. It is constructed by introducing the Schwinger boson representation \cite{AAH,FM,KK,KHH,XKHK,XKHK2,XK}. This method uses a pair of bosonic creation and annihilation operators (similar to the treatment of the harmonic oscillator problem) to realize the $SU(2)$ Lie algebra.

Define a pair of independent canonical bosonic operators $a_{l}$ and $b_{l}$ for each vertex (or site) $l$:
\begin{eqnarray}
	[\,a_{k}\,,\,a^{\dagger}_{l}\,]=[\,b_{k}\,,\,b^{\dagger}_{l}\,]=\delta_{kl} \label{commutator}
\end{eqnarray}
with all other commutators vanishing:
\begin{eqnarray}
	[\,a_{k}\,,\,a_{l}\,]=[\,b_{k}\,,\,b_{l}\,]=[\,a_{k}\,,\,b_{l}\,]=[\,a_{k}\,,\,b^{\dagger}_{l}\,]=0, \qquad \forall\ k, l. \label{commvani}
\end{eqnarray}
Spin operators are represented as 
\begin{eqnarray}
	S^{+}_{l}=a^{\dagger}_{l}b_{l}, \qquad S^{-}_{l}=b^{\dagger}_{l}a_{l}, \qquad S^{z}_{l}=\frac{1}{2}(a^{\dagger}_{l}a_{l}-b^{\dagger}_{l}b_{l}). \label{spinrep}
\end{eqnarray}
It is straightforward to verify that the $SU(2)$ Lie algebra is realized. To reproduce the dimension of the spin-$S_{l}$ Hilbert space at vertex $l$, a constraint on the total boson occupation number is required:
\begin{eqnarray}
	\hat{S}_{l}\equiv\frac{1}{2}(a^{\dagger}_{l}a_{l}+b^{\dagger}_{l}b_{l})=S_{l}. \label{constraintspin}
\end{eqnarray}
\textit{i.e.} any physical spin state $|\psi\rangle_{l}$ at vertex $l$ must satisfy $\hat{S}_{l}|\psi\rangle_{l}=S_{l}|\psi\rangle_{l}$. In this framework the spin state $|S_{l},m_{l}\rangle_{l}$ such that $\boldsymbol{S}^{2}_{l}|S_{l},m_{l}\rangle_{l}=S_{l}(S_{l}+1)|S_{l},m_{l}\rangle_{l}$ and $S^{z}_{l}|S_{l},m_{l}\rangle_{l}=m_{l}|S_{l},m_{l}\rangle_{l}$ is represented by
\begin{eqnarray}
	|S_{l},m_{l}\rangle_{l}=\frac{(a^{\dagger}_{l})^{S_{l}+m_{l}}(b^{\dagger}_{l})^{S_{l}-m_{l}}}{\sqrt{(S_{l}+m_{l})!(S_{l}-m_{l})!}}\,|\mathrm{vac}\rangle_{l}, \label{spinstatesch}
\end{eqnarray}
where the \textit{vacuum} $\left|\mathrm{vac}\right\rangle_{l}$ is annihilated by any of the annihilation operators:
\begin{eqnarray}
	a_{l}\left|\mathrm{vac}\right\rangle_{l}=b_{l}\left|\mathrm{vac}\right\rangle_{l}=0. \label{vacuum}
\end{eqnarray}
As a result, the VBS ground state in the Schwinger representation is constructed as
\begin{eqnarray}
	|\mathrm{VBS}\rangle=
 \prod_{\langle kl\rangle}
\left(a^{\dagger}_{k}b^{\dagger}_{l}-b^{\dagger}_{k}a^{\dagger}_{l}\right)^{M_{kl}}|\mathrm{vac}\rangle. \label{generalizedvbs}
\end{eqnarray}
It worth mentioning that this representation shows that for a full graph (each vertex is connected to every other vertex by definition) the VBS state coincides with the Laughlin wave function \cite{AAH,HR0,HZS}. In (\ref{generalizedvbs}) the product runs over all bonds (edges) and the vacuum $|\mathrm{vac}\rangle$ is the direct product of vacuums of each vertex
\begin{eqnarray}
	|\mathrm{vac}\rangle=\bigotimes_{l}|\mathrm{vac}\rangle_{l},
\end{eqnarray}
which is destroyed by any annihilation operators $a_{l}$ or $b_{l}$, $\forall\ l$. (Note that $[\,a^{\dagger}_{k}\,,\,b^{\dagger}_{l}\,]=0$, $\forall\ k,l$.)

To prove that (\ref{generalizedvbs}) is the ground state we need only to verify for any vertex $l$ and bond (edge) $\langle kl\rangle$:
\begin{enumerate}
	\item The total power of $a^{\dagger}_{l}$ and $b^{\dagger}_{l}$ is $2S_{l}$, so that we have spin-$S_{l}$ at vertex $l$;
	\item The $z$-component of the bond spin satisfies
	\begin{eqnarray}
	-\frac{1}{2}(\sum_{l^{\prime}\neq l}M_{l^{\prime}k}+\sum_{k^{\prime}\neq k}M_{k^{\prime}l})\leq J^{z}_{kl}\equiv S^{z}_{k}+S^{z}_{l}\leq \frac{1}{2}(\sum_{l^{\prime}\neq l}M_{l^{\prime}k}+\sum_{k^{\prime}\neq k}M_{k^{\prime}l})\nonumber\\
	\end{eqnarray}
	by a binomial expansion. Consequently, the maximum value of the bond spin $J_{kl}$ is $(\sum_{l^{\prime}\neq l}M_{l^{\prime}k}+\sum_{k^{\prime}\neq k}M_{k^{\prime}l})/2=S_{k}+S_{l}-M_{kl}$ (from $SU(2)$ invariance, see \S$\,\ref{sec:larges}\,$ and \cite{AAH}).
\end{enumerate}
Therefore, the state $|\mathrm{VBS}\rangle$ defined in (\ref{generalizedvbs}) has spin-$S_{l}$ at vertex $l$ and no projection onto the $J_{kl}>S_{k}+S_{l}-M_{kl}$ subspace for any bond (edge). As a consequence,
\begin{eqnarray}
	H|\mathrm{VBS}\rangle=0, \qquad \pi_{J}(k,l)|\mathrm{VBS}\rangle=0,\nonumber\\
	S_{k}+S_{l}-M_{kl}+1\leq J\leq S_{k}+S_{l}, \qquad \forall\ \langle kl\rangle. \label{hgvbs=0}
\end{eqnarray}
The introduction of Schwinger bosons can be used to construct a spin coherent state basis (as expected due to the similarity with the harmonic oscillator) in which spins at each vertex behave as classical unit vectors, see \S$\,\ref{sec:cohe}\,$ and \cite{AAH,FM,KK,KHH,XKHK,XKHK2}. The coherent state basis converts algebraic computations into classical integrals which becomes extremely useful in later sections.

\subsection{The Uniqueness Condition}
\label{sec:unique}

As stated in previous sections, the condition for the existence of a unique VBS ground state is $2S_{l}=z_{l}$ for the basic model and $2S_{l}=\sum_{k}M_{kl}$ for the generalized model (the former being a special case of the latter). This uniqueness condition for the AKLT model defined on a finite graph or lattice was proved in \cite{KK}. 


Now let us turn to the proof  of the uniqueness condition $2S_{l}=\sum_{k}M_{kl}$ , \textit{i.e.} the equation
\begin{eqnarray}
	H|\Psi\rangle=0 \label{condgrou}
\end{eqnarray}
with $H$ the AKLT Hamiltonian (\ref{generalizedh}) has exactly one solution under the condition (\ref{condition1}) or (\ref{condition2}). Note that this expression (\ref{condgrou}) is equivalent to 
\begin{eqnarray}
	\pi_{J}(k,l)|\Psi\rangle=0, \qquad \forall\ \langle k,l\rangle, \qquad S_{k}+S_{l}-M_{kl}+1\leq J\leq S_{k}+S_{l} \label{condgrou1}
\end{eqnarray}
because of the positive semi-definiteness of every projector $\pi_{J}$ and the positive coefficients $C_{J}$.
In order to prove the uniqueness condition, we first prove the following lemma.

\begin{lemma}
\label{lemma1}
All solutions of the equation
\begin{eqnarray}
	\pi_{J}(k,l)|\psi\rangle=0, \qquad S_{k}+S_{l}-M_{kl}+1\leq J\leq S_{k}+S_{l} \label{pipsi0}
\end{eqnarray}
for fixed $k$ and $l$ can be represented in the following form
\begin{eqnarray}
	|\psi\rangle=f(a^{\dagger},b^{\dagger})(a^{\dagger}_{k}b^{\dagger}_{l}-a^{\dagger}_{l}b^{\dagger}_{k})^{M_{kl}}|\mathrm{vac}\rangle. \label{psicom}
\end{eqnarray}
Here $f(a^{\dagger},b^{\dagger})$ is some polynomial in $a^{\dagger}_{k}$, $b^{\dagger}_{k}$ and $a^{\dagger}_{l}$, $b^{\dagger}_{l}$.
\end{lemma}

\begin{proof}
For convenience, let us apply the Weyl representation of the $SU(2)$ Lie algebra. Consider the space of polynomials in pairs of variables $x_{l}$ and $y_{l}$ with coefficients in $\mathbf{C}$. We represent operator $a^{\dagger}_{l}$ as multiplication on $x_{l}$ and $b^{\dagger}_{l}$ as multiplication on $y_{l}$. At site $l$ we have
\begin{eqnarray}
	S^{+}_{l}=x_{l}\frac{\partial}{\partial y_{l}}, \qquad\qquad\qquad S^{-}_{l}=y_{l}\frac{\partial}{\partial x_{l}} \nonumber\\
	2S^{z}_{l}=x_{l}\frac{\partial}{\partial x_{l}}-y_{l}\frac{\partial}{\partial y_{l}}, \qquad 2\hat{S}_{l}=x_{l}\frac{\partial}{\partial x_{l}}+y_{l}\frac{\partial}{\partial y_{l}}.
\end{eqnarray}
A basis of the $(2S_{l}+1)$-dimensional irreducible representation of spin-$S_{l}$ can be chosen in such a way:
\begin{eqnarray}
	\mathbf{V}_{S_{l}}=\{x^{S_{l}+m_{l}}_{l}y^{S_{l}-m_{l}}_{l}\ |\ m=-S,\ldots,S\}.
\end{eqnarray}
These monomials with total power $2S_{l}$ are clearly eigenvectors of $S^{z}_{l}$ and $\hat{S}_{l}$. Now let us consider the tensor product of two irreducible representations $\mathbf{V}_{S_{l}}\otimes\mathbf{V}_{S_{k}}$. Define the bond spin $\boldsymbol{J}_{kl}\equiv\boldsymbol{S}_{k}+\boldsymbol{S}_{l}$, then
\begin{eqnarray}
	J^{+}_{kl}&=&x_{k}\frac{\partial}{\partial y_{k}}+x_{l}\frac{\partial}{\partial y_{l}}, \nonumber\\
	J^{-}_{kl}&=&y_{k}\frac{\partial}{\partial x_{k}}+y_{l}\frac{\partial}{\partial x_{l}}, \nonumber\\
	2J^{z}_{kl}&=&x_{k}\frac{\partial}{\partial x_{k}}+x_{l}\frac{\partial}{\partial x_{l}}-y_{k}\frac{\partial}{\partial y_{k}}-y_{l}\frac{\partial}{\partial y_{l}}, \nonumber\\
	2\hat{J}_{kl}&=&x_{k}\frac{\partial}{\partial x_{k}}+x_{l}\frac{\partial}{\partial x_{l}}+y_{k}\frac{\partial}{\partial y_{k}}+y_{l}\frac{\partial}{\partial y_{l}}.
\end{eqnarray}
The tensor product of irreducible representation can be reduced to a direct sum of irreducible representations
\begin{eqnarray}
	\mathbf{V}_{S_{k}}\otimes\mathbf{V}_{S_{l}}=\bigoplus^{S_{k}+S_{l}}_{J=|S_{k}-S_{l}|}\mathbf{V}_{J}.
\end{eqnarray}
Now we construct the highest vector (polynomial) $v_{J}$ of irreducible representation $\mathbf{V}_{J}$ with fixed $J$:
\begin{eqnarray}
	J^{+}_{kl}v_{J}=0, \qquad J^{z}_{kl}v_{J}=Jv_{J}, \qquad \hat{J}_{kl}v_{J}=(S_{k}+S_{l})v_{J}. \label{sativj}
\end{eqnarray}
It must have a total power $2(S_{k}+S_{l})$, so that the form can be taken as
\begin{eqnarray}
	v_{J}=\sum_{m_{k}+m_{l}=J}C_{m_{k}m_{l}}x^{S_{l}+m_{l}}_{l}y^{S_{l}-m_{l}}_{l}x^{S_{k}+m_{k}}_{k}y^{S_{k}-m_{k}}_{k}. \label{vjform}
\end{eqnarray}
This form already satisfies the second and third equations in (\ref{sativj}). After rearranging terms (relabeling indices), the first equation of (\ref{sativj}) becomes
\begin{eqnarray}
	J^{+}_{kl}v_{J}&=&\sum^{J-1}_{m_{k}=0}\left[(S_{k}-m_{k})C_{m_{k},J-m_{k}}+(S_{l}-J+m_{k}+1)C_{m_{k}+1,J-m_{k}-1}\right]
	\nonumber\\
	&&\cdot x^{S_{k}+m_{k}+1}_{k}y^{S_{k}-m_{k}-1}_{k}x^{S_{l}+J-m_{k}}_{l}y^{S_{l}-J+m_{k}}_{l}. \label{jplusvj}
\end{eqnarray}  
Because of the linear independence of the monomials appearing in (\ref{jplusvj}), the coefficients must vanish, which yields the following recurrence relation
\begin{eqnarray}
	C_{m_{k}+1,J-m_{k}-1}=-\frac{S_{k}-m_{k}}{S_{l}-J+m_{k}+1}\,C_{m_{k},J-m_{k}}. \label{recurcc}
\end{eqnarray}
The solution to (\ref{recurcc}) in terms of $C_{0,J}$ is
\begin{eqnarray}
	C_{m_{k},J-m_{k}}=\frac{(-1)^{S_{k}-m_{k}}\left(\begin{array}{c} S_{k}+S_{l}-J \\ S_{k}-m_{k} \end{array}\right)}{(-1)^{S_{k}}\left(\begin{array}{c} S_{k}+S_{l}-J \\ S_{k} \end{array}\right)}\,C_{0,J}. \label{solutioncc}
\end{eqnarray}
Therefore by substituting (\ref{solutioncc}) into (\ref{vjform}) and recognizing a binomial expansion, the form of $v_{J}$ is found to be
\begin{eqnarray}
	v_{J}&=&\frac{C_{0,J}}{(-1)^{S_{k}}\left(\begin{array}{c} S_{k}+S_{l}-J \\ S_{k} \end{array}\right)}\,x^{S_{k}-S_{l}+J}_{k}x^{S_{l}-S_{k}+J}_{l}(x_{k}y_{l}-x_{l}y_{k})^{S_{k}+S_{l}-J} \nonumber\\ \\
	&\propto& x^{2S_{k}-M}_{k}x^{2S_{l}-M}_{l}(x_{k}y_{l}-x_{l}y_{k})^{M}, \qquad M=S_{k}+S_{l}-J. \label{vjsolu}
\end{eqnarray}
The over all constant factor is irrelevant. All other vectors of representation $\mathbf{V}_{J}$ can be obtained from the highest vector $v_{J}$ by applications of operator $J^{-}_{kl}$. Notice that $J^{-}_{kl}$ commutes with the factor $(x_{k}y_{l}-x_{l}y_{k})$
\begin{eqnarray}
	[\,J^{-}_{kl}\,,\,x_{k}y_{l}-x_{l}y_{k}\,]=0. \label{commjminus}
\end{eqnarray}
So that all vectors of representation $\mathbf{V}_{J}$ are divisible by
\begin{eqnarray}
	(x_{k}y_{l}-x_{l}y_{k})^{M}, \qquad M=S_{k}+S_{l}-J. \label{divisor}
\end{eqnarray}
In other words, any vector (polynomial in $x_{k}$, $y_{k}$ and $x_{l}$, $y_{l}$) in the vector space spanned by $\mathbf{V}_{J}$ has a common factor (\ref{divisor}). As a consequence, if there is no projection on the states with bond spin values
\begin{eqnarray}
	S_{k}+S_{l}-M_{kl}+1\leq J\leq S_{k}+S_{l} \label{noproj}
\end{eqnarray}
after summation of spins $\boldsymbol{S}_{k}$ and $\boldsymbol{S}_{l}$ (\textit{i.e.} no projection on $\mathbf{V}_{J}$ with $S_{k}+S_{l}-M_{kl}+1\leq J\leq S_{k}+S_{l}$), then a factor
\begin{eqnarray}
	(x_{k}y_{l}-x_{l}y_{k})^{M_{kl}} \label{factorcom}
\end{eqnarray}
can be isolated. \textit{i.e.} Any vector in $\sum^{S_{k}+S_{l}-M_{kl}}_{J=|S_{k}-S_{l}|}\mathbf{V}_{J}$ would have a common factor (\ref{factorcom}). Moreover, this fact is independent of whether we are using the Weyl representation or the Schwinger representation of the Lie algebra. Therefore, any solution to (\ref{pipsi0}) must take the form of (\ref{psicom}) with the factor $(a^{\dagger}_{k}b^{\dagger}_{l}-a^{\dagger}_{l}b^{\dagger}_{k})^{M_{kl}}$ isolated. Thus we have proved \textbf{Lemma} \ref{lemma1}.
\end{proof}

Now let us use \textbf{Lemma} \ref{lemma1} to prove the uniqueness condition (\ref{condition1}) or (\ref{condition2}). Note that (\ref{pipsi0}) is valid for each bond $\langle kl\rangle$, consequently any ground state $|\Psi\rangle$ of the Hamiltonian satisfying (\ref{condgrou}) and (\ref{condgrou1}) can be presented in the form
\begin{eqnarray}
	|\Psi\rangle=\prod_{\langle kl\rangle}F(a^{\dagger},b^{\dagger})(a^{\dagger}_{k}b^{\dagger}_{l}-a^{\dagger}_{l}b^{\dagger}_{k})^{M_{kl}}|\mathrm{vac}\rangle, \label{groufac}
\end{eqnarray} 
where $F(a^{\dagger},b^{\dagger})$ is some polynomial in $a^{\dagger}$'s and $b^{\dagger}$'s. Now we have to make sure that in (\ref{groufac}) each vertex (site) should have the correct spin value. By applying $2\hat{S}_{l}=(a^{\dagger}_{l}a_{l}+b^{\dagger}_{l}b_{l})$ to the state $|\Psi\rangle$, we realize that the explicit factor in (\ref{groufac}) contribute to $2S_{l}$ (denoting the eigenvalue of $2\hat{S}_{l}$) exactly the value $\sum_{\langle kl\rangle}M_{kl}$ which is the sum of powers of $a^{\dagger}_{l}$ and $b^{\dagger}_{l}$. A comparison with expression (\ref{condition1}) or (\ref{condition2}) shows that if we require this condition $2S_{l}=\sum_{\langle kl\rangle}M_{kl}$, then each site would already have the correct spin value with the presence in (\ref{groufac}) of the explicit factor only. Therefore the degree of the polynomial in variables $a^{\dagger}_{l}$ and $b^{\dagger}_{l}$ is zero. This is true for every site $l$. Therefore the polynomial $F(a^{\dagger},b^{\dagger})$ is a constant which can be removed. So that we have proved that the uniqueness condition (\ref{condition1}) or (\ref{condition2}) guarantees the existence (through explicit construction in \S$\,\ref{sec:vbssch}\,$) and uniqueness of an energy ground state -- the VBS state.

\section{The Subsystem and Measures of Entanglement}
\label{sec:subandmeas}

The VBS states constructed in previous sections \S$\,\ref{sec:basic}\,$ and \S$\,\ref{sec:generalized}\,$ as ground states of AKLT models are highly entangled states. The quantification of entanglement is our main subject of study.

\subsection{The Block Density Matrix and the Block Hamiltonian}
\label{sec:block}

\subsubsection{The Block Density Matrix and Entropies}
\label{sec:rhoands}

The VBS state (see (\ref{basicvbs}) and (\ref{generalizedvbs})) has non-trivial entanglement properties. The density matrix of the VBS state is a projector (a pure state density matrix):
\begin{eqnarray}
	\boldsymbol{\rho}=\frac{|\mathrm{VBS}\rangle\langle\mathrm{VBS}|}{\langle\mathrm{VBS}|\mathrm{VBS}\rangle}. \label{purematrix}
\end{eqnarray}
In order to analyze the entanglement, let us cut the original graph (lattice) into two subgraphs (sublattices) $B$ and $E$. That is, we cut through some edges (bonds) such that the resulting graph (or lattice) $B\cup E$ becomes disconnected (no edge between $B$ and $E$). We may call one of them $B$, the \textit{block}, and the other one $E$ the \textit{environment}. The distinction is somewhat arbitrary and the two subsystems are equivalent in measuring entanglement.

Let us focus on the block (subsystem $B$). It is described by the density matrix $\boldsymbol{\rho}_{B}$ of the block (obtained by tracing out all degrees of freedom of the environment $E$ from the density matrix $\boldsymbol{\rho}$ (\ref{purematrix})):
\begin{eqnarray}
	\boldsymbol{\rho}_{B}=\mathrm{tr}_{E}\left[\,\boldsymbol{\rho}\,\right]. \label{rhoa}
\end{eqnarray}
In (\ref{rhoa}) and below we use subscript $B$ for \textit{block} and $E$ for \textit{environment}. After tracing out all degrees of freedom outside the block the density matrix $\boldsymbol{\rho}_{B}$ is, in general, a mixed state density matrix (unless the pure state density matrix $\boldsymbol{\rho}$ of the whole system projects on a product state, which is obviously not our case of the VBS state). Formula (\ref{rhoa}) is the definition of the block (subsystem) density matrix and it satisfies all three requirements of a density matrix:
\begin{enumerate}
	\item The trace $\mathrm{tr}_{B}\left[\,\boldsymbol\rho_{B}\,\right]=1$ and hermiticity $\boldsymbol{\rho}^{\dagger}_{B}=\boldsymbol{\rho}_{B}$ follow immediately from those of $\boldsymbol{\rho}$;
	\item The positive semi-definiteness is seen by picking up an arbitrary state $|\psi\rangle_{B}$ of the block and realizing that
\begin{eqnarray}
	{}_{B}\langle \psi|\boldsymbol{\rho}_{B}|\psi\rangle_{B}&=&\mathrm{tr}_{B}\left[\,\boldsymbol{\rho}_{B}|\psi\rangle_{B}\langle\psi|\,\right]\nonumber\\
	&=&\mathrm{tr}_{B}\left[\,(\mathrm{tr}_{E}\boldsymbol{\rho})|\psi\rangle_{B}\langle\psi|\,\right]\nonumber\\
	&=&\mathrm{tr}\left[\,\boldsymbol{\rho}\,|\psi\rangle_{B}\langle \psi|\otimes I_{E}\,\right]\geq 0,
\end{eqnarray}
because of the positive semi-definiteness of $\boldsymbol{\rho}$ ($I_{E}$ is the identity of the environment).
\end{enumerate}
The density matrix $\boldsymbol{\rho}_{B}$ is a central quantity in description of the subsystem (block). It contains all correlation functions in the VBS ground state as matrix entries \cite{AAH,JK,KHH,XKHK}. (The relation between elements of the density matrix and correlation functions is given in \S$\,\ref{sec:denandcorr}\,$.) It is essential in measuring the entanglement which is our main subject.

The entanglement can be measured or \textit{quantified} by the von Neumann entropy
\begin{eqnarray}
	S_{\mathrm{v\ N}}=-\mathrm{tr_{B}}\left[\,\boldsymbol{\rho}_{B}\ln\boldsymbol{\rho}_{B}\,\right] =-\sum_{\lambda\neq 0}\lambda\ln\lambda \label{vnentropy}
\end{eqnarray}
or the R\'enyi entropy
\begin{eqnarray}
	S_{\mathrm{R}}(\alpha)=\frac{1}{1-\alpha}\ln\left\{\mathrm{tr}_{B}\left[\,\boldsymbol{\rho}^{\alpha}_{B}\,\right]\right\}=\frac{1}{1-\alpha}\ln\left(\sum_{\lambda\neq 0}\lambda^{\alpha}\right), \qquad \alpha>0.
\end{eqnarray}
Here $\lambda$'s are (non-zero) eigenvalues of the density matrix $\boldsymbol{\rho}_{B}$. The corresponding eigenvector is denoted by $|\lambda\rangle$. \textit{i.e.}
\begin{eqnarray}
	\boldsymbol{\rho}_{B}|\lambda\rangle=\lambda|\lambda\rangle, \qquad \lambda\neq 0.
\end{eqnarray}
The R\'enyi entropy depends on an arbitrary parameter $\alpha$. If we know the R\'enyi entropy at any $\alpha$, then we know all eigenvalues of the density matrix. Note that the R\'enyi entropy can be regarded a generalization of the von Neumann entropy and reduces to the latter in the limit $\alpha\to 1$. The von Neumann entropy is a proper extension of the Gibbs entropy (in statistical mechanics) and the Shannon entropy (in information theory) to the quantum case. (The Shannon entropy measures the uncertainty associated with a classical probability distribution. Whereas in quantum case a density operator replaces a classical distribution.) It was shown by using the Schmidt decomposition (Section $2.5$ of \cite{NC}) that non-zero eigenvalues of the density matrix of subsystem $B$ (block) is equal to those of the density matrix of subsystem $E$ (environment). So that the two subsystems are equivalent in measuring entanglement in terms of entanglement entropies, \textit{i.e.} $S_{\mathrm{v\ N}}[B]=S_{\mathrm{v\ N}}[E]$ and $S_{\mathrm{R}}[B]=S_{\mathrm{R}}[E]$. This fact has been used in obtaining entanglement entropies of $1$-dimensional VBS states as in \cite{FKR,KHH} instead of diagonalizing the density matrix directly. We will study the entropies in detail in following sections.

\subsubsection{The Block Hamiltonian}
\label{sec:blockhamil}

The AKLT block density matrix $\boldsymbol{\rho}_{B}$ possesses certain characteristic properties which distinguish the VBS states from others. We shall show in \S$\,\ref{sec:den0spec}\,$ that the spectrum of the density matrix $\boldsymbol{\rho}_{B}$ contains a lot of zero eigenvalues. In order to understand this and give the subsystem (block) a more complete description, we first introduce the Hamiltonian of the subsystem (called the \textit{block} Hamiltonian).

The \textit{block} Hamiltonian $H_{B}$ is the sum of Hamiltonian densities $H(k,l)$ with both $k\in B$ and $l\in B$, \textit{i.e.} nearest neighbor interactions (bond terms) within the block $B$:
\begin{eqnarray}
	H_{B}=\sum_{\langle kl\rangle\in B}H(k,l),\qquad k\in B, \quad l\in B. \label{subsystemh}
\end{eqnarray}
Here $H(k,l)$ is given in (\ref{basichd}) for the basic model and (\ref{generalizedhd}) for the generalized model, for $k$ and $l$ connected by an edge (bond). In (\ref{subsystemh}) no cut edges are present (boundary edges between subgraphs $B$ and $E$ removed). In other words, the block Hamiltonian is the \textit{internal} interactions of the subsystem $B$. This Hamiltonian has degenerate ground states because uniqueness conditions (\ref{basiccond}) and (\ref{condition1}) are no longer valid.

\begin{figure}
	\centering
		\includegraphics[width=2.9in]{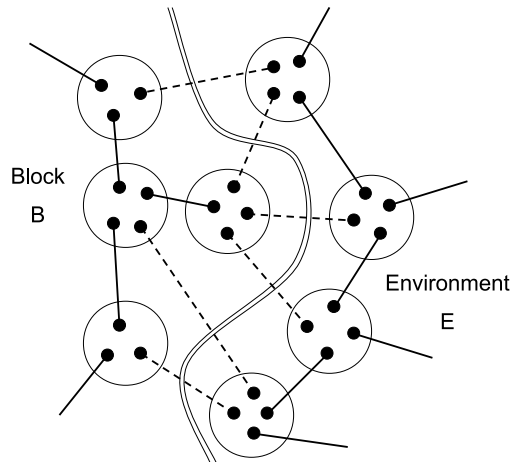}
	\caption{Example of the cutting for the basic model. The curved double line represents the boundary between the two subgraphs. We have the block $B$ on the left and the environment $E$ on the right. Solid lines represent edges (bonds) while dashed lines represent cut edges (cut bonds). Each dashed line connects two dots. All vertices in the figure belong to the boundary of $B$ or $E$ because of the presence of one or more cut incident edges (dashed lines).}
	\label{figure2}
\end{figure}

Let us discuss the degeneracy of ground states of the block Hamiltonian (\ref{subsystemh}). Let us denote by $N_{\partial B}$ the number of vertices on the boundary $\partial B$ of the block $B$. The boundary consists of vertices (sites) with several incident edges (bonds) being cut, see Figure \ref{figure2}.
The degeneracy of ground states of $H_{B}$ (which is abbreviated by $deg$) is given by the Katsura's formula
\begin{eqnarray}
	deg=\prod_{l\in \partial B}\left[\left(\sum_{k \in \partial E}M_{kl}\right)+1\right], \qquad \langle kl\rangle\in\{\mathrm{cut\ edges}\}. \label{generalizeddeg}
\end{eqnarray}
Here $\partial B$ denotes vertices (sites) on the boundary of the block $B$ and $\partial E$ are vertices (sites) on the boundary of the environment $E$. In (\ref{generalizeddeg}) we have $N_{\partial B}$ terms in the product. Formula (\ref{generalizeddeg}) is valid for both the basic and the generalized model. For the basic model all $M_{kl}=1$, including those corresponding to cut edges. Take, for example, a particularly simple case that each vertex on the boundary of the block $\partial B$ was connected to exactly one vertex on the boundary of the environment $\partial E$. Then the degeneracy $deg=2^{N_{\partial B}}$. A general proof of formula (\ref{generalizeddeg}) is given in the next section \S$\,\ref{sec:deg}\,$. The subspace spanned by the degenerate ground states is called the \textit{ground space}, with the dimension given by $deg$ in (\ref{generalizeddeg}). We emphasize at this point that the block $B$ should contain more than one vertices, otherwise we have a trivial case that the block Hamiltonian vanishes $H_{B}=0$ and the whole Hilbert space become the \textit{ground space}. We discuss the density matrix for a single vertex block at the end of \S$\,\ref{sec:den0spec}\,$. The spectrum of the density matrix $\boldsymbol{\rho}_{B}$ is closely related to the block Hamiltonian. The density matrix is a projector onto the ground space multiplied by another matrix. We shall prove this statement for an arbitrary graph or lattice in \S$\,\ref{sec:den0spec}\,$.

\subsection{The Degeneracy of Ground States of the Block Hamiltonian}
\label{sec:deg}

We prove Katsura's formula (\ref{generalizeddeg}) for the degeneracy of ground states of the block Hamiltonian. The proof applies to both the basic and the generalized models. The block Hamiltonian is defined in (\ref{subsystemh}). Let us first look at the uniqueness condition (\ref{condition1}). (The condition (\ref{basiccond}) for the basic model is a special case of this general one.) For an arbitrary vertex (site) $l$ in the block $B$, the condition can be written as
\begin{eqnarray}
	2S_{l}=\sum_{k}M_{kl}=\sum_{k\in B}M_{kl}+\sum_{k\in \partial E}M_{kl}, \qquad l\in B. \label{split}
\end{eqnarray}
Note that the sum over vertices $k\in \partial E$ is \textit{outside} the block $B$. These terms are only present for boundary vertices $l\in \partial B$. Expression (\ref{split}) is valid for \textit{any} vertex in the block (for a bulk vertex the last summation vanishes). Next we define \textit{the block VBS state}
\begin{eqnarray}
	|\mathrm{VBS}_{N_{B}}\rangle=\prod_{\langle kl\rangle\in B}
\left(a^{\dagger}_{k}b^{\dagger}_{l}-b^{\dagger}_{k}a^{\dagger}_{l}\right)^{M_{kl}}|\mathrm{vac}\rangle, \qquad k\in B,\quad l\in B. \label{blockvbs}
\end{eqnarray}
Here edge (bond) $\langle kl\rangle$ lies completely inside the block $B$. Now an \textit{arbitrary} ground state of the block Hamiltonian $H_{B}$ takes the following form (see \textbf{Lemma} \ref{lemma1} in \S$\,\ref{sec:unique}\,$ which explains the appearance of the factor $(a^{\dagger}_{k}b^{\dagger}_{l}-b^{\dagger}_{k}a^{\dagger}_{l})^{M_{kl}}$ in (\ref{blockvbs}) above):
\begin{eqnarray}
	|\mathrm{G}\rangle=\left[\prod^{N_{\partial B}\ \mathrm{terms}}_{l\ \in\ \partial B}f(a^{\dagger}_{l},b^{\dagger}_{l})\right]|\mathrm{VBS}_{N_{B}}\rangle, \label{ground}
\end{eqnarray}
where $f(a^{\dagger}_{l},b^{\dagger}_{l})$ is a polynomial (it may depend on the vertex $l$) in $a^{\dagger}_{l}$ and $b^{\dagger}_{l}$ and the product runs over all boundary vertices (with the total number denoted by $N_{\partial B}$). The degree of this polynomial is equal to $\sum_{k\in \partial E}M_{kl}$. (Each term in the polynomial must have the same total power $\sum_{k\in \partial E}M_{kl}$ of $a^{\dagger}_{l}$ and $b^{\dagger}_{l}$.) It is straightforward to verify that $|\mathrm{G}\rangle$ in (\ref{ground}) is a ground state: 
\begin{enumerate}
	\item The power of $a^{\dagger}_{l}$ and $b^{\dagger}_{l}$ in $|\mathrm{VBS}_{N_{B}}\rangle$ is $\sum_{k\in B}M_{kl}$ (see (\ref{blockvbs})), so that the total power of $a^{\dagger}_{l}$ and $b^{\dagger}_{l}$ in (\ref{ground}) is $\sum_{k\in B}M_{kl}+\sum_{k\in \partial E}M_{kl}=2S_{l}$ according to (\ref{split}). Therefore, we have the correct power $2S_{l}$ of the bosonic operators $a^{\dagger}_{l}$ and $b^{\dagger}_{l}$ for each vertex $l$ in the block $B$ (constraint (\ref{constraintspin}) is satisfied);
	\item There is no projection on any bond (edge) spin value greater than or equal to $S_{k}+S_{l}-M_{kl}+1$ because of the construction of the block VBS state (\ref{blockvbs}). (One could use the same reasoning as in \S$\,\ref{sec:generalized}\,$).
\end{enumerate}
Therefore the degeneracy $deg$ of the ground states of $H_{B}$ is equal to the number of linearly independent states of the form (\ref{ground}). Since $a^{\dagger}_{l}$'s and $b^{\dagger}_{l}$'s are bosonic and commute, the number of linearly independent polynomials $f(a^{\dagger}_{l},a^{\dagger}_{l})$ for an arbitrary $l$ is equal to its degree plus one, \textit{i.e.} 
\begin{eqnarray}
	\left(\sum_{k\in \partial E}M_{kl}\right)+1,\qquad \forall\ l\in \partial B. \label{degreenum}
\end{eqnarray}
So that the total number of linearly independent polynomials of the form 
\begin{eqnarray}
	\prod^{N_{\partial B}\ \mathrm{terms}}_{l\ \in\ \partial B}f(a^{\dagger}_{l},b^{\dagger}_{l})
\end{eqnarray}
is the product of these numbers (\ref{degreenum}) for each $l\in \partial B$. Finally, the ground state degeneracy of the block Hamiltonian $H_{B}$ is (Katsura's formula)
\begin{eqnarray}
	deg=\prod_{l\in\partial B}\left[\left(\sum_{k\in \partial E}M_{kl}\right)+1\right]. \label{kastura}
\end{eqnarray}
In the case of the basic model all $M_{kl}=1$, formula (\ref{kastura}) has a graphical illustration, see Figure \ref{figure2}. We count the number \# of all cut edges (dashed lines) incident to one boundary vertex of the block, then add one to the number \#. The degeneracy is the product of these $(\# +1)$'s for each boundary vertex.

\subsection{General Properties of the Density Matrix}
\label{sec:den0spec}

The reduced density matrix $\boldsymbol{\rho}_{B}$ from a VBS state has important and special spectrum structures which are universal among AKLT models. Let us denote by $N_{B}$ the number of vertices in the block $B$. Then the dimension $dim$ of the Hilbert space of the block $B$ is equal to $\prod_{l}(2S_{l}+1)$ with $l\in B$, which is also the dimension of the density matrix $\boldsymbol{\rho}_{B}$. The value is
\begin{eqnarray}
	dim=\prod_{l\in B}\left[z_{l}+1\right], \label{basicdima}
\end{eqnarray}
for the basic model and
\begin{eqnarray}
	dim=\prod_{l\in B}\left[\left(\sum_{k\in (B\cup\partial E)}M_{kl}\right)+1\right], \label{generalizeddima}
\end{eqnarray}
for the generalized model. In both expressions (\ref{basicdima}) and (\ref{generalizeddima}) we have $N_{B}$ factors in the product. Take, for example, a particularly simple basic model in which each vertex is connected with the same number $z$ of vertices, including those corresponding to boundary vertices (such as in the case of a lattice). Then the dimension $dim=(z+1)^{N_{B}}$. The density matrix $\boldsymbol{\rho}_{B}$ would have $dim$ number of eigenvalues. However, most of the eigenvalues are vanishing and $\boldsymbol{\rho}_{B}$ is a projector onto a much smaller subspace multiplied by another matrix. To prove this statement, we define a \textit{support} to be the subspace of the Hilbert space of the block $B$ with non-zero eigenvalues, \textit{i.e.} it is spanned by eigenstates of $\boldsymbol{\rho}_{B}$ with non-zero eigenvalues. The dimension of the support is denoted by $D$. Then We have the following theorem on the structure of the density matrix $\boldsymbol{\rho}_{B}$ (Assuming that the block have more than one vertices, \textit{i.e.} $N_{B}\geq 2$, so that $H_{B}$ is not equal to zero identically):

\begin{theorem}
\label{theorem1}
The \textit{support} of $\boldsymbol{\rho}_{B}$ (\ref{rhoa}) is a subspace of the \textit{ground space} of the block Hamiltonian $H_{B}$ (\ref{subsystemh}).
\end{theorem}

\begin{proof}
To prove the theorem, we recall that $H=\sum_{\langle kl\rangle\in B}H(k,l)$ and each $H(k,l)$ is a sum of projectors (\ref{generalizedhd}). We have $H(k,l)\geq 0$. Then the construction of the VBS ground state (\ref{basicvbs}) and (\ref{generalizedvbs}) guarantees that there is no projection onto the subspace with higher bond spins ($J\geq S_{k}+S_{l}-M_{kl}+1$) for \textit{any} bond (edge). (See \S$\,\ref{sec:unique}\,$ for the proof.) Therefore,
\begin{eqnarray}
	H(k,l)|\mathrm{VBS}\rangle=0, \qquad \forall\ \langle kl\rangle. \label{noprojection}
\end{eqnarray}
In particular, this is true for bonds (edges) \textit{inside} the block $B$, \textit{i.e.} both $k\in B$ and $l\in B$. Now, from the definition of $\boldsymbol{\rho}_{B}$ in ($\ref{rhoa}$), we have
\begin{eqnarray}
	H(k,l)\boldsymbol{\rho}_{B}&=&H(k,l)\mathrm{tr}_{E}\left[\,\boldsymbol{\rho}\,\right]
\nonumber \\
&=&\frac{H(k,l)\mathrm{tr}_{E}\left[\,|\mathrm{VBS}\rangle\langle \mathrm{VBS}|\,\right]}{\langle \mathrm{VBS}|\mathrm{VBS}\rangle}
\nonumber \\
&=&\frac{\mathrm{tr}_{E}\left[\,H(k,l)|\mathrm{VBS}\rangle\langle \mathrm{VBS}|\,\right]}{\langle \mathrm{VBS}|\mathrm{VBS}\rangle} \nonumber\\
&=&0, \qquad k\in B, \quad l\in B. \label{proof}
\end{eqnarray}
In the last step of (\ref{proof}) we have used (\ref{noprojection}) and the fact that bond (edge) $\langle kl\rangle$ lies completely inside the block $B$ so that $H(k,l)$ commutes with the tracing operation in the environment $E$. Equation (\ref{proof}) is true for any bond (edge) in $B$, so that
\begin{eqnarray}
	H_{B}\boldsymbol{\rho}_{B}=\sum_{\langle kl\rangle\in B}H(k,l)\boldsymbol{\rho}_{B}=0, \qquad k\in B, \quad l\in B. \label{harhoa}
\end{eqnarray}
If we diagonalize the density matrix $\boldsymbol{\rho}_{B}$
\begin{eqnarray}
	\boldsymbol{\rho}_{B}=\sum_{\lambda\neq 0}\lambda\,|\lambda\rangle\langle \lambda|, \label{diag}
\end{eqnarray}
where $|\lambda\rangle$ is the eigenstate corresponding to eigenvalue $\lambda$. Then (\ref{harhoa}) can be re-written as
\begin{eqnarray}
	H_{B}\sum_{\lambda\neq 0}\lambda|\lambda\rangle\langle \lambda|=\sum_{\lambda\neq 0}\lambda H_{B}|\lambda\rangle\langle \lambda|=0, \label{eigenstate}
\end{eqnarray}
Note that $\{|\lambda\rangle\}$ is a linearly independent set. Therefore the solution of (\ref{eigenstate}) means that
\begin{eqnarray}
	H_{B}|\lambda\rangle=0, \qquad \lambda\neq 0. \label{solution}
\end{eqnarray}
Expression (\ref{solution}) states that any eigenstate of $\boldsymbol{\rho}_{B}$ (with non-zero eigenvalue) is a ground state of $H_{B}$. As a result, we have proved \textbf{Theorem} \ref{theorem1} that the \textit{support} of $\boldsymbol{\rho}_{B}$ is a subspace of the \textit{ground space} of $H_{B}$, so that $D\leq deg$. The density matrix takes the form of a projector multiplied by another matrix (a constant matrix depending on non-vanishing eigenvalues) and the projector projects on the \textit{ground space}.
\end{proof}

Also, it is clear from expressions (\ref{generalizeddeg}) and (\ref{basicdima}), (\ref{generalizeddima}) that $deg \leq dim$ ($\partial B\subseteq B$ so that $N_{\partial B}\leq N_{B}$). Usually, $deg$ is much smaller than $dim$ because the former involves only contributions from boundary vertices (sites) of the block while the latter also involves contributions from all bulk vertices (sites). Then as a corollary of \textbf{Theorem} \ref{theorem1}, we have
\begin{eqnarray}
	 D\leq deg\leq dim .
\end{eqnarray}

If the block $B$ consists of only one vertex with a spin-$S$, then we \textit{conjecture} that it is in the maximally entangled state. The \textit{support} has dimension $D=2S+1$.

\section{The One--dimensional Spin--1 Model}
\label{sec:1dspin1}

One of the most simple models is defined on a $1$-dimensional lattice with spin-$1$'s in the bulk and spin-$1/2$'s at both ends. We shall denote by $\boldsymbol{S}_{j}$ the vector spin operator at site $j$ ($j=0,1,\ldots,N+1$). The Hamiltonian is
\begin{eqnarray}
	H=\frac{1}{2}\sum^{N-1}_{j=1} \left(\boldsymbol{S}_{j}\cdot\boldsymbol{S}_{j+1}+\frac{1}{3}\left(\boldsymbol{S}_{j}\cdot\boldsymbol{S}_{j+1}\right)^{2}+\frac{2}{3}\right)+\pi_{\frac{3}{2}}(0,1)+\pi_{\frac{3}{2}}(N,N+1).\nonumber\\ \label{uniq1}
\end{eqnarray} 	
Each bulk term is a projector $\pi_{2}$ onto the states with bond spin-$2$. The boundary terms $\pi_{3/2}$ describe interactions of a spin-$1/2$ on the boundary and a spin-$1$ in the bulk. Each term is a projector onto the states with bond spin-$3/2$:
\begin{eqnarray}
	\pi_{\frac{3}{2}}(0,1)=\frac{2}{3}\left(1+\boldsymbol{S}_{0}\cdot\boldsymbol{S}_{1}\right),  \qquad   		 \pi_{\frac{3}{2}}(N,N+1)=\frac{2}{3}\left(1+\boldsymbol{S}_{N}\cdot\boldsymbol{S}_{N+1}\right). \label{boun1}
\end{eqnarray}
The choice of boundary terms guarantees the uniqueness of the ground state. As mentioned before, if we have spin-$1$ at every site in (\ref{uniq1}) instead, the ground state would become $4$-fold degenerate.

In this section we study the entanglement of the unique VBS ground state of this $1$-dimensional spin-$1$ model. As to be shown below, the density matrix $\boldsymbol{\rho}_{L}$ of a block of $L$ contiguous spins is diagonalizable. It has four non-zero eigenvalues:
\begin{eqnarray}
	\Lambda_{\alpha}=\left\{ \begin{array}{cc}
         \frac{1}{4}(1+3(-\frac{1}{3})^{L}), & \alpha=0;\\ \\
         \frac{1}{4}(1-(-\frac{1}{3})^{L}), & \alpha=1, 2, 3. \end{array}\right. \label{eigval1rep}
\end{eqnarray}
These eigenvalues depend on the length $L$ of the block subsystem and are independent of the size of the whole spin chain. The von Neumann entropy and the R\'enyi entropy derived from these eigenvalues are
\begin{eqnarray}
	S_{\mathrm{v\ N}}&=&\ln4-\frac{1}{4}(1+3(-\frac{1}{3})^{L})\ln(1+3(-\frac{1}{3})^{L})\nonumber\\
	&&-\frac{3}{4}(1-(-\frac{1}{3})^{L})\ln(1-(-\frac{1}{3})^{L})\nonumber\\ \nonumber\\
	S_{\mathrm{R}}(\alpha)&=&\frac{1}{1-\alpha}\ln\left\{\left[\frac{1}{4}(1+3(-\frac{1}{3})^{L})\right]^{\alpha}+3\left[\frac{1}{4}(1-(-\frac{1}{3})^{L})\right]^{\alpha}\right\}.	
	\label{entropies}
\end{eqnarray}
Note that the parameter $\alpha$ in the R\'enyi entropy should not be confused with the label $\alpha$ for the eigenvalues in (\ref{eigval1rep}).

\subsection{The VBS Ground State}
\label{sec:1dspin1VBS}

Given the Hamiltonian (\ref{uniq1}), we are going to use the pictorial method (see \S$\,\ref{sec:graphvbs}\,$) to construct the unique VBS ground state. In order to represent the state, we first introduce the following notation for convenience \cite{FKR}: 
\begin{eqnarray}
	|\alpha\rangle \equiv (-1)^{1+\delta_{\alpha, 0}}I\otimes\sigma_{\alpha}|0\rangle, \qquad \alpha=0, 1, 2, 3 \label{alph}
\end{eqnarray}
where $\sigma_{0}\equiv I$ (2-dimensional identity), $\sigma_{\alpha=1, 2, 3}$ are standard Pauli matrices and $|0\rangle\equiv -(|\uparrow\downarrow\rangle-|\downarrow\uparrow\rangle)/\sqrt{2}$ is the singlet state (antisymmetric projection) of two spin-$1/2$'s. (It corresponds to the antisymmetrized state $|\Phi\rangle$ in \S$\,\ref{sec:graphvbs}\,$.) These four states (\ref{alph}) (called \textit{maximally entangled states}) form an orthonormal basis of the Hilbert space of two spin-$1/2$ operators.

\begin{figure}
	\centering
		\includegraphics[width=5in]{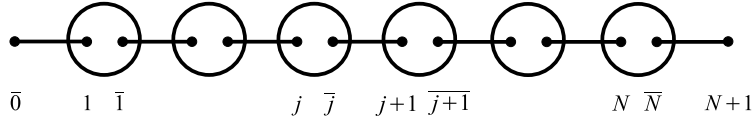}
	\caption{Graphical representation of the VSB ground state for the $1$-dimensional spin-$1$ model: Each spin-$1$ is realized by a pair of spin-$1/2$'s which are represented by small black dots in the figure. The pair of spin-$1/2$ states at site $j$ are labeled $j$, $\bar{j}$. The solid lines connecting two neighboring dots ($\bar{j}$ and $j+1$) represent anti-symmetrization of two spin-$1/2$'s; The large circles enclosing two dots ($j$ and $\bar{j}$) represent symmetrization at each site. The boundary spin-$1/2$'s are labeled $\bar{0}$ and $N+1$ in consistency with the prescription.}
	\label{fig:Figure3}
\end{figure}

The spin-$1$ state at each site is represented by a symmetric projection of two spin-$1/2$ states given by (\ref{alph}) for $\alpha=1, 2, 3$. Let us take the $j^{\mathrm{th}}$ site for example, see Figure \ref{fig:Figure3}.
The two spin-$1/2$'s are labeled by $(j, \bar{j})$ (from left to right, respectively). Then the spin-$1$ states are prepared by projecting these two spin-$1/2$'s (4-dimensional space) onto a symmetric 3-dimensional subspace spanned by
\begin{eqnarray}
	|1\rangle_{j\bar{j}} &=& \frac{1}{\sqrt{2}}(|\uparrow\rangle_{j}|\uparrow\rangle_{\bar{j}}-|\downarrow\rangle_{j}|\downarrow\rangle_{\bar{j}}), \nonumber \\
	|2\rangle_{j\bar{j}} &=& \frac{-i}{\sqrt{2}}(|\uparrow\rangle_{j}|\uparrow\rangle_{\bar{j}}+|\downarrow\rangle_{j}|\downarrow\rangle_{\bar{j}}), \nonumber \\
	|3\rangle_{j\bar{j}} &=& \frac{-1}{\sqrt{2}}(|\uparrow\rangle_{j}|\downarrow\rangle_{\bar{j}}+|\downarrow\rangle_{j}|\uparrow\rangle_{\bar{j}}). \label{3j}
\end{eqnarray}  
Thus the two ending spin-$1/2$'s are labeled as site $\bar{0}$ and $N+1$, consistently (Figure \ref{fig:Figure3}). The unique VBS ground state in this representation is \cite{AKLT1,AKLT2,FKR}
\begin{eqnarray}
	|\mathrm{VBS}\rangle=\left(\bigotimes^{N}_{j=1}\mathbf{P}_{j\bar{j}}\right)|0\rangle_{\bar{0}1}|0\rangle_{\bar{1}2}\cdots|0\rangle_{\bar{N}N+1}. \label{grou1}
\end{eqnarray}
Here $\mathbf{P}_{j\bar{j}}$ projects two spin-$1/2$ states onto a symmetric subspace, which describes spin-$1$. Using basis (\ref{alph}), we have
\begin{eqnarray}
	\mathbf{P}_{j\bar{j}}=\sum^{3}_{\alpha=1}|\alpha\rangle_{j\bar{j}}\langle \alpha|. \label{proj}
\end{eqnarray}
This projector $\mathbf{P}_{j\bar{j}}$ serves the same purpose as the symmetrization operator $\mathbf{P}(l)$ in \S$\,\ref{sec:graphvbs}\,$ and their results acting on a product state of spin-$1/2$'s only differ by a normalization. We use the projector $\mathbf{P}_{j\bar{j}}$ for convenience here.

A crucial step (see \cite{FKR}) is that the ground state (\ref{grou1}) can be expressed in a different form using the following identity
\begin{eqnarray}
	|0\rangle_{\bar{A}B}|0\rangle_{\bar{B}C}=\frac{-1}{2}\sum^{3}_{\alpha=0}|\alpha\rangle_{B\bar{B}}\left[I_{\bar{A}}\otimes\left(\sigma_{\alpha}\right)_{C}\right]|0\rangle_{\bar{A}C} \label{maxi}
\end{eqnarray}
for arbitrary labels (indices) $A$, $B$ and $C$. This identity (\ref{maxi}) can be verified by direct calculation and comparison. Repeatedly using relation (\ref{maxi}), the product of $|0\rangle$'s in (\ref{grou1}) can be re-written as
\begin{eqnarray}
	&&|0\rangle_{\bar{0}1}|0\rangle_{\bar{1}2}\cdots|0\rangle_{\bar{N}N+1} \label{prod} \\
	&=&\left(\frac{-1}{2}\right)^{N}\sum^{3}_{\alpha_{1}, \cdots, \alpha_{N}=0}|\alpha_{1}\rangle\cdots|\alpha_{N}\rangle\left[ I_{\bar{0}}\otimes\left(\sigma_{\alpha_{N}}\cdots\sigma_{\alpha_{1}}\right)_{N+1}\right]|0\rangle_{\bar{0}N+1}. \nonumber
\end{eqnarray}
Then by projecting onto the symmetric subspace spanned by $|\alpha=1, 2, 3\rangle$, the ground VBS state (\ref{grou1}) takes the form \cite{VMC}
\begin{eqnarray}
	|\mathrm{VBS}\rangle=\frac{1}{3^{N/2}}\sum^{3}_{\alpha_{1}, \cdots, \alpha_{N}=1}|\alpha_{1}\rangle\cdots|\alpha_{N}\rangle\left[I_{\bar{0}}\otimes\left(\sigma_{\alpha_{N}}\cdots\sigma_{\alpha_{1}}\right)_{N+1}\right]|0\rangle_{\bar{0}N+1}. \label{grou}
\end{eqnarray}
Note that this ground state (\ref{grou}) is normalized and we have re-written the overall phase for it has no physical content.

\subsection{The Block Density Matrix}
\label{sec:1ds1den}

Given the ground state in the form (\ref{grou}), we obtain the density matrix of a block of $L$
contiguous bulk spins starting at site $k$ by tracing out spin degrees of freedom outside the block using basis (\ref{alph}):
\begin{eqnarray}
	\boldsymbol{\rho}_{L}\equiv \mathrm{tr}_{\bar{0}, 1, \ldots, k-1, k+L, \ldots, N, N+1}\left[\,|\mathrm{VBS}\rangle\langle \mathrm{VBS}|\,\right]. \label{den1}
\end{eqnarray}
(Note that we use subscript $L$ to emphasize the dependence of the density matrix on the size of the block instead of using the general $B$ as representing `block'.) In taking the partial trace, we will encounter the following expression in calculation
\begin{eqnarray}
	I_{n}=\sum^{3}_{\sigma_{\alpha_{1}},\cdots,\sigma_{\alpha_{n}}=1} I\otimes\left(\sigma_{\alpha_{n}}\cdots\sigma_{\alpha_{1}}\right)|0\rangle\langle 0|\, I\otimes\left(\sigma_{\alpha_{1}}\cdots\sigma_{\alpha_{n}}\right), \qquad n\geq1 \label{itera}
\end{eqnarray}
given $I_{0}=|0\rangle\langle0|$. To solve this (\ref{itera}), we introduce iterative coefficients $A_{n}$ and $B_{n}$ and write
\begin{eqnarray}
	I_{n}=A_{n}|0\rangle\langle0|+B_{n}\sum^{3}_{\beta=1}|\beta\rangle\langle\beta|, \qquad n\geq1.
\end{eqnarray}
Then from (\ref{itera}) we could write down the expression of $I_{n+1}$ in terms of $A_{n}$ and $B_{n}$. Comparison of coefficients yields the following iteration relation
\begin{eqnarray}
	A_{n+1}=3B_{n}, \qquad B_{n+1}=A_{n}+2B_{n}, \qquad n\geq1 \label{anbn}
\end{eqnarray}
with $A_{0}=1$ and $B_{0}=0$. The solution to (\ref{anbn}) is
\begin{eqnarray}
	A_{n}=\frac{1}{4}\left(3^{n}+3(-1)^{n}\right), \qquad B_{n}=\frac{1}{4}\left(3^{n}-(-1)^{n}\right).
\end{eqnarray}
As a result, we have found that
\begin{eqnarray}
	I_{n}=\frac{1}{4}\left(3^{n}+3(-1)^{n}\right)|0\rangle\langle0|+\frac{1}{4}\left(3^{n}-(-1)^{n}\right)\sum^{3}_{\beta=1}|\beta\rangle\langle\beta|. \label{inexpp}
\end{eqnarray}
Using (\ref{inexpp}), it is straightforward to take the partial trace in (\ref{den1}). The result is independent of the starting site $k$ and the total length $N$ (see \cite{FKR}). (So that the density matrix is translational invariant though the whole spin chain Hamiltonian does not have complete translational invariance because of the boundary conditions.) We choose $k=1$ (\textit{i.e.} re-label the indices of sites for notational convenience) so that the density matrix reads \cite{FKR}
\begin{eqnarray}
	\boldsymbol{\rho}_{L} \label{matr1}=
	\frac{1}{3^{L}}\sum^{3}_{\alpha, \alpha^{\prime}=1}|\alpha_{1}\rangle \langle\alpha^{\prime}_{1}|\cdots|\alpha_{L}\rangle \langle\alpha^{\prime}_{L}|\,\langle 0|I\otimes (\sigma_{\alpha^{\prime}_{1}}\cdots\sigma_{\alpha^{\prime}_{L}})I\otimes (\sigma_{\alpha_{L}}\cdots\sigma_{\alpha_{1}})|0\rangle.
\nonumber\\
\end{eqnarray}

\subsection{Ground States of the Block Hamiltonian}
\label{sec:1ds1gsbh}

The block in $1$-dimension is $L$ contiguous bulk spins. The block Hamiltonian $H_{B}$ by definition (\ref{subsystemh}) reads
\begin{eqnarray}
	H_{B}=\frac{1}{2}\sum^{L-1}_{j=1} \left(\boldsymbol{S}_{j}\cdot\boldsymbol{S}_{j+1}+\frac{1}{3}\left(\boldsymbol{S}_{j}\cdot\boldsymbol{S}_{j+1}\right)^{2}+\frac{2}{3}\right). \label{degel1}
\end{eqnarray}
Any linear combination of states of the following form
\begin{eqnarray}
	|\mathrm{G}; \chi_{1}, \chi_{\bar{L}}\rangle
\equiv\left(\bigotimes^{L}_{j=1}\mathbf{P}_{j\bar{j}}\right)|\chi_{1}\rangle_{1}|0\rangle_{\bar{1}2}|0\rangle_{\bar{2}3}\cdots|0\rangle_{\overline{L-1}L}|\chi_{\bar{L}}\rangle_{\bar{L}} \label{dgini}
\end{eqnarray}
is a ground state of the block Hamiltonian (\ref{degel1}). In (\ref{dgini}) we have made notation $|\chi\rangle \equiv |\uparrow \mathrm{or} \downarrow\rangle$ represents the two spin-$1/2$ states and $\mathbf{P}_{j\bar{j}}$ is defined in (\ref{proj}). Let us make a particular linear combination of these $|\mathrm{G}; \chi_{1}, \chi_{\bar{L}}\rangle$ states using (\ref{alph}) and write the four ($\alpha=0,1,2,3$) linearly independent ground states of the block Hamiltonian (\ref{degel1}) as follows
\begin{eqnarray}
	|\mathrm{VBS}; \alpha\rangle
\equiv\left(\bigotimes^{L}_{j=1}\mathbf{P}_{j\bar{j}}\right)|\alpha\rangle_{\bar{L}1}|0\rangle_{\bar{1}2}|0\rangle_{\bar{2}3}\cdots|0\rangle_{\overline{L-1}L}. \label{dgalph}
\end{eqnarray}
Note that we have changed the label G to VBS and these 4 states in (\ref{dgalph}) are called \textit{degenerate VBS states}. Now we go through the same steps as from (\ref{grou1}) to (\ref{grou}), the resultant form of the four ground states ($\alpha=0, 1, 2, 3$) is
\begin{eqnarray}
	|\mathrm{VBS}; \alpha\rangle=
	\sum^{3}_{\alpha_{1}, \cdots, \alpha_{L}=1}|\alpha_{1}\rangle\cdots|\alpha_{L}\rangle\,\langle\alpha_{L}|\sigma_{\alpha}\otimes\left(\sigma_{\alpha_{L-1}}\cdots\sigma_{\alpha_{1}}\right)|0\rangle.
	\label{galph}
\end{eqnarray}
Again we have re-written the overall phase for simplicity. These four states are orthogonal, and the normalization is given by (the calculation is similar to that of $A_{n}$ and $B_{n}$ in (\ref{anbn}))
\begin{eqnarray}
	\langle \mathrm{VBS}; \alpha|\mathrm{VBS}; \alpha\rangle=\left\{ \begin{array}{cc}
         \frac{1}{4}(3^{L}+3(-1)^{L}), & \alpha=0;\\ \\
         \frac{1}{4}(3^{L}-(-1)^{L}), & \alpha=1, 2, 3. \end{array}\right. \label{norm}
\end{eqnarray}

\subsection{Spectrum of the Density Matrix}
\label{sec:1ds1specden}

According to \textbf{Theorem} \ref{theorem1}, the eigenvectors corresponding to non-zero eigenvalues of the density matrix (\ref{matr1}) are degenerate ground states of the block Hamiltonian (\ref{degel1}). These are exactly the degenerate VBS states found in (\ref{galph}). Let us apply $\boldsymbol{\rho}_{L}$ to $|\mathrm{VBS}; \alpha\rangle$ and use orthogonality of the $|\alpha\rangle$ states. Then we obtain
\begin{eqnarray}
	\boldsymbol{\rho}_{L}|\mathrm{VBS}; \alpha\rangle=\frac{1}{3^{L}}\sum^{3}_{\alpha_{1}, \cdots, \alpha_{L}=1}|\alpha_{1}\rangle\cdots|\alpha_{L}\rangle\,C_{\alpha_{1}\cdots\alpha_{L}} \label{rhoga}
\end{eqnarray}
with coefficient
\begin{eqnarray}
	C_{\alpha_{1}\cdots\alpha_{L}}&=&\sum^{3}_{\alpha^{\prime}_{1}, \cdots, \alpha^{\prime}_{L}=1}
\langle\alpha^{\prime}_{L}|\sigma_{\alpha}\otimes(\sigma_{\alpha^{\prime}_{L-1}}\cdots\sigma_{\alpha^{\prime}_{1}})|0\rangle \label{coef} \\
	&&
\cdot\langle 0|I\otimes (\sigma_{\alpha^{\prime}_{1}}\cdots\sigma_{\alpha^{\prime}_{L}}) 
I\otimes (\sigma_{\alpha_{L}}\cdots\sigma_{\alpha_{1}})|0\rangle.
\nonumber
\end{eqnarray}
Using the same method of induction as in obtaining $A_{n}$ and $B_{n}$ in (\ref{anbn}), we have
\begin{eqnarray}
	\sum^{3}_{\alpha^{\prime}_{1}, \cdots, \alpha^{\prime}_{L-1}=1}
(I\otimes\sigma_{\alpha^{\prime}_{L-1}}\cdots\sigma_{\alpha^{\prime}_{1}})|0\rangle
\langle 0|(I\otimes \sigma_{\alpha^{\prime}_{1}}\cdots\sigma_{\alpha^{\prime}_{L-1}})=
\sum^{3}_{\beta=0}A_{\beta}|\beta\rangle\langle \beta| \label{indu}
\end{eqnarray}
with
\begin{eqnarray}
	A_{\beta}=\left\{ \begin{array}{cc}
         \frac{1}{4}(3^{L-1}+3(-1)^{L-1}), & \beta=0;\\ \\
         \frac{1}{4}(3^{L-1}-(-1)^{L-1}), & \beta=1, 2, 3. \end{array}\right. \label{indu2}
\end{eqnarray}
Therefore the coefficient $C_{\alpha_{1}\cdots\alpha_{L}}$ defined in (\ref{coef}) can be simplified as
\begin{eqnarray}
	C_{\alpha_{1}\cdots\alpha_{L}}
	=\sum^{3}_{\alpha^{\prime}_{L}=1, \beta=0}
	A_{\beta}\langle\alpha^{\prime}_{L}|\sigma_{\alpha}\otimes I|\beta\rangle
	\langle \beta|I\otimes (\sigma_{\alpha^{\prime}_{L}}\sigma_{\alpha_{L}}) 
	I\otimes (\sigma_{\alpha_{L-1}}\cdots\sigma_{\alpha_{1}})|0\rangle. \nonumber\\
\label{coefind}
\end{eqnarray}
Straightforward calculation using multiplication rules of Pauli matrices shows that (\ref{coefind}) can be further simplified as
\begin{eqnarray}
	&&C_{\alpha_{1}\cdots\alpha_{L}}=
	3A_{1}\delta_{\alpha,0}\langle\alpha_{L}|I\otimes (\sigma_{\alpha_{L-1}}\cdots\sigma_{\alpha_{1}})|0\rangle \label{coefsimp} \\
	&&+(A_{0}+2A_{1})(1-\delta_{\alpha,0})(\delta_{\alpha\alpha_{L}}\langle 0|-\mathrm{i}\sum^{3}_{\beta=1}\epsilon_{\alpha\alpha_{L}\beta}\langle \beta|)I\otimes(\sigma_{\alpha_{L-1}}\cdots\sigma_{\alpha_{1}})|0\rangle \nonumber
\end{eqnarray}
where $\epsilon_{\alpha\alpha_{L}\beta}$ is the totally antisymmetric tensor of three indices with $\epsilon_{123}=1$. By realizing that 
\begin{eqnarray}
	\delta_{\alpha\alpha_{L}}\langle 0|-\mathrm{i}\sum^{3}_{\beta=1}\epsilon_{\alpha\alpha_{L}\beta}\langle \beta|=\langle 0|\sigma_{\alpha_{L}}\sigma_{\alpha}\otimes I=\langle \alpha_{L}|\sigma_{\alpha}\otimes I, \label{real}
\end{eqnarray}
we have reached the final form of the coefficient $C_{\alpha_{1}\cdots\alpha_{L}}$ such that
\begin{eqnarray}
	C_{\alpha_{1}\cdots\alpha_{L}}=\left[3A_{1}\delta_{\alpha,0}+(A_{0}+2A_{1})(1-\delta_{\alpha,0})\right]\langle\alpha_{L}|\sigma_{\alpha}\otimes (\sigma_{\alpha_{L-1}}\cdots\sigma_{\alpha_{1}})|0\rangle. \label{codffin}
\end{eqnarray}
As a result, we plug (\ref{codffin}) into (\ref{rhoga}) and find that
\begin{eqnarray}
	\boldsymbol{\rho}_{L}|\mathrm{VBS}; \alpha\rangle&=&\frac{3A_{1}\delta_{\alpha,0}+(A_{0}+2A_{1})(1-\delta_{\alpha,0})}{3^{L}}\label{eigeqn} \\
	&&\cdot\sum^{3}_{\alpha_{1}, \cdots, \alpha_{L}=1}|\alpha_{1}\rangle\cdots|\alpha_{L}\rangle\,\langle\alpha_{L}|\sigma_{\alpha}\otimes (\sigma_{\alpha_{L-1}}\cdots\sigma_{\alpha_{1}})|0\rangle. \nonumber
\end{eqnarray}
By comparing with (\ref{galph}), we find that (\ref{eigeqn}) is exactly the statement that $|\mathrm{VBS}; \alpha\rangle$ $(\alpha=0, 1, 2, 3)$ are eigenvectors of the density matrix $\boldsymbol{\rho}_{L}$:
\begin{eqnarray}
	\boldsymbol{\rho}_{L}|\mathrm{VBS}; \alpha\rangle=\Lambda_{\alpha}|\mathrm{VBS}; \alpha\rangle, \qquad \alpha=0, 1, 2, 3 \label{eigen}
\end{eqnarray}
with eigenvalues
\begin{eqnarray}
	\Lambda_{\alpha}=\frac{3A_{1}\delta_{\alpha,0}+(A_{0}+2A_{1})(1-\delta_{\alpha,0})}{3^{L}}
	=\left\{ \begin{array}{cc}
         \frac{1}{4}(1+3(-\frac{1}{3})^{L}), & \alpha=0;\\ \\
         \frac{1}{4}(1-(-\frac{1}{3})^{L}), & \alpha=1, 2, 3. \end{array}\right. \label{eigval1}
\end{eqnarray}
These numbers obtained in (\ref{eigval1}) are exactly the eigenvalues found in \cite{FKR,KHH} for spin-$1$, and are consistent with our later explicit expression for eigenvalues in the more general case, see (\ref{eivaex}) in \S$\,\ref{sec:eigval1}\,$.

We can also prove explicitly that any other eigenvectors of $\boldsymbol{\rho}_{L}$ orthogonal to the set $\{|\mathrm{VBS}; \alpha\rangle\}$ have zero eigenvalue. Note that a complete basis of the Hilbert space $\mathbf{H}$ of the block of spins can be chosen as
\begin{eqnarray}
	\{|\alpha_{1}\rangle\cdots|\alpha_{L}\rangle\}, \qquad \alpha=1, 2, 3. \label{hlba}
\end{eqnarray}
The subspace $\mathbf{H}_{4}$ with non-zero eigenvalues is panned by $\{|\mathrm{VBS}; \alpha\rangle\}$, as we have already shown. The Hilbert space can be reduced into a direct sum 
\begin{eqnarray}
	\mathbf{H}=\mathbf{H}_{4}\oplus\mathbf{H}_{\Phi}. \label{splithil}
\end{eqnarray}
We will show that the subspace $\mathbf{H}_{\Phi}$ orthogonal to $\mathbf{H}_{4}$ is a subspace of vanishing eigenvalues. Mathematically, this means that for an arbitrary basis vector $|\beta_{1}\rangle\cdots|\beta_{L}\rangle$, we shall have
\begin{eqnarray}
\boldsymbol{\rho}_{L}(\mathbf{I}-\mathbf{P}_{4})|\beta_{1}\rangle\cdots|\beta_{L}\rangle=0, \label{zero}
\end{eqnarray}
where $\mathbf{I}$ is the identity of $\mathbf{H}$ and $\mathbf{P}_{4}$ is the projector onto $\mathbf{H}_{4}$:
\begin{eqnarray}
	\mathbf{I}&\equiv& \sum^{3}_{\alpha_{1},\cdots,\alpha_{L}=1}|\alpha_{1}\rangle\cdots|\alpha_{L}\rangle\langle\alpha_{1}|\cdots\langle\alpha_{L}|, \nonumber\\
	\mathbf{P}_{4}&\equiv& \sum^{3}_{\alpha=1}\frac{|\mathrm{VBS}; \alpha\rangle\langle\mathrm{VBS}; \alpha|}{\langle \mathrm{VBS}; \alpha|\mathrm{VBS}; \alpha\rangle}. \label{ip}
\end{eqnarray}
By taking expressions (\ref{matr1}), (\ref{ip}), (\ref{eigen}), and realizing that
\begin{eqnarray}
	\sum^{3}_{\alpha=0}\frac{3^{L}\Lambda_{\alpha}}{\langle \mathrm{VBS}; \alpha|\mathrm{VBS}; \alpha\rangle}\,|\alpha\rangle \langle \alpha|=\sum^{3}_{\alpha=0}|\alpha\rangle \langle \alpha|=I\otimes I \label{iden},
\end{eqnarray}
we find the left hand side of (\ref{zero}) being equal to
\begin{eqnarray}
	&&\boldsymbol{\rho}_{L}(\mathbf{I}-\mathbf{P}_{4})|\beta_{1}\rangle\cdots|\beta_{L}\rangle \label{comm1} \\
	&=&\frac{1}{3^{L}}\sum^{3}_{\alpha_{1}\cdots\alpha_{L}=1}
	|\alpha_{1}\rangle\cdots|\alpha_{L}\rangle\,\langle 0|\,[\,I\otimes(\sigma_{\beta_{1}}\cdots\sigma_{\beta_{L}})\,,\,I\otimes(\sigma_{\alpha_{L}}\cdots\sigma_{\alpha_{1}})\,]\,|0\rangle. \nonumber
\end{eqnarray}
We use multiplication rules of Pauli matrices to write the two terms within the commutator in (\ref{comm1}) as
\begin{eqnarray}
	I\otimes(\sigma_{\beta_{1}}\cdots\sigma_{\beta_{L}})=\mathrm{e}^{\mathrm{i}\theta(\beta)}I\otimes\sigma_{\beta}, \qquad \beta=0, 1, 2, 3; \nonumber \\
	I\otimes(\sigma_{\alpha_{L}}\cdots\sigma_{\alpha_{1}})=\mathrm{e}^{\mathrm{i}\theta(\alpha)}I\otimes\sigma_{\alpha}, \qquad \alpha=0, 1, 2, 3. \label{ab}
\end{eqnarray}
Here $\mathrm{e}^{\mathrm{i}\theta(\beta)}$ and $\mathrm{e}^{\mathrm{i}\theta(\alpha)}$ are two phase factors. Then the commutator is
\begin{eqnarray}
	[\,I\otimes(\sigma_{\beta_{1}}\cdots\sigma_{\beta_{L}})\,,\,I\otimes(\sigma_{\alpha_{L}}\cdots\sigma_{\alpha_{1}})\,]=\mathrm{e}^{\mathrm{i}(\theta(\beta)+\theta(\alpha))}I\otimes [\,\sigma_{\beta}\,,\,\sigma_{\alpha}\,]. \label{comm2}
\end{eqnarray}
There are two possibilities:
\begin{enumerate}
	\item $\alpha=\beta$ or at least one of the two is equal to zero, then $\sigma_{\beta}$ and $\sigma_{\alpha}$ commutes;
	\item $\alpha\neq\beta\neq 0$, then $[\,\sigma_{\beta}\,,\,\sigma_{\alpha}\,]=2\mathrm{i}\sum^{3}_{\gamma=1}\epsilon_{\beta\alpha\gamma}\sigma_{\gamma}$, but we still have $\langle 0|I\otimes\sigma_{\gamma}|0\rangle=\langle 0|\gamma\rangle=0$.
\end{enumerate}
Therefore, the factor $\langle 0|\,[I\,\otimes(\sigma_{\beta_{1}}\cdots\sigma_{\beta_{L}})\,,\,I\otimes(\sigma_{\alpha_{L}}\cdots\sigma_{\alpha_{1}})\,]\,|0\rangle$ in (\ref{comm1}) is identically zero. So that we have proved (\ref{zero}). Therefore $\mathbf{H}_{\Phi}$ is a subspace with only zero eigenvalues.

\subsection{The Large Block Limit}
\label{sec:1ds1limit}

It is interesting to study the large block limit that $L\to\infty$. We recognized from (\ref{eigval1}) that all four eigenvalues approach the same limit
\begin{eqnarray}
	\Lambda_{\alpha}=\frac{1}{4}, \qquad L\to\infty.
\end{eqnarray}
As a result, the von Neumann entropy coincides with the R\'enyi entropy in the numerical value and both equal to
\begin{eqnarray}
	S_{\mathrm{v\ N}}=S_{\mathrm{R}}(\alpha)=\ln4, \qquad L\to\infty.
\end{eqnarray}
The limiting density matrix $\boldsymbol{\rho}_{\infty}$ ($\lim_{L\to\infty}\boldsymbol{\rho}_{L}\equiv\boldsymbol{\rho}_{\infty}$) is proportional to the projector $\mathbf{P}_{4}$ (\ref{ip}) which projects on the $4$-degenerate ground states (the ground space) of the block Hamiltonian, \textit{i.e.}
\begin{eqnarray}
	\boldsymbol{\rho}_{L}\to\boldsymbol{\rho}_{\infty}=\frac{1}{4}\,\mathbf{P}_{4}, \qquad L\to\infty. \label{1dsirholimit}
\end{eqnarray}

\section{The One--dimensional Spin-S Homogeneous Model}
\label{sec:1dssh}

In $1$-dimension, if all bulk spins take the same \textbf{integer} value $S$, the model is called the \textit{homogeneous} model. The system consists of a linear chain of $N$ spin-$S$'s in the bulk, and two spin-$S/2$'s on the boundaries. Let $\boldsymbol{S}_{j}$ denotes the vector spin operator at site $j$ ($j=0,1,\ldots,N+1$). The Hamiltonian is
\begin{eqnarray}
	H=\sum^{N-1}_{j=1}\sum^{2S}_{J=S+1} C_{J}(j,j+1)\,\pi_{J}(j,j+1)+H(0,1)+H(N, N+1), \label{uniq}
\end{eqnarray}
where the projector $\pi_{J}(j,j+1)$ projects the bond spin $\boldsymbol{J}_{j,j+1}\equiv\boldsymbol{S}_{j}+\boldsymbol{S}_{j+1}$ onto the subspace with total spin $J$ ($J=S+1, \ldots, 2S$). Physically formation of bond spins with these values would increase energy. The boundary terms describe interactions between a spin-$S/2$ and a spin-$S$:
\begin{eqnarray}
	H(0,1)&=&\sum^{3S/2}_{J=S/2+1} C_{J}(0,1)\,\pi_{J}(0, 1),\nonumber\\  H(N, N+1)&=&\sum^{3S/2}_{J=S/2+1} C_{J}(N,N+1)\,\pi_{J}(N, N+1). \label{boun}
\end{eqnarray}
Coefficients $C_{J}(j,j+1)$ can take arbitrary positive values. This model is a special case of the generalized model in $1$-dimension with all multiplicity number $M_{j,j+1}=S$. 

We study the entanglement of the unique VBS ground state of the Hamiltonian (\ref{uniq}) in this section. The density matrix of a block of spins and the entropies were calculated by H. Katsura, T. Hirano, Y. Hatsugai and their collaborators in \cite{KHH,XKHK}.

\subsection{The VBS Ground State}
\label{sec:1dsshvbs}

\subsubsection{The Construction of the VBS State}
\label{sec:convbs}

According to the general approach in \S$\,\ref{sec:vbssch}\,$, the unique VBS ground state of the Hamiltonian (\ref{uniq}) is constructed in the Schwinger representation as \cite{AAH}
\begin{eqnarray}
	|\mathrm{VBS}\rangle \equiv
 \prod^{N}_{j=0}
\left(a^{\dagger}_{j}b^{\dagger}_{j+1}-b^{\dagger}_{j}a^{\dagger}_{j+1}\right)^{S}|\mathrm{vac}\rangle, \label{vbs}
\end{eqnarray}
where $a^{\dagger}$, $b^{\dagger}$ are bosonic creation operators and $\left|\mathrm{vac}\right\rangle$ is destroyed by any of the annihilation operators $a$, $b$. Recall that these operators satisfy $[\,a_{i}\,,\,a^{\dagger}_{j}\,]=[\,b_{i}\,,\,b^{\dagger}_{j}\,]=\delta_{ij}$ with all other commutators vanishing. The spin operators are represented as $S^{+}_{j}=a^{\dagger}_{j}b_{j}$, $S^{-}_{j}=b^{\dagger}_{j}a_{j}$, $S^{z}_{j}=(a^{\dagger}_{j}a_{j}-b^{\dagger}_{j}b_{j})/2$. To reproduce the dimension of the spin-$S$ Hilbert space at each site, an additional constraint on the total boson occupation number is required, namely $(a^{\dagger}_{j}a_{j}+b^{\dagger}_{j}b_{j})/2=S$. More details and properties of the VBS state in the Schwinger representation can be found in \S$\,\ref{sec:vbssch}\,$ and \cite{AAH,A,KK}. The pure state density matrix of the VBS ground state (\ref{vbs}) is
\begin{eqnarray}
	\boldsymbol{\rho}=\frac{|\mathrm{VBS}\rangle\langle \mathrm{VBS}|}{\langle \mathrm{VBS}|\mathrm{VBS}\rangle}. \label{pure}
\end{eqnarray}
We will discuss the normalization $\langle \mathrm{VBS}|\mathrm{VBS}\rangle$ of the VBS state after introducing the coherent state basis.

\subsubsection{The Coherent State Basis}
\label{sec:cohe}

In order to calculate the normalization of the VBS state (\ref{vbs}) and later the density matrix of the block, it is convenient to introduce a spin coherent state basis. As shown in \cite{AAH}, this basis represents the quantum spins in the model in terms of classical unit vectors. We first introduce spinor coordinates
\begin{eqnarray}
	\left(u, v\right)\equiv\left(\cos\frac{\theta}{2}\,\mathrm{e}^{\mathrm{i}\frac{\phi}{2}}, \sin\frac{\theta}{2}\,\mathrm{e}^{-\mathrm{i}\frac{\phi}{2}}\right), \qquad 0\leq\theta\leq\pi, \quad 0\leq\phi\leq 2\pi. \label{spin}
\end{eqnarray}
Then for a point $\hat{\Omega}\equiv (\sin\theta\cos\phi, \sin\theta\sin\phi, \cos\theta)$ on the unit sphere, the spin-$S$ coherent state is defined as
\begin{eqnarray}
	|\hat{\Omega}\rangle\equiv\frac{\left(ua^{\dagger}+vb^{\dagger}\right)^{2S}}{\sqrt{\left(2S\right)!}}\,|\mathrm{vac}\rangle. \label{cohebas}
\end{eqnarray}
Here we have fixed the overall phase (a $U(1)$ gauge degree of freedom) since it has no physical content. Note that (\ref{cohebas}) is covariant under $SU(2)$ transforms (see \S$\,\ref{sec:larges}\,$). The set of coherent states is complete (but not orthogonal) such that \cite{FM}
\begin{eqnarray}
	\frac{2S+1}{4\pi}\int \mathrm{d}\hat{\Omega}\,|\hat{\Omega}\rangle \langle \hat{\Omega}|=\sum^{S}_{m=-S}|S, m\rangle\langle S, m|=I_{2S+1}, \label{comp}
\end{eqnarray}
where $|S, m\rangle$ denote the eigenstate of $\boldsymbol{S}^{2}$ and $S_{z}$, and $I_{2S+1}$ is the identity of the $(2S+1)$-dimensional Hilbert space for spin-$S$. To prove (\ref{comp}), we expand the expression (\ref{cohebas}) (see also (\ref{spinstatesch}))
\begin{eqnarray}
	|\hat{\Omega}\rangle=\sum^{S}_{m=-S}\sqrt{\frac{(2S)!}{(S+m)!(S-m)!}}\,\,u^{S+m}v^{S-m}|S,m\rangle. \label{omegaexp}
\end{eqnarray}
Then, by substituting (\ref{omegaexp}) into (\ref{comp}) and realizing that
\begin{eqnarray}
	\int\mathrm{d}\hat{\Omega}\,u^{S+m}v^{S-m}u^{\ast S+m'}v^{\ast S-m'}=\frac{(S+m)!(S-m)!}{(2S+1)!}\,4\pi\delta_{mm'},
\end{eqnarray}
the completeness relation (\ref{comp}) is then established. This relation (\ref{comp}) can be used in taking trace of an arbitrary operator.

\subsubsection{Normalization of the VBS State}
\label{sec:norm}

The VBS state $|\mathrm{VBS}\rangle$ defined in (\ref{vbs}) is not normalized. Using the coherent state formalism (\ref{cohebas}) and the completeness relation (\ref{comp}), we express the norm square as
\begin{eqnarray}
	&&\langle \mathrm{VBS}|\mathrm{VBS}\rangle \label{norm1}\\
	&=&\left[\frac{(S+1)!}{4\pi}\right]^{2}\left[\frac{(2S+1)!}{4\pi}\right]^{N}\int \left(\prod^{N+1}_{j=0}\mathrm{d}\hat{\Omega}_{j}\right)\prod^{N}_{j=0}\left[\frac{1}{2}(1-\hat{\Omega}_{j}\cdot\hat{\Omega}_{j+1})\right]^{S} \nonumber
\end{eqnarray}
where we have used
\begin{eqnarray}
	\langle 0|a^{S+l}b^{S-l}|\hat{\Omega}\rangle=\sqrt{(2S)!}\,\,u^{S+l}v^{S-l}.
\end{eqnarray}
In order to carry out the integral in (\ref{norm1}), we consider the expansion of the function $\left[\frac{1}{2}(1-x)\right]^{S}$ in terms of Legendre polynomials 
\begin{eqnarray}
	P_{l}(x)=\frac{1}{2^{l}l!}\left(\frac{\mathrm{d}}{\mathrm{d}x}\right)^{l}(x^{2}-1)^{l}
\end{eqnarray}
as follows
\begin{eqnarray}
	\left[\frac{1}{2}(1-x)\right]^{S}=\sum^{S}_{l=0}C_{l}P_{l}(x).
\end{eqnarray}
The coefficient $C_{l}$ is derived by using the orthogonality of $P_{l}$ and repeatedly integrating by parts
\begin{eqnarray}
	C_{l}&=&\frac{2l+1}{2}\int^{1}_{-1}\mathrm{d}x P_{l}(x)\left[\frac{1}{2}(1-x)\right]^{S} \nonumber\\
	&=&\frac{2l+1}{2}\int^{1}_{-1}\mathrm{d}x\frac{1}{2^{l}l!}\left(\frac{\mathrm{d}}{\mathrm{d}x}\right)^{l}(x^{2}-1)^{l}\left[\frac{1}{2}(1-x)\right]^{S} \nonumber\\
	&=&\frac{(2l+1)S!}{2^{S+l+1}l!(S-l)!}\int^{1}_{-1}\mathrm{d}x (x^{2}-1)^{l}(1-x)^{S-l} \nonumber\\
	&=&\frac{(-1)^{l}(2l+1)S!}{2^{S+l+1}l!(S-l)!}\int^{1}_{-1}\mathrm{d}x (1-x)^{S}(1+x)^{l} \nonumber\\
	&=&\frac{(-1)^{l}(2l+1)S!S!}{(S-l)!(S+l+1)!}. \label{ccoeff}
\end{eqnarray}
Having expansion coefficients (\ref{ccoeff}) and by replacing $x$ with $\hat{\Omega}_{j}\cdot\hat{\Omega}_{j+1}$, the factor $\left[\frac{1}{2}(1-\hat{\Omega}_{j}\cdot\hat{\Omega}_{j+1})\right]^{S}$ under the integral of (\ref{norm1}) can be expanded in terms of Legendre polynomials and further in terms of spherical harmonics by further using
\begin{eqnarray}
	P_{l}(\hat{\Omega}_{j}\cdot\hat{\Omega}_{j+1})=\frac{4\pi}{2l+1}\sum^{l}_{m=-l}Y_{lm}(\hat{\Omega}_{j})Y^{\ast}_{lm}(\hat{\Omega}_{j+1}).
\end{eqnarray}
The final result is \cite{FM,KHH} 
\begin{eqnarray}
	&&\left[\frac{1}{2}(1-\hat{\Omega}_{j}\cdot\hat{\Omega}_{j+1})\right]^{S} \nonumber\\
	&=&\frac{1}{S+1}\sum^{S}_{l=0}(2l+1)\lambda(l,S)P_{l}(\hat{\Omega}_{j}\cdot\hat{\Omega}_{j+1})\nonumber \\
	&=&\frac{4\pi}{S+1}\sum^{S}_{l=0}\lambda(l,S)\sum^{l}_{m=-l}Y_{lm}(\hat{\Omega}_{j})Y^{\ast}_{lm}(\hat{\Omega}_{j+1}) \label{expansion}
\end{eqnarray}
with coefficients $\lambda(l,S)$ given by
\begin{eqnarray}
	\lambda(l,S)\equiv\frac{(-1)^{l}S!(S+1)!}{(S-l)!(S+l+1)!}. \label{lamb}
\end{eqnarray}
Now we expand $\left[\frac{1}{2}(1-\hat{\Omega}_{j}\cdot\hat{\Omega}_{j+1})\right]^{S}$ in terms of spherical harmonics as in (\ref{expansion}), then integrate from $\hat{\Omega}_{0}$ to $\hat{\Omega}_{N+1}$. We notice by using the orthogonality of spherical harmonics that each integral contributes a factor of $4\pi/(S+1)$ except the last one. For example,
\begin{eqnarray}
	&&\int \mathrm{d}\hat{\Omega}_{0}\left[\frac{1}{2}(1-\hat{\Omega}_{0}\cdot\hat{\Omega}_{1})\right]^{S} \nonumber\\	&=&\frac{4\pi}{S+1}\sum^{S}_{l=0}\lambda(l,S)\sum^{l}_{m=-l}\sqrt{4\pi}\ Y^{\ast}_{lm}(\hat{\Omega}_{1})\int \mathrm{d}\hat{\Omega}_{0}\,Y_{lm}(\hat{\Omega}_{0})Y^{\ast}_{00}(\hat{\Omega}_{0}) 
\nonumber \\
	&=&\frac{4\pi}{S+1}\sqrt{4\pi}\ Y^{\ast}_{00}(\hat{\Omega}_{1})= \frac{4\pi}{S+1}. \label{exam}
\end{eqnarray}
The last integral over $\hat{\Omega}_{N+1}$ contributes simply a factor of $4\pi$. Consequently, the norm square (\ref{norm1}) is equal to
\begin{eqnarray}
	\langle \mathrm{VBS}|\mathrm{VBS}\rangle=\left[\frac{(2S+1)!}{S+1}\right]^{N}S!(S+1)!. \label{normvbs}
\end{eqnarray}

\subsection{The Block Density Matrix}
\label{sec:1dssden}

We take a block of $L$ contiguous bulk spins as a subsystem. Now we calculate the block density matrix in the VBS state (\ref{vbs}). By definition, this is achieved by taking the pure state density matrix  (\ref{pure}) and tracing out all spin degrees of freedom outside the block:
\begin{eqnarray}
	\boldsymbol{\rho}_{L}\equiv \mathrm{tr}_{0, 1, \ldots, k-1, k+L, \ldots, N, N+1}\ \left[\,\boldsymbol{\rho}\,\right],\qquad 1\leq k,\quad k+L-1\leq N. \label{trac}
\end{eqnarray}
Here the block of length $L$ starts from site $k$ and ends at site $k+L-1$. $\boldsymbol{\rho}_{L}$ is no longer a pure state density matrix because of entanglement of the block with the environment (sites outside the block of the spin chain). It was shown in \cite{JK,XKHK} that entries of the density matrix are multi-point correlation functions in the ground state. We give the proof of this statement for our spin-$S$ case in \S$\,\ref{sec:denandcorr}\,$. 

Using the coherent state basis (\ref{cohebas}) and completeness relation (\ref{comp}), $\boldsymbol{\rho}_{L}$ can be written as \cite{KHH}
\begin{eqnarray}
	&&\boldsymbol{\rho}_{L}=
	\label{rougs} \\
	&&\frac{
\displaystyle\int\left[\prod^{k-1}_{j=0}\prod^{N+1}_{j=k+L}\mathrm{d}\hat{\Omega}_{j}\right]\prod^{k-2}_{j=0}\prod^{N}_{j=k+L}\left[\frac{1}{2}(1-\hat{\Omega}_{j}\cdot\hat{\Omega}_{j+1})\right]^{S}
	B^{\dagger}|\mathrm{VBS}_{L}\rangle\langle \mathrm{VBS}_{L}|B}
	{\displaystyle\left[\frac{(2S+1)!}{4\pi}\right]^{L}
\int\left[\prod^{N+1}_{j=0}\mathrm{d}\hat{\Omega}_{j}\right]\prod^{N}_{j=0}\left[\frac{1}{2}(1-\hat{\Omega}_{j}\cdot\hat{\Omega}_{j+1})\right]^{S}}. \nonumber 
\end{eqnarray}
Here the boundary operator $B$ and block VBS state $\left|\mathrm{VBS}_{L}\right\rangle$ are defined as
\begin{eqnarray}
	&&B\equiv \left(u_{k-1}b_{k}-v_{k-1}a_{k}\right)^{S}\left(a_{k+L-1}v_{k+L}-b_{k+L-1}u_{k+L}\right)^{S},\label{bope} \\
	&&|\mathrm{VBS}_{L}\rangle \equiv
 \prod^{k+L-2}_{j=k}
\left(a^{\dagger}_{j}b^{\dagger}_{j+1}-b^{\dagger}_{j}a^{\dagger}_{j+1}\right)^{S}|\mathrm{vac}\rangle, \label{vbsl}
\end{eqnarray}
respectively. Note that both $B$ and $|\mathrm{VBS}_{L}\rangle$ are $SU(2)$ covariant (see \S$\,\ref{sec:larges}\,$). The expression (\ref{rougs}) can be simplified. We can perform the integrals over $\hat{\Omega}_{j}$ ($j=0, 1, \ldots, k-2, k+L+1, \ldots, N, N+1$) in the numerator and all integrals in the denominator (see \S$\,\ref{sec:norm}\,$). After integrating over these variables, the density matrix $\boldsymbol{\rho}_{L}$ turns out to be independent of both the starting site $k$ and the total length $N$ of the spin chain. This property has been proved in \cite{FKR} for spin $S=1$ (using a different representation, namely the maximally entangled states, see \S$\,\ref{sec:1ds1den}\,$) and generalized in \cite{KHH} for generic spin-$S$. Therefore, we can choose $k=1$ (a relabeling for convenience) and the density matrix takes the form
\begin{eqnarray}
	\boldsymbol{\rho}_{L}=\left[\frac{S+1}{(2S+1)!}\right]^{L}\frac{(S+1)}{(4\pi)^{2}}
	\int \mathrm{d}\hat{\Omega}_{0}\mathrm{d}\hat{\Omega}_{L+1}\,B^{\dagger}|\mathrm{VBS}_{L}\rangle\langle \mathrm{VBS}_{L}|B \label{matrix}
\end{eqnarray}
with
\begin{eqnarray}
&&B^{\dagger}=\left(u^{\ast}_{0}b^{\dagger}_{1}-v^{\ast}_{0}a^{\dagger}_{1}\right)^{S}\left(a^{\dagger}_{L}v^{\ast}_{L+1}-b^{\dagger}_{L}u^{\ast}_{L+1}\right)^{S}, \label{bope1} \\
	&&|\mathrm{VBS}_{L}\rangle=\prod^{L-1}_{j=1}
\left(a^{\dagger}_{j}b^{\dagger}_{j+1}-b^{\dagger}_{j}a^{\dagger}_{j+1}\right)^{S}|\mathrm{vac}\rangle. \label{vbsl1}	
\end{eqnarray}
The state $|\mathrm{VBS}_{L}\rangle$ is called the block VBS state. The last two integral of (\ref{matrix}) can be performed, but we keep its present form for later use.

\subsection{Ground States of the Block Hamiltonian}
\label{sec:gsbhs}

\subsubsection{Degenerate Ground States}

In order to describe the eigenvectors and spectrum of the density matrix (\ref{matrix}), we first study the zero-energy ground states of the block Hamiltonian. The block Hamiltonian is a collection of interacting terms within the block, \textit{i.e.}
\begin{eqnarray}
	H_{B}=\sum^{L-1}_{j=1}\sum^{2S}_{J=S+1} C_{J}\pi_{J}(j, j+1). \label{degels}
\end{eqnarray}
Now we define a set of $S+1$ operators covariant under $SU(2)$ (see \S$\,\ref{sec:larges}\,$)
\begin{eqnarray}
 A^{\dagger}_{J}\equiv
\left(ua^{\dagger}_{1}+vb^{\dagger}_{1}\right)^{J}  \left(a^{\dagger}_{1}b^{\dagger}_{L}-b^{\dagger}_{1}a^{\dagger}_{L}\right)^{S-J}
\left(ua^{\dagger}_{L}+vb^{\dagger}_{L}\right)^{J}, \quad 0\leq J\leq S. \label{aope}
\end{eqnarray}
These operators act on the direct product of Hilbert spaces of spins at site $1$ and site $L$. Then the set of ground states of (\ref{degels}) can be chosen as
\begin{eqnarray}
	|\mathrm{G}; J, \hat{\Omega}\rangle \equiv
A^{\dagger}_{J}|\mathrm{VBS}_{L}\rangle, \qquad J=0, \ldots, S. \label{eiges}
\end{eqnarray}
Any state $|\mathrm{G}; J, \hat{\Omega}\rangle$ of this set for fixed $J$ and $\hat{\Omega}$ is a zero-energy ground state of (\ref{degels}). To prove this we need only to verify:
\begin{enumerate}
	\item The total power of $a^{\dagger}_{1}$ and $b^{\dagger}_{1}$ is $2S$, so that we have spin-$S$ at the first site;
	\item The bond spin value satisfies $-S\leq J^{z}_{1,2}\equiv S^{z}_{1}+S^{z}_{2}\leq S$ by a binomial expansion, so that the maximum value of the bond spin $J_{1,2}$ is $S$ (from $SU(2)$ invariance, see \S$\,\ref{sec:larges}\,$ and \cite{AAH}).
\end{enumerate}
These properties are true for any other site $j$ and bond $\langle j, j+1\rangle$, respectively. Therefore, the state $|\mathrm{G}; J, \hat{\Omega}\rangle$ defined in (\ref{eiges}) has spin-$S$ at each site and no projection onto the $J_{j, j+1}>S$ subspace for any bond.

\subsubsection{Degenerate VBS States}

The set of states $\{|\mathrm{G}; J, \hat{\Omega}\rangle\}$ depend on a discrete parameter $J$ as well as a continuous unit vector $\hat{\Omega}$. States with the same $J$ value are not mutually orthogonal. It is possible also to introduce an orthogonal basis in description of the degenerate zero-energy ground states. This new basis could be used in determining the rank and the completeness of the set $\{|\mathrm{G}; J, \hat{\Omega}\rangle\}$. For notational convenience, we define
\begin{eqnarray}
	X_{JM}\equiv\frac{u^{J+M}v^{J-M}}{\sqrt{(J+M)!(J-M)!}}, \qquad
	\psi^{\dagger}_{Sm}\equiv\frac{(a^{\dagger})^{S+m}(b^{\dagger})^{S-m}}{\sqrt{(S+m)!(S-m)!}}. \label{2def}
\end{eqnarray}
These two variables transform conjugately with respect to one another under $SU(2)$. (See \S$\,\ref{sec:larges}\,$ for more details of transformation properties.) Variable $X_{JM}$ has the following orthogonality relation
\begin{eqnarray}
	\int \mathrm{d}\hat{\Omega}\,X^{\ast}_{JM}X_{JM^{\prime}}=\frac{4\pi}{(2J+1)!}\,\delta_{MM^{\prime}}. \label{orth}
\end{eqnarray}
Operator $\psi^{\dagger}_{Sm}$ is a spin state creation operator such that
\begin{eqnarray}
	\psi^{\dagger}_{Sm}|\mathrm{vac}\rangle=|S, m\rangle. \label{crea1}
\end{eqnarray}
With the introduction of these variables (\ref{2def}), the operator $A^{\dagger}_{J}$ defined in (\ref{aope}) can be expanded as (see Chapter $9$ of \cite{Ha})
\begin{eqnarray}
	A^{\dagger}_{J}&=&\sqrt{\frac{(S+J+1)!(S-J)!J!J!}{2J+1}} \label{aexp} \\
	&&\cdot\sum^{J}_{M=-J}X_{JM}\sum^{m_{1}+m_{L}=M}_{m_{1}, m_{L}}(S/2, m_{1}; S/2, m_{2}|J, M)\,\psi^{\dagger}_{S/2,m_{1}}\otimes\psi^{\dagger}_{S/2,m_{L}}, \nonumber
\end{eqnarray}
where $(S/2, m_{1}; S/2, m_{2}|J, M)$ are the Clebsch-Gordan coefficients. Note that $\psi^{\dagger}_{S/2,m_{1}}$ and $\psi^{\dagger}_{S/2,m_{L}}$ are defined in the Hilbert spaces of spins at site $1$ and site $L$, respectively. We realize that the particular form of the sum over $m_{1}$ and $m_{L}$ in (\ref{aexp}) can be identified as a single spin state creation operator
\begin{eqnarray}
	\Psi^{\dagger}_{JM}\equiv\sum^{m_{1}+m_{L}=M}_{m_{1}, m_{L}}(S/2, m_{1}; S/2, m_{2}|J, M)\,\psi^{\dagger}_{S/2,m_{1}}\otimes\psi^{\dagger}_{S/2,m_{L}}. \label{crea2}
\end{eqnarray}
This operator $\Psi^{\dagger}_{JM}$ acts on the direct product of two Hilbert spaces of spins at site $1$ and site $L$. It has the property that
\begin{eqnarray}
	\Psi^{\dagger}_{JM}|\mathrm{vac}\rangle_{1}\otimes|\mathrm{vac}\rangle_{L}=|J, M\rangle_{1,L}. \label{crea3}
\end{eqnarray}
If we define a set of \textit{degenerate VBS states} $\{|\mathrm{VBS}_L(J,M)\rangle\}$ such that
\begin{eqnarray}
	|\mathrm{VBS}_L(J,M)\rangle\equiv \Psi^{\dagger}_{JM}|\mathrm{VBS}_L\rangle, \quad J=0,...,S, \quad M=-J, ...,J, \label{devb}
\end{eqnarray}
then these $(S+1)^{2}$ states (\ref{devb}) are not only linearly independent but also mutually orthogonal.

\subsubsection{The Orthogonality}

To show the orthogonality of the degenerate VBS states (\ref{devb}), it is convenient to introduce the total spin operators of the subsystem (block):
\begin{equation}
S^{+}_{\mathrm{tot}}=\sum^{L}_{j=1} a^{\dagger}_{j} b_{j}, \qquad
S^{-}_{\mathrm{tot}}=\sum^{L}_{j=1} b^{\dagger}_{j} a_{j}, \qquad
S^{z}_{\mathrm{tot}}=\frac{1}{2}\sum^{L}_{j=1} (a^{\dagger}_{j} a_{j}-b^{\dagger}_{j} b_{j}).
\end{equation}
First we show that the set of operators $\{ S^{+}_{\mathrm{tot}}, S^{-}_{\mathrm{tot}}, S^{z}_{\mathrm{tot}} \}$ commute with the product of valence bonds, \textit{i.e.}
\begin{equation}
[\,S^{\pm}_{\mathrm{tot}}\,,\,\prod_{j=1}^{L-1} (a^{\dagger}_{j} b^{\dagger}_{j+1}-b^{\dagger}_{j} a^{\dagger}_{j+1})^{S}\,]=0, \quad 
[\,S^{z}_{\mathrm{tot}}\,,\,\prod_{j=1}^{L-1} (a^{\dagger}_{j} b^{\dagger}_{j+1}-b^{\dagger}_{j} a^{\dagger}_{j+1})^{S}\,]=0.
\label{commutere}
\end{equation}
These commutation relations (\ref{commutere}) can be shown in similar ways. Take the commutator with $S^{+}_{\mbox{\scriptsize{tot}}}$ first. We re-write the commutator as
\begin{eqnarray}
&&[\,S^{+}_{\mathrm{tot}}\,,\,\prod_{j=1}^{L-1} (a^{\dagger}_{j} b^{\dagger}_{j+1}-b^{\dagger}_{j} a^{\dagger}_{j+1})^{S}\,] \\
&=& \sum_{j=1}^{L-1} (a^{\dagger}_{1} b^{\dagger}_{2}-b^{\dagger}_{1} a^{\dagger}_{2})^{S} \cdots 
[\,S^{+}_{\mbox{\scriptsize{tot}}}\,,\,(a^{\dagger}_{j} b^{\dagger}_{j+1}-b^{\dagger}_{j} a^{\dagger}_{j+1})^{S}\,] \cdots (a^{\dagger}_{L-1} b^{\dagger}_{L}-b^{\dagger}_{L-1} a^{\dagger}_{L})^{S} \nonumber \\
&=& \sum_{j=1}^{L-1} (a^{\dagger}_{1} b^{\dagger}_{2}-b^{\dagger}_{1} a^{\dagger}_{2})^{S} \cdots 
[\,S^{+}_{j} + S^{+}_{j+1}\,,\,(a^{\dagger}_{j} b^{\dagger}_{j+1}-b^{\dagger}_{j} a^{\dagger}_{j+1})^{S}\,]
\cdots\nonumber\\ &&\cdots(a^{\dagger}_{L-1} b^{\dagger}_{L}-b^{\dagger}_{L-1} a^{\dagger}_{L})^{S}.
\nonumber
\end{eqnarray}
Then using commutators $[a_{i}, a^{\dagger}_{j}]=\delta_{ij}$ and $[b_{i}, b^{\dagger}_{j}]=\delta_{ij}$, we find that
\begin{eqnarray}
&& [\,S^{+}_{j} + S^{+}_{j+1}\,,\,(a^{\dagger}_{j} b^{\dagger}_{j+1}-b^{\dagger}_{j} a^{\dagger}_{j+1})^{S}\,]
\nonumber \\
&=& [\,a^{\dagger}_{j} b_{j}+a^{\dagger}_{j+1} b_{j+1}\,,\,(a^{\dagger}_{j} b^{\dagger}_{j+1}-b^{\dagger}_{j} a^{\dagger}_{j+1})^{S}\,] \nonumber \\
&=& a^{\dagger}_{j} [\,b_{j}\,,\,(a^{\dagger}_{j} b^{\dagger}_{j+1}-b^{\dagger}_{j} a^{\dagger}_{j+1})^{S}\,]+
      a^{\dagger}_{j+1} [\,b_{j+1}\,,\,(a^{\dagger}_{j} b^{\dagger}_{j+1}-b^{\dagger}_{j} a^{\dagger}_{j+1})^{S}\,]
      \nonumber \\
&=& a^{\dagger}_{j} (-S)a^{\dagger}_{j+1} (a^{\dagger}_{j} b^{\dagger}_{j+1}-b^{\dagger}_{j} a^{\dagger}_{j+1})^{S-1}
    +a^{\dagger}_{j+1} S a^{\dagger}_{j} (a^{\dagger}_{j} b^{\dagger}_{j+1}-b^{\dagger}_{j} a^{\dagger}_{j+1})^{S-1}
\nonumber \\
&=&0. \label{comm}
\end{eqnarray}
Therefore $[\,S^{+}_{\mathrm{tot}}\,,\,\prod_{j=1}^{L-1} (a^{\dagger}_{j} b^{\dagger}_{j+1}-b^{\dagger}_{j} a^{\dagger}_{j+1})^{S}\,]=0$. In (\ref{comm}) we have used
\begin{eqnarray}
	[\,b_{j}\,,\,(a^{\dagger}_{j} b^{\dagger}_{j+1}-b^{\dagger}_{j} a^{\dagger}_{j+1})^{S}\,]=-S a^{\dagger}_{j+1} (a^{\dagger}_{j} b^{\dagger}_{j+1}-b^{\dagger}_{j} a^{\dagger}_{j+1})^{S-1}.
\end{eqnarray}
In a similar way, we find that the commutator with $S^{-}_{\mathrm{tot}}$ also vanishes. Next we consider the commutator with $S^{z}_{\mathrm{tot}}$:
\begin{eqnarray}
&&[\,S^{z}_{\mathrm{tot}}\,,\,\prod_{j=1}^{L-1} (a^{\dagger}_{j} b^{\dagger}_{j+1}-b^{\dagger}_{j} a^{\dagger}_{j+1})^{S}\,] \label{comSz} \\
&=& \sum_{j=1}^{L-1} (a^{\dagger}_{1} b^{\dagger}_{2}-b^{\dagger}_{1} a^{\dagger}_{2})^{S}
\cdots [\,S^{z}_{j} + S^{z}_{j+1}\,,\,(a^{\dagger}_{j} b^{\dagger}_{j+1}-b^{\dagger}_{j} a^{\dagger}_{j+1})^{S}\,]\cdots\nonumber\\
&&\cdots(a^{\dagger}_{L-1} b^{\dagger}_{L}-b^{\dagger}_{L-1} a^{\dagger}_{L})^{S}.
\nonumber
\end{eqnarray}
In the right hand side of (\ref{comSz}), the commutator involved also vanishes because
\begin{eqnarray}
&&[\,S^{z}_{j} + S^{z}_{j+1}\,,\,(a^{\dagger}_{j} b^{\dagger}_{j+1}-b^{\dagger}_{j} a^{\dagger}_{j+1})^{S}\,] 
\nonumber \\
&=& \frac{1}{2}[\,a^{\dagger}_{j} a_{j} -b^{\dagger}_{j} b_{j} +a^{\dagger}_{j+1} a_{j+1} -b^{\dagger}_{j+1} b_{j+1}\,,\,(a^{\dagger}_{j} b^{\dagger}_{j+1}-b^{\dagger}_{j} a^{\dagger}_{j+1})^{S}\,] 
\nonumber \\
&=& \frac{1}{2}a^{\dagger}_{j}  [\,a_{j}\,,\,(a^{\dagger}_{j} b^{\dagger}_{j+1}-b^{\dagger}_{j} a^{\dagger}_{j+1})^{S}\,]
     -\frac{1}{2}b^{\dagger}_{j}  [\,b_{j}\,,\,(a^{\dagger}_{j} b^{\dagger}_{j+1}-b^{\dagger}_{j} a^{\dagger}_{j+1})^{S}\,] 
\nonumber \\
&&+ \frac{1}{2}a^{\dagger}_{j+1} [\,a_{j+1}\,,\,(a^{\dagger}_{j} b^{\dagger}_{j+1}-b^{\dagger}_{j} a^{\dagger}_{j+1})^{S}\,]
     -\frac{1}{2}b^{\dagger}_{j+1} [\,b_{j+1}\,,\,(a^{\dagger}_{j} b^{\dagger}_{j+1}-b^{\dagger}_{j} a^{\dagger}_{j+1})^{S}\,]
\nonumber \\
&=& 0  \label{com1}
\end{eqnarray}
Substituting (\ref{com1}) into (\ref{comSz}), we obtain $[\,S^{z}_{\mathrm{tot}}\,,\,\prod_{j=1}^{L-1} (a^{\dagger}_{j} b^{\dagger}_{j+1}-b^{\dagger}_{j} a^{\dagger}_{j+1})^{S}\,]=0$.
Now we shall show that the state $|\mathrm{VBS}_L(J,M)\rangle$ is a common eigenstate of $S^{z}_{\mathrm{tot}}$ and the total spin square $\boldsymbol{S}^2_{\mathrm{tot}}=\frac{1}{2}(S^{+}_{\mathrm{tot}}S^{-}_{\mathrm{tot}}+S^{-}_{\mathrm{tot}}S^{+}_{\mathrm{tot}}) + (S^{z}_{\mathrm{tot}})^{2}$ with eigenvalues $M$ and $J(J+1)$, respectively. 
Using the commutation relations (\ref{commutere}), we can show that
\begin{eqnarray}
S^{\pm}_{\mathrm{tot}}|\mathrm{VBS}_L(J,M)\rangle = \prod_{j=1}^{L-1} (a^{\dagger}_{j} b^{\dagger}_{j+1}-b^{\dagger}_{j} a^{\dagger}_{j+1})^{S} (S^{\pm}_{1} + S^{\pm}_{L})|J,M \rangle_{1,L} |\mathrm{vac}\rangle_{2, ..., L-1} \nonumber \\
S^{z}_{\mathrm{tot}}|\mathrm{VBS}_L(J,M)\rangle = \prod_{j=1}^{L-1} (a^{\dagger}_{j} b^{\dagger}_{j+1}-b^{\dagger}_{j} a^{\dagger}_{j+1})^{S} (S^{z}_{1} + S^{z}_{L})|J,M \rangle_{1,L} |\mathrm{vac}\rangle_{2,..., L-1}.\nonumber\\
\end{eqnarray}
Then from the definition of the state $|\mathrm{VBS}_L(J,M)\rangle$ and the following relations:
\begin{eqnarray}
(S^{\pm}_1+S^{\pm}_{L})|J,M \rangle_{1,L}&=& \sqrt{(J \mp M) (J \pm M +1)}\,\,|J, M\pm1 \rangle, 
\nonumber \\
(S^{z}_{1}+S^{z}_{L})|J,M \rangle_{1,L} &=& M |J,M \rangle_{1,L},
\end{eqnarray}
we obtain 
\begin{eqnarray}
S^{\pm}_{\mathrm{tot}}|\mathrm{VBS}_L(J,M)\rangle &=& \sqrt{(J \mp M) (J \pm M +1)}\,\,|\mathrm{VBS}_L(J,M\pm1)\rangle, \nonumber \\
S^z_{\mathrm{tot}}|\mathrm{VBS}_L(J,M)\rangle &=& M |\mathrm{VBS}_L(J,M)\rangle
\end{eqnarray}
and hence $\boldsymbol{S}^{2}_{\mathrm{tot}}|\mathrm{VBS}_L(J,M)\rangle=J(J+1)|\mathrm{VBS}_L(J,M)\rangle$.
It is now proved that $|\mathrm{VBS}_L(J,M)\rangle$ is a common eigenstate of Hermitian operators $S^{z}_{\mbox{\scriptsize{tot}}}$ and $\boldsymbol{S}^{2}_{\mbox{\scriptsize{tot}}}$ with eigenvalues $M$ and $J(J+1)$, respectively. Therefore the states with different eigenvalues $(J, M)$ are orthogonal to each other. Thus we have proved the orthogonality of the set $\{|\mathrm{VBS}_L(J,M)\rangle\,|\,J=0,\ldots,S; \ M=-J,\ldots,J\}$.

\subsubsection{Completeness and Equivalence}

It is obvious from (\ref{aexp}) that any ground state $|\mathrm{G}; J, \hat{\Omega}\rangle$ can be written as a linear superposition over these degenerate VBS states:
\begin{eqnarray}
	|\mathrm{G};J,\hat{\Omega}\rangle=\sqrt{\frac{(S+J+1)!(S-J)!J!J!}{2J+1}}\sum^{J}_{M=-J}X_{JM}|\mathrm{VBS}_{L}(J, M)\rangle, \label{line}
\end{eqnarray}
and \textit{vice versa}. Now we can derive the completeness relation of the set $\{|\mathrm{G}; J, \hat{\Omega}\rangle\}$ using (\ref{orth}), (\ref{aexp}) and (\ref{crea2}):
\begin{eqnarray}
	&&\int \mathrm{d}\hat{\Omega}\,|\mathrm{G}; J, \hat{\Omega}\rangle\langle \mathrm{G}; J, \hat{\Omega}| \label{compga} \\
	&=&\frac{4\pi}{(2J+1)!}\frac{(S+J+1)!(S-J)!J!J!}{2J+1}\sum^{J}_{M=-J}\Psi^{\dagger}_{JM}|\mathrm{VBS}_{L}\rangle\langle \mathrm{VBS}_{L}|\Psi_{JM}. \nonumber
\end{eqnarray}
The set of states $\{\Psi^{\dagger}_{JM}|\mathrm{VBS}_{L}\rangle\ |\ M=-J, \ldots, J\}$  are linearly independent. So that the rank of $\{|\mathrm{G}; J, \hat{\Omega}\rangle\}$ with fixed $J$ value is $2J+1$, which can be obtained from the completeness relation (\ref{compga}) (see \cite{Ha}). Thus the total number of linearly independent states of the set $\{|\mathrm{G}; J, \hat{\Omega}\rangle\}$ is $\sum^{S}_{J=0}(2J+1)=(S+1)^2$, which is exactly the degeneracy of the ground states of (\ref{degels}). So that $\{|\mathrm{G}; J, \hat{\Omega}\rangle\}$ forms a complete set of zero-energy ground states. The set $\{|\mathrm{VBS}_L(J,M)\rangle\}$ differs from $\{|\mathrm{G}; J, \hat{\Omega}\rangle\}$ by a change of basis, therefore it also forms a complete set of zero-energy ground states. These two sets (\ref{eiges}) and (\ref{devb}) are equivalent in description of the degenerate ground states of the block Hamiltonian (\ref{degels}). (More details such as the expansion (\ref{aexp}) \textit{etc.} can be found in Chapter $9$ of \cite{Ha}.)

\subsection{Eigenvectors of the Density Matrix}
\label{sec:eigvecs}

Eigenvalues of the density matrix $\boldsymbol{\rho}_{L}$ can be derived indirectly, as in \cite{FKR} for spin-$1$ (see \S$\,\ref{sec:1ds1specden}\,$ for comparison) and in \cite{KHH} for spin-$S$. The basic idea is the following: Because the density matrix is independent of both the total length of the spin chain and the starting site of the block, we can add boundary spins directly to the ends of the block. It was shown in \cite{FKR,KHH} by a Schmidt decomposition (see Section $2.5$ of \cite{NC}) that non-zero eigenvalues of the density matrix (\ref{matrix}) are equal to those of the density matrix of the two boundary spins. All other eigenvalues of the density matrix (\ref{matrix}) are zero. This fact reveals the structure of the density matrix as a projector (up to a multiplicative `scaling' matrix) onto a subspace of dimension $(S+1)^{2}$.

Now we propose a theorem on the eigenvectors of the density matrix $\boldsymbol{\rho}_{L}$ given by (\ref{matrix}). The explicit construction of eigenvectors allows us to diagonalize the density matrix directly. The set of eigenvectors also spans the subspace that the density matrix projects onto.

\begin{theorem}
\label{theorem2}
Eigenvectors of the density matrix $\boldsymbol{\rho}_{L}$ (\ref{matrix}) with non-zero eigenvalues are given by the set $\{\,|\mathrm{G}; J, \hat{\Omega}\rangle\,\}$ (\ref{eiges}), or, equivalently, by the set $\{\,|\mathrm{VBS}_L(J,M)\rangle\,\}$ (\ref{devb}). \textit{i.e.} They are zero-energy ground states of the block Hamiltonian $H_{B}$ (\ref{degels}). (We shall emphasize that the eigenvectors $|\mathrm{G}; J, \hat{\Omega}\rangle$ and $|\mathrm{VBS}_L(J,M)\rangle$ correspond to the vectors $|\lambda\rangle$ in \S$\,\ref{sec:rhoands}\,$ and in \textbf{Theorem} \ref{theorem1} of \S$\,\ref{sec:den0spec}\,$.)
\end{theorem}

\begin{proof}
We prove this theorem by showing that the density matrix $\boldsymbol{\rho}_{L}$ (\ref{matrix}) can be written as a projector in diagonal form onto the orthogonal degenerate VBS states $\{\,|\mathrm{VBS}_{L}(J, M)\rangle\,\}$ introduced in (\ref{devb}). An alternative proof taking a different approach is given in the next section \S$\,\ref{sec:alterproof}\,$.

First, it is realized from the definition of spinor coordinates (\ref{spin}) that if we change variables $(u, v)$ to $(\mathrm{i}v^{\ast}, -\mathrm{i}u^{\ast})$, then the unit vector $\hat{\Omega}$ is inverted about the origin to $-\hat{\Omega}$. So that we have \cite{KHH}
\begin{eqnarray}
	(u^{\ast}b^{\dagger}-v^{\ast}a^{\dagger})^{S}|\mathrm{vac}\rangle=\mathrm{i}^{S}\sqrt{S!}\,\,|-\hat{\Omega}\rangle, \label{inve}
\end{eqnarray}
where $|-\hat{\Omega}\rangle$ means a spin-$S/2$ coherent state for a point opposite to $\hat{\Omega}$ on the unit sphere. Therefore, taking expressions of the boundary operator $B^{\dagger}$ (\ref{bope1}) and the block VBS state $|\mathrm{VBS}_{L}\rangle$ (\ref{vbsl1}), we have
\begin{eqnarray}
	&&B^{\dagger}|\mathrm{VBS}_{L}\rangle \label{bvbs}= \\
	&&S!\prod^{L-1}_{j=1}
\left(a^{\dagger}_{j}b^{\dagger}_{j+1}-b^{\dagger}_{j}a^{\dagger}_{j+1}\right)^{S}|-\hat{\Omega}_{0}\rangle_{1}\otimes|\mathrm{vac}\rangle_{2}\otimes\cdots\otimes|\mathrm{vac}\rangle_{L-1}\otimes|-\hat{\Omega}_{L+1}\rangle_{L}. \nonumber
\end{eqnarray}
Consequently the density matrix $\boldsymbol{\rho}_{L}$ (\ref{matrix}) can be re-written as
\begin{eqnarray}
	\boldsymbol{\rho}_{L}&&=\left[\frac{S+1}{(2S+1)!}\right]^{L}\frac{S!S!}{S+1}\prod^{L-1}_{j=1}
\left(a^{\dagger}_{j}b^{\dagger}_{j+1}-b^{\dagger}_{j}a^{\dagger}_{j+1}\right)^{S} \label{rewr} \\
&&\cdot I^{(1)}_{S+1}\otimes|\mathrm{vac}\rangle_{2}\langle \mathrm{vac}|\otimes\cdots\otimes|\mathrm{vac}\rangle_{L-1}\langle \mathrm{vac}|\otimes I^{(L)}_{S+1}\prod^{L-1}_{j=1}\left(a_{j}b_{j+1}-b_{j}a_{j+1}\right)^{S}, \nonumber
\end{eqnarray}
where $I^{(1)}_{S+1}$ and $I^{(L)}_{S+1}$ are $(S+1)$-dimensional identities associated with site $1$ and site $L$, respectively. In obtaining (\ref{rewr}), we have changed integral variables from $\hat{\Omega}_{0}$ , $\hat{\Omega}_{L+1}$ to $-\hat{\Omega}_{0}$, $-\hat{\Omega}_{L+1}$ and performed these two integrals using the completeness relation (\ref{comp}). Next we notice that (see \S$\,\ref{sec:gsbhs}\,$)
\begin{eqnarray}
	I^{(1)}_{S+1}\otimes I^{(L)}_{S+1}&=&\sum^{S}_{J=0}\sum^{J}_{M=-J}|J, M\rangle_{1,L}\langle J, M| \label{iide} \\
&=&\sum^{S}_{J=0}\sum^{J}_{M=-J}\Psi^{\dagger}_{JM}|\mathrm{vac}\rangle_{1}\langle \mathrm{vac}|\otimes|\mathrm{vac}\rangle_{L}\langle \mathrm{vac}|\Psi_{JM}. \nonumber
\end{eqnarray}
As a result, combining (\ref{rewr}) and (\ref{iide}), recalling definitions of $|\mathrm{VBS}_{L}\rangle$ (\ref{vbsl1}) and $|\mathrm{VBS}_{L}(J, M)\rangle$ (\ref{devb}), the density matrix $\boldsymbol{\rho}_{L}$ takes the following final form
\begin{eqnarray}
	\boldsymbol{\rho}_{L}&=&\left[\frac{S+1}{(2S+1)!}\right]^{L}\frac{S!S!}{S+1}\sum^{S}_{J=0}\sum^{J}_{M=-J}\Psi^{\dagger}_{JM}|\mathrm{VBS}_{L}\rangle\langle \mathrm{VBS}_{L}|\Psi_{JM} \label{fipr} \\
&=&\left[\frac{S+1}{(2S+1)!}\right]^{L}\frac{S!S!}{S+1}\sum^{S}_{J=0}\sum^{J}_{M=-J}|\mathrm{VBS}_{L}(J, M)\rangle\langle \mathrm{VBS}_{L}(J, M)|. \nonumber
\end{eqnarray}
The set of degenerate VBS states $\{|\mathrm{VBS}_{L}(J, M)\rangle\}$ with $J=0,\ldots,S$ and $M=-J,\ldots,J$ forms an orthogonal basis (see \S$\,\ref{sec:gsbhs}\,$). These $(S+1)^{2}$ states also forms a complete set of zero-energy ground states of the block Hamiltonian (\ref{degels}) (see \S$\,\ref{sec:gsbhs}\,$). So that in expression (\ref{fipr}) we have put the density matrix as a projector in diagonal form over an orthogonal basis. Each degenerate VBS state $|\mathrm{VBS}_{L}(J, M)\rangle$ is an eigenvector of the density matrix, so as any of the state $|\mbox{G}; J, \hat{\Omega}\rangle$ (because of the degeneracy of corresponding eigenvalues of the density matrix, see \S$\,\ref{sec:eigval1}\,$ and \S$\,\ref{sec:eigval2}\,$ that the eigenvalues depend only on $J$). Thus we have proved \textbf{Theorem} \ref{theorem2}.
\end{proof}

\subsection{An Alternative Proof of \textbf{Theorem} \ref{theorem2}}
\label{sec:alterproof}

It was shown in \S$\,\ref{sec:eigvecs}\,$ that the density matrix takes a diagonal form in the basis of zero-energy ground states of the block Hamiltonian (\ref{degels}). In this section, we show the same result by taking a different approach. This alternative proof of \textbf{Theorem} \ref{theorem2} does not involve the coherent state basis. 

The proof uses the fact that the density matrix is independent of the starting site and the total length of the chain (see \S$\,\ref{sec:1dssden}\,$). So that we could change the configuration of the whole system by adding the two ending spins directly to the block without affecting the form of the block density matrix. The new system now has $L+2$ sites with the block starting at site $1$ and ending at site $L$. Let us start with the ground VBS state of the Hamiltonian (\ref{uniq}) of the system with $N=L$:
\begin{eqnarray}
|\mathrm{VBS}\rangle&\equiv&\prod_{j=0}^{L}\left(
				      a_j^{\dagger}b_{j+1}^{\dagger}-b_j^{\dagger}a_{j+1}^{\dagger}\right)^{S}|\mathrm{vac}\rangle. \label{vbsn=l}
\end{eqnarray}
In order to calculate the density matrix $\boldsymbol{\rho}_L=
\mathrm{tr}_{0,L+1}\left[\,\boldsymbol{\rho}\,\right]$, where $\boldsymbol{\rho}$ is defined in (\ref{pure}),
we introduce a useful identity:
\begin{eqnarray}
{}_{0,L+1}\langle J,M|\left(|s\rangle_{0,1}\otimes|s\rangle_{L,L+1}\right)
&=&
\frac{(-1)^{S-J+M}}{(S+1)}\,|J,-M\rangle_{1,L},\label{EigenTrans}
\end{eqnarray}
where $|J,M\rangle_{0,L+1}$ is identical to the spin state defined in (\ref{crea3}) except for
site indices.
$|s\rangle_{i,j}$ in (\ref{EigenTrans}) is the normalized singlet state with $S$ valence bonds defined as
\begin{eqnarray}
 |s\rangle_{i,j}&=&\frac{1}{S!\sqrt{S+1}}\left(
a^{\dagger}_{i}b^{\dagger}_{j}-b^{\dagger}_{i}a^{\dagger}_{j}\right)^{S}|\mathrm{vac}\rangle_{i}\otimes|\mathrm{vac}\rangle_{j}
\nonumber \\ 
&=&\frac{(-1)^{\frac{S}{2}}}{\sqrt{S+1}}\sum_{m=-S/2}^{S/2}(-1)^{m}|S/2,-m \rangle_{i}\otimes|S/2,m \rangle_{j}. \label{singlet}
\end{eqnarray}
Identity (\ref{EigenTrans}) is derived using properties of the singlet state (\ref{singlet}) and Clebsch-Gordan coefficients as follows:
\begin{eqnarray}
&&{}_{0,L+1}\langle J,M|\left(|s\rangle_{0,1}\otimes|s\rangle_{L,L+1}\right) \nonumber\\
&=&
\sum^{m_{0}+m_{L+1}=M}_{m_{0},m_{L+1}}(J,M|S/2,m_{0};S/2,m_{L+1})
{}_{0}\langle S/2,m_{0}| {}_{L+1}\langle S/2,m_{L+1}|
\nonumber\\&&\cdot
\frac{(-1)^{\frac{S}{2}}}{\sqrt{S+1}}\sum_{m_{1}=-S/2}^{S/2}(-1)^{m_{1}}|S/2,-m_{1}\rangle_{0}|S/2,m_{1}\rangle_{1}
\nonumber\\&&\cdot
\frac{(-1)^{\frac{S}{2}}}{\sqrt{S+1}}\sum_{m_{L}=-S/2}^{S/2}(-1)^{m_{L}}|S/2,-m_{L}\rangle_{L}|S/2,m_{L}\rangle_{L+1}
\nonumber\\&=&
\frac{1}{S+1}
\sum^{m_{0}+m_{L+1}=M}_{m_0,m_{L+1}}(-1)^{m_0+m_{L+1}}
(J,M|S/2,m_0;S/2,m_{L+1})
\nonumber\\&&\cdot
|S/2,-m_0\rangle_1
|S/2,-m_{L+1}\rangle_L.
\end{eqnarray}
Here the Clebsch-Gordan coefficient is defined by
\begin{equation}
(J,M|S/2,m_0;S/2,m_{L+1})= {}_{i,j}\langle J,M|\left(|S/2,m_{0} \rangle_{i}\otimes|S/2,m_{L+1} \rangle_{j}\right).
\end{equation}
Then using the symmetry property of Clebsch-Gordan coefficients
\begin{eqnarray}
(J,M|S/2,m_{0};S/2,m_{L+1})
=(-1)^{S-J}
(J,-M|S/2,-m_{0};S/2,-m_{L+1})\nonumber\\
\end{eqnarray}
and the completeness of the basis $\{\,|S/2,m_{0}\rangle_{0}\otimes|S/2,m_{L+1}\rangle_{L+1}\,\}$,
we obtain the identity (\ref{EigenTrans}).

With the help of identity (\ref{EigenTrans}), we calculate the partial inner product of the VBS state (\ref{vbsn=l}) with the state $|J,M\rangle_{0,L+1}$, which is involved in taking trace of boundary spins. The VBS state $|\mbox{VBS}\rangle$ is decomposed into the bulk part and boundary parts, then
making use of (\ref{EigenTrans}), we have
\begin{eqnarray}
&&{}_{0,L+1}\langle J,M|\mathrm{VBS}\rangle\nonumber\\
&=&
{}_{0,L+1}\langle J,M|
\prod_{j=0}^{L}\left(
				      a_j^{\dagger}b_{j+1}^{\dagger}-b_j^{\dagger}a_{j+1}^{\dagger}\right)^{S}|\mathrm{vac}\rangle
\nonumber\\&=& S! (S+1)!
\prod_{j=1}^{L-1}\left(
				      a_j^{\dagger}b_{j+1}^{\dagger}-b_j^{\dagger}a_{j+1}^{\dagger}\right)^{S}
{}_{0,L+1}\langle J,M|s\rangle_{0,1} |s\rangle_{L,L+1} |\mathrm{vac}\rangle_{2\cdots L-1}
\nonumber\\
&=&(S!)^2
\prod_{j=1}^{L-1}\left(
				      a_j^{\dagger}b_{j+1}^{\dagger}-b_j^{\dagger}a_{j+1}^{\dagger}\right)^{S}
(-1)^{S-J+M}
|J,-M\rangle_{1,L} |\mathrm{vac}\rangle_{2\cdots L-1}
\nonumber\\
&=&
(-1)^{S-J+M}(S!)^2
|\mathrm{VBS}_L(J,-M)\rangle. \label{partial}
\end{eqnarray}
We see that the $(S+1)^2$ degenerate VBS states $\{\,|\mbox{VBS}_L(J,M)\rangle\,\}$ defined in (\ref{devb}) appear in the partial inner product (\ref{partial}). As discussed in \S$\,\ref{sec:gsbhs}\,$, they form a complete set of zero-energy ground states of the block Hamiltonian (\ref{degels}).

Now, it is straightforward to evaluate the density matrix as
\begin{eqnarray}
\mathrm{tr}_{0,L+1}\left[\,\boldsymbol{\rho}\,\right]
&=&
\sum_{J,M}
\frac{{}_{0,L+1}\langle J,M
|\mbox{VBS}\rangle\langle\mbox{VBS}|
 J,M\rangle_{0,L+1}
}{\langle\mbox{VBS}|\mbox{VBS}\rangle}
\nonumber\\&=&
\frac{(S!)^4}{\langle\mbox{VBS}|\mbox{VBS}\rangle}
\sum_{J,M}|\mbox{VBS}_{L}(J,-M)\rangle
\langle\mbox{VBS}_{L}(J,-M)|. \nonumber\\
\end{eqnarray}
This expression is identical to (\ref{fipr}) as we change dummy index from $M$ to $-M$. Therefore, in this approach again we arrive at \textbf{Theorem} \ref{theorem2} that the density matrix is proportional to a projector onto a subspace spanned by the $(S+1)^2$ ground states of the block Hamiltonian (\ref{degels}). Normalization $\langle\mathrm{VBS}|\mathrm{VBS}\rangle$
has been obtained in \S$\,\ref{sec:norm}\,$. States $|\mbox{VBS}_{L}(J,M)\rangle$ have been shown to be mutually orthogonal in \S$\,\ref{sec:gsbhs}\,$.

\subsection{Eigenvalues of the Density Matrix (normalization of degenerate VBS states)}
\label{sec:eigval1}

As the next step in analyzing the spectrum of the density matrix, now we study the eigenvalues. Based on the diagonalized form (\ref{fipr}), it is clear that eigenvalues of the density matrix $\boldsymbol{\rho}_{L}$ can be derived from the normalization of degenerate VBS states. We obtain an explicit expression for eigenvalues in terms of Wigner $3j$-symbols in this section. 

First, the following property is important:
Normalization of the degenerate VBS state $|\mathrm{VBS}_L(J,M)\rangle$ depends only on $J$ and is independent of $M$. This point is important in proving that any $|\mathrm{G};J,\hat{\Omega}\rangle$ is an eigenvector of $\boldsymbol{\rho}_{L}$ because it can be written as a superposition of $|\mathrm{VBS}_L(J,M)\rangle$'s with the same $J$ value (\ref{line}). With the introduction of total spin operators of the block $S^{\pm}_{\mathrm{tot}}$, $S^{z}_{\mathrm{tot}}$ and $\boldsymbol{S}^{2}_{\mathrm{tot}}$ (see \S$\,\ref{sec:gsbhs}\,$), we prove the statement as follows:
\begin{eqnarray}
&& \langle \mathrm{VBS}_{L}(J,M\pm 1)|\mathrm{VBS}_{L}(J,M\pm 1)\rangle
\nonumber \\
&=& \frac{1}{(J\mp M)(J \pm M+1)}\langle\mathrm{VBS}_{L}(J,M)|S^{\mp}_{\mathrm{tot}}S^{\pm}_{\mathrm{tot}}|\mathrm{VBS}_{L}(J,M)\rangle
\nonumber \\
&=& \frac{1}{(J\mp M)(J \pm M+1)}\langle\mathrm{VBS}_{L}(J,M)|(\boldsymbol{S}^{2}_{\mathrm{tot}}-(S^{z}_{\mathrm{tot}})^{2}\mp S^{z}_{\mathrm{tot}})|\mathrm{VBS}_{L}(J,M)\rangle
\nonumber \\
&=& \langle\mathrm{VBS}_{L}(J,M)|\mathrm{VBS}_{L}(J,M)\rangle.
\end{eqnarray}
Here we have used the fact that $|\mathrm{VBS}_{L}(J,M)\rangle$ is the common eigenstate of $\boldsymbol{S}^{2}_{\mathrm{tot}}$ and $S^{z}_{\mathrm{tot}}$ with eigenvalues $J(J+1)$ and $M$, respectively (see \S$\,\ref{sec:gsbhs}\,$). 

It is also realized that the normalization of $|\mathrm{VBS}_{L}(J,M)\rangle$ can be calculated from integrating the inner product of $|\mathrm{G};J,{\hat\Omega}\rangle$ with itself over the unit vector $\hat{\Omega}$ such that
\begin{eqnarray}
&&\frac{1}{4\pi}\int \mathrm{d}\hat{\Omega}\,\langle\mathrm{G};J,\hat{\Omega}|\mathrm{G};J,\hat{\Omega}\rangle \nonumber\\
&=& \frac{(S+J+1)!(S-J)!J!J!}{(2J+1)!}\langle\mathrm{VBS}_L(J,M)|\mathrm{VBS}_L(J,M)\rangle. \label{novb}
\end{eqnarray}
In obtaining this relation (\ref{novb}) we have used expansion (\ref{line}) and orthogonality (\ref{orth}) in \S$\,\ref{sec:gsbhs}\,$.

Let us consider the integral involved in (\ref{novb}). Using the coherent state basis (\ref{cohebas}) and completeness relation (\ref{comp}) as before, we obtain
\begin{eqnarray}
&&\frac{1}{4\pi}\int \mathrm{d}\hat{\Omega}\,\langle\mathrm{G};J,\hat{\Omega}|\mathrm{G};J,{\hat\Omega}\rangle
\label{inne} \\
&=& \frac{1}{4\pi}\left[\frac{(2S+1)!}{4\pi}\right]^{L} \int \mathrm{d}\hat{\Omega} \int \left[\prod^{L}_{j=1} \mathrm{d}\hat{\Omega}_{j}\right] \prod^{L-1}_{j=1}\left[\frac{1}{2}(1-\hat{\Omega}_{j}\cdot\hat{\Omega}_{j+1})\right]^{S} \nonumber \\
&&\cdot\left[\frac{1}{2}(1-\hat{\Omega}_{1}\cdot\hat{\Omega}_{L})\right]^{S-J}
\left[\frac{1}{2}(1+\hat{\Omega}_{1} \cdot\hat{\Omega})\right]^{J}
\left[\frac{1}{2}(1+\hat{\Omega} \cdot\hat{\Omega}_{L})\right]^{J}. \nonumber
\end{eqnarray}
Now we expand $\left[\frac{1}{2}(1-\hat{\Omega}_i \cdot\hat{\Omega}_j)\right]^{J}$ in terms of spherical harmonics as in (\ref{expansion}), then integrate over ${\hat\Omega}$ and from ${\hat\Omega}_2$ to ${\hat\Omega}_{L-1}$, the right hand side of (\ref{inne}) is equal to
\begin{eqnarray}
&& \frac{4\pi ((2S+1)!)^{L}}{(S+1)^{L-1}(S-J+1)(J+1)^{2}}
\sum_{l_{1}=0}^{S} \sum_{l_{L}=0}^{S-J} \sum_{l=0}^{J}
\sum_{m_{1}=-l_{1}}^{l_{1}} \sum_{m_{L}=-l_{L}}^{l_{L}} \sum_{m=-l}^{l}  \nonumber \\
&& \int \mathrm{d}\hat{\Omega}_{1} \int \mathrm{d}\hat{\Omega}_{L}\,
\lambda^{L-1}(l_{1},S) \lambda(l_{L},S-J) \lambda^{2}(l,J) \nonumber \\
&&\cdot Y_{l_{1},m_{1}}({\hat\Omega}_{1})Y_{l_{L},m_{L}}({\hat\Omega}_{1})Y_{l,m}({\hat\Omega}_1)
Y^{\ast}_{l_{1},m_{1}}({\hat\Omega}_{L})Y^{\ast}_{l_{L},m_{L}}({\hat\Omega}_{L})Y^{\ast}_{l,m}({\hat\Omega}_{L}).\label{pre1}
\end{eqnarray}
Here we apply the following useful formula:
\begin{eqnarray}
&&\int \mathrm{d}\hat{\Omega}\, Y_{l_{1},m_{1}}({\hat\Omega})Y_{l_{L},m_{L}}({\hat\Omega})Y_{l,m}({\hat\Omega})
\nonumber \\
&=&\sqrt{\frac{(2l_1+1)(2l_L+1)(2l+1)}{4\pi}}
\left(\begin{array}{c c c}
l_{1} & l_{L} & l \\
0   &  0  & 0 
\end{array}\right)
\left(\begin{array}{c c c}
l_{1} & l_{L} & l \\
m_{1} & m_{L} & m 
\end{array}\right), \nonumber\\ \label{wign}
\end{eqnarray}
where $\left(\begin{array}{c c c}
l_{1} & l_{L} & l \\
m_{1} & m_{L} & m 
\end{array}\right)$ is the Wigner $3j$-symbol.
Using formula (\ref{wign}), we carry out the remaining integrals in (\ref{pre1}) and obtain
\begin{eqnarray}
&& \frac{((2S+1)!)^{L}}{(S+1)^{L-1}(S-J+1)(J+1)^{2}} \sum_{l_{1}=0}^S \sum_{l_{L}=0}^{S-J} \sum_{l=0}^J
\sum_{m_{1}=-l_{1}}^{l_{1}} \sum_{m_{L}=-l_{L}}^{l_{L}} \sum_{m=-l}^{l} 
\nonumber \\
&& (2l_{1}+1)(2l_{L}+1)(2l+1)
\lambda^{L-1}(l_{1},S) \lambda(l_{L},S-J) \lambda^{2}(l,J) \nonumber \\
&&\cdot \left(\begin{array}{c c c}
l_{1} & l_{L} & l \\
0   &  0  & 0 
\end{array}\right)^{2}
\left(\begin{array}{c c c}
l_{1} & l_{L} & l \\
m_{1} & m_{L} & m 
\end{array}\right)^{2}.
\label{wigner}
\end{eqnarray}
These $3j$-symbols obey the following orthogonality relation:
\begin{eqnarray}
	\sum_{m_{1},m_{L}}(2l+1)
	\left(\begin{array}{c c c}
	l_{1} & l_{L} & l \\
	m_{1} & m_{L} & m 
	\end{array}\right)
	\left(\begin{array}{c c c}
	l_{1} & l_{L} & l^{\prime} \\
	m_{1} & m_{L} & m^{\prime} 
	\end{array}\right)=\delta_{ll^{\prime}} \delta_{mm^{\prime}}. \label{3jor}
\end{eqnarray}
Using this orthogonality (\ref{3jor}), we can recast expression (\ref{wigner}) as
\begin{eqnarray}
&&\frac{((2S+1)!)^{L}}{(S+1)^{L-1}(S-J+1)(J+1)^{2}} \sum_{l_{1}=0}^{S} \sum_{l_{L}=0}^{S-J} \sum_{l=0}^{J} 
\label{pre2} \\
&&(2l_{1}+1)(2l_{L}+1)(2l+1) \lambda^{L-1}(l_{1},S) \lambda(l_{L},S-J) \lambda^{2}(l,J)
\left(\begin{array}{c c c}
l_{1} & l_{L} & l \\
0   &  0  & 0 
\end{array}\right)^{2}.\nonumber
\end{eqnarray}
The explicit value of $\left(\begin{array}{c c c}
l_{1} & l_{L} & l \\
0   &  0  & 0 
\end{array} \right)$ is given by
\begin{eqnarray}
&&\left(\begin{array}{c c c}
l_{1} & l_{L} & l \\
0   &  0  & 0 
\end{array} \right) \nonumber\\
&=&(-1)^{g} \sqrt{\frac{(2g-2l_{1})!(2g-2l_{L})!(2g-2l)!}{(2g+1)!}}\,\frac{g!}{(g-l_{1})!(g-l_{L})!(g-l)!},
\nonumber \\ \label{3jva}
\end{eqnarray}
if $l_{1}+l_{L}+l=2g$ ($g \in \mathbf{N}$), otherwise zero. 
Finally, the normalization of degenerate VBS states $|\mathrm{VBS}_{L}(J,M)\rangle$ is obtained as
\begin{eqnarray}
&&\langle\mathrm{VBS}_{L}(J,M)|\mathrm{VBS}_{L}(J,M)\rangle \label{renovb} \\
&=&\frac{(2J+1)!((2S+1)!)^{L}}{(S+1)^{L-1}(S+J+1)!(S-J+1)!(J+1)!(J+1)!}\sum_{l_{1}=0}^{S} \sum_{l_{L}=0}^{S-J} \sum_{l=0}^{J}
\nonumber \\
&&(2l_{1}+1)(2l_{L}+1)(2l+1) \lambda^{L-1}(l_{1},S) \lambda(l_{L},S-J) \lambda^{2}(l,J)
\left(\begin{array}{c c c}
l_{1} & l_{L} & l \\
0   &  0  & 0 
\end{array}\right)^{2}. \nonumber
\end{eqnarray}

Combining results of (\ref{fipr}) and (\ref{renovb}), we arrive at the following theorem on eigenvalues:

\begin{theorem}
\label{theorem3}
Eigenvalues $\Lambda(J)$ $(J=0,\ldots,S)$ of the density matrix $\boldsymbol{\rho}_{L}$ are independent of $\hat{\Omega}$ and $M$ in defining eigenvectors (see (\ref{eiges}) and (\ref{devb})). An explicit expression is given by the following triple sum
	\begin{eqnarray}
	&&\Lambda(J) \label{eivaex} \\	&=&\left[\frac{S+1}{(2S+1)!}\right]^{L}\frac{S!S!}{S+1}\langle\mathrm{VBS}_{L}(J,M)|\mathrm{VBS}_{L}(J,M)\rangle \nonumber\\
	&=&\frac{(2J+1)!S!S!}{(S+J+1)!(S-J+1)!(J+1)!(J+1)!}\sum^{S}_{l_{1}=0} \sum^{S-J}_{l_{L}=0} \sum^{J}_{l=0}
	\nonumber \\
	&&(2l_{1}+1)(2l_{L}+1)(2l+1) \lambda^{L-1}(l_{1},S) \lambda(l_{L},S-J) \lambda^{2}(l,J)
	\left(\begin{array}{ccc}
	l_{1} & l_{L} & l \\
	0   &  0  & 0 
	\end{array}\right)^{2}. \nonumber
	\end{eqnarray}
\end{theorem}

Although not straightforward to verify, this expression (\ref{eivaex}) should be consistent with eigenvalues given through the recurrence expression (\ref{eiva}) in the next section \S$\,\ref{sec:eigval2}\,$ and the expression $\Lambda_{\alpha}$ in \S$\,\ref{sec:1ds1specden}\,$ as a special case. We could check the case when $S=1$ that
\begin{eqnarray}
&&\langle\mathrm{VBS}_{L}(0,0)|\mathrm{VBS}_{L}(0,0)\rangle=\frac{1}{2}(3^{L}+3(-1)^{L}),
\nonumber \\
&&\langle\mathrm{VBS}_{L}(1,M)|\mathrm{VBS}_{L}(1,M)\rangle=\frac{1}{2}(3^{L}-(-1)^{L}),
\end{eqnarray}
where we have used the selection rule of the Wigner $3j$-symbol. From (\ref{normvbs}) we find that $\langle\mathrm{VBS}|\mathrm{VBS}\rangle=2\cdot 3^{L}$, so that we obtain the correct eigenvalues of the density matrix from the above result (\ref{eivaex}) (see \S$\,\ref{sec:1ds1specden}\,$ for comparison).

We shall emphasize at this point that given eigenvalues (\ref{eivaex}), both von Neumann entropy 
\begin{eqnarray}
	S_{\mathrm{v\ N}}=-\mathrm{tr}\left[\,\boldsymbol{\rho}_{L}\ln{\boldsymbol{\rho}_{L}}\,\right]=-\sum^{S}_{J=0}(2J+1)\Lambda(J)\ln\Lambda(J) \label{vonn}
\end{eqnarray}
and R\'enyi entropy 
\begin{eqnarray}
	S_{R}=\frac{1}{1-\alpha}\ln \left\{\mathrm{tr}\left[\,\boldsymbol{\rho}^{\alpha}_{L}\,\right]\right\}=\frac{1}{1-\alpha}\ln\left\{\sum^{S}_{J=0}(2J+1)\Lambda^{\alpha}(J)\right\} \label{renyi}
\end{eqnarray}
can be expressed directly.

\subsection{Eigenvalues of the Density Matrix (recurrence formula)}
\label{sec:eigval2}

Having constructed eigenvectors, there are more than one way to specify the corresponding eigenvalues. An explicit expression of eigenvalues has been obtained in \S$\,\ref{sec:eigval1}\,$. In this section we express eigenvalues through a conjectured recurrence formula as appeared in \cite{FM,KHH}. Let us apply the density matrix $\boldsymbol{\rho}_{L}$ (\ref{matrix}) to the state $|\mathrm{G}; J, \hat{\Omega}\rangle$ (\ref{eiges}) and obtain
\begin{eqnarray}
	&&\boldsymbol{\rho}_{L}|\mathrm{G}; J, \hat{\Omega}\rangle \nonumber\\
	&=&\left[\frac{S+1}{(2S+1)!}\right]^{L}\frac{S+1}{(4\pi)^{2}}
	\int \mathrm{d}\hat{\Omega}_{0}\mathrm{d}\hat{\Omega}_{L+1}\,B^{\dagger}|\mathrm{VBS}_{L}\rangle\langle \mathrm{VBS}_{L}|BA^{\dagger}_{J}|\mathrm{VBS}_{L}\rangle. \nonumber\\ \label{appl1}
\end{eqnarray}
Using the coherent state basis (\ref{cohebas}) and completeness relation (\ref{comp}), the factor $\langle \mathrm{VBS}_{L}|BA^{\dagger}_{J}|\mathrm{VBS}_{L}\rangle$
in (\ref{appl1}) can be re-written as
\begin{eqnarray}
	&&\langle \mathrm{VBS}_{L}|BA^{\dagger}_{J}|\mathrm{VBS}_{L}\rangle \label{appl2} \\
	&=&\displaystyle\left[\frac{(2S+1)!}{4\pi}\right]^{L}\int \left(\prod^{L}_{j=1}\mathrm{d}\hat{\Omega}_{j}\right)\prod^{L-1}_{j=1}\left[\frac{1}{2}(1-\hat{\Omega}_{j}\cdot\hat{\Omega}_{j+1})\right]^{S}
	\left(u_{0}v_{1}-v_{0}u_{1}\right)^{S} \nonumber \\ &&\cdot\left(uu^{\ast}_{1}+vv^{\ast}_{1}\right)^{J}\left(u^{\ast}_{1}v^{\ast}_{L}-v^{\ast}_{1}u^{\ast}_{L}\right)^{S-J}\left(uu^{\ast}_{L}+vv^{\ast}_{L}\right)^{J}\left(u_{L}v_{L+1}-v_{L}u_{L+1}\right)^{S}. \nonumber
\end{eqnarray}
The factor $\left[\frac{1}{2}(1-\hat{\Omega}_{j}\cdot\hat{\Omega}_{j+1})\right]^{S}$ under the integral of (\ref{appl2}) can be expanded in terms of Legendre polynomials and further in terms of spherical harmonics as discussed in \S$\,\ref{sec:norm}\,$ (see also \cite{FM,KHH}). Using the expansion (\ref{expansion}) and orthogonality of spherical harmonics, the integrals over $\hat{\Omega}_{j}$ with $j=2, \ldots, L-1$ in (\ref{appl2}) can be performed. The result is
\begin{eqnarray}
	\langle \mathrm{VBS}_{L}|BA^{\dagger}_{J}|\mathrm{VBS}_{L}\rangle &&= \frac{S+1}{(4\pi)^{2}}\left[\frac{(2S+1)!}{S+1}\right]^{L}\sum^{S}_{l=0}(2l+1)\lambda^{L-1}(l,S) \nonumber\\
	&&\cdot\int \mathrm{d}\hat{\Omega}_{1}\mathrm{d}\hat{\Omega}_{L}\,P_{l}(\hat{\Omega}_{1}\cdot\hat{\Omega}_{L})\left(u_{0}v_{1}-v_{0}u_{1}\right)^{S}\left(uu^{\ast}_{1}+vv^{\ast}_{1}\right)^{J} \nonumber \\
&&\left(u^{\ast}_{1}v^{\ast}_{L}-v^{\ast}_{1}u^{\ast}_{L}\right)^{S-J}\left(uu^{\ast}_{L}+vv^{\ast}_{L}\right)^{J}\left(u_{L}v_{L+1}-v_{L}u_{L+1}\right)^{S}. \nonumber\\ \label{resu}
\end{eqnarray}
We plug the expression (\ref{resu}) into (\ref{appl1}). Using transformation properties under $SU(2)$ and a binomial expansion (see \S$\,\ref{sec:larges}\,$), the integral over $\hat{\Omega}_{0}$ yields that
\begin{eqnarray}
	\int \mathrm{d}\hat{\Omega}_{0}\left(u^{\ast}_{0}b^{\dagger}_{1}-v^{\ast}_{0}a^{\dagger}_{1}\right)^{S}\left(u_{0}v_{1}-v_{0}u_{1}\right)^{S}=\frac{4\pi}{S+1}\left(u_{1}a^{\dagger}_{1}+v_{1}b^{\dagger}_{1}\right)^{S} \label{int0}
\end{eqnarray}
Similarly we can perform the integral over $\hat{\Omega}_{L+1}$. As a result, the following expression is obtained from (\ref{appl1}):
\begin{eqnarray}
\boldsymbol{\rho}_{L}|\mathrm{G}; J, \hat{\Omega}\rangle =  \frac{1}{(4\pi)^{2}}\sum^{S}_{l=0}(2l+1)\lambda^{L-1}(l,S)
K^{\dagger}_{l}(\hat{\Omega})
\left|\mathrm{VBS}_{L}\right\rangle. \label{sum}
\end{eqnarray}
The operator $K^{\dagger}_{l}(\hat{\Omega})$ involved in (\ref{sum}) is defined as
\begin{eqnarray}
	K^{\dagger}_{l}(\hat{\Omega})&\equiv&\int \mathrm{d}\hat{\Omega}_{1} \mathrm{d}\hat{\Omega}_{L} 
	\left(u_{1}a^{\dagger}_{1}+v_{1}b^{\dagger}_{1}\right)^{S} \left(uu^{\ast}_{1}+vv^{\ast}_{1}\right)^{J}
\left(u^{\ast}_{1}v^{\ast}_{L}-v^{\ast}_{1}u^{\ast}_{L}\right)^{S-J}	
\nonumber \\ &&\cdot
\left(uu^{\ast}_{L}+vv^{\ast}_{L}\right)^{J}
\left(u_{L}a^{\dagger}_{L}+v_{L}b^{\dagger}_{L}\right)^{S}
P_{l}(\hat{\Omega}_{1} \cdot \hat{\Omega}_{L}). \label{inte}
\end{eqnarray} 
It is expressed as an integral depending on the order $l$ of the Legendre polynomial $P_{l}(\hat{\Omega}_{1} \cdot \hat{\Omega}_{L})$. $K^{\dagger}_{l}(\hat{\Omega})$ can be calculated from the lowest few orders (see \S$\,\ref{sec:larges}\,$ for example). It becomes increasingly difficult to perform the integral as order $l$ increases. Based on the eigenvalue expressions of the density matrix obtained in \cite{FKR,KHH}, we make a conjecture on the explicit form of the operator $K^{\dagger}_{l}(\hat{\Omega})$ for generic order $l$:

\textbf{Conjecture 1}
	\begin{eqnarray}
	K^{\dagger}_{l}(\hat{\Omega})=\left(\frac{4\pi}{S+1}\right)^{2}I_{l}\left(\frac{1}{2}J(J+1)-\frac{1}{2}S(\frac{1}{2}S+1)\right)A^{\dagger}_{J}. \label{conj2}
\end{eqnarray}
Here the polynomial $I_{l}\left(x\right)$ satisfy the recurrence relation
\begin{eqnarray}
	I_{l+1}(x)=
	\frac{2l+1}{\left(S+l+2\right)^{2}}\left(\frac{4x}{l+1}+l\right)I_{l}\left(x\right)
	-\frac{l}{l+1}\left(\frac{S-l+1}{S+l+2}\right)^{2}I_{l-1}(x)\nonumber\\ \label{recu}
\end{eqnarray}
with $I_{0}=1$ and $I_{1}=\frac{x}{(\frac{S}{2}+1)^{2}}$.

Note that it is important that $K^{\dagger}_{l}(\hat{\Omega})\propto A^{\dagger}_{J}$ defined in (\ref{aope}) and $I_{l}(x)$ has the same order as the Legendre polynomial $P_{l}(x)$. The recurrence relation (\ref{recu}) was proposed in \cite{FM} and used in \cite{KHH} to obtain the eigenvalues of the density matrix. 
(The original definition of $I_l(x)$ differs from our definition in (\ref{recu}) by a factor of $(2l+1)/4\pi$.)
\textbf{Conjecture 1} is an alternative form of \textbf{Theorem} \ref{theorem2} together with \textbf{Theorem} \ref{theorem3}, which also gives eigenvalues through the recurrence relation (\ref{recu}). Indeed, expressions (\ref{sum}), altogether with (\ref{conj2}) and (\ref{recu}) yields that
\begin{eqnarray}
	&&\boldsymbol{\rho}_{L}|\mathrm{G}; J, \hat{\Omega}\rangle \label{appl} \\ 		&=&\frac{1}{(S+1)^{2}}\sum^{S}_{l=0}(2l+1)\lambda^{L-1}(l,S)
I_{l}\left(\frac{1}{2}J(J+1)-\frac{1}{2}S(\frac{1}{2}S+1)\right)|\mathrm{G}; J, \hat{\Omega}\rangle. \nonumber
\end{eqnarray}
Non-zero eigenvalues ($J=0, 1, \ldots, S$) are seen from (\ref{appl}) as
\begin{eqnarray}
	\Lambda(J)\equiv\frac{1}{(S+1)^{2}}\sum^{S}_{l=0}(2l+1)\lambda^{L-1}(l,S)
I_{l}\left(\frac{1}{2}J(J+1)-\frac{1}{2}S(\frac{1}{2}S+1)\right).\nonumber\\ \label{eiva}
\end{eqnarray}
Since all other eigenvalues of the density matrix are vanishing, then we conclude again that the density matrix $\boldsymbol{\rho}_{L}$ (\ref{matrix}) is a projector onto a subspace of dimension $(S+1)^{2}$. This subspace is spanned by the set of vectors $\{|\mathrm{G}; J, \hat{\Omega}\rangle\}$ (\ref{eiges}). (The rank of the set is equal to $(S+1)^2$.) Furthermore, we observe from (\ref{eiva}) again that non-zero eigenvalues $\Lambda(J)$ depend only on $J$, not on $\hat{\Omega}$. Therefore, $\{|\mbox{G}; J, \hat{\Omega}\rangle\}$ with fixed $J$ value spans a degenerate subspace with the same eigenvalue.

\subsection{The Large Block Limit}
\label{sec:larges}

In the limit $L\rightarrow\infty$, that is when the size of the block becomes large, we learned from \cite{FKR,Had,KHH} that the von Neumann entropy reaches the saturated value $S_{\mathrm{v\ N}}=\ln\left(S+1\right)^{2}$. This fact implies that the density matrix (denoted by $\boldsymbol{\rho}_{\infty}$ in the limit) can only take the form (see \cite{NC} for a general proof)
\begin{eqnarray}
	\boldsymbol{\rho}_{\infty}=\frac{1}{(S+1)^{2}}I_{(S+1)^{2}}\oplus \Phi_{\infty}, \label{denl}
\end{eqnarray}
where $I_{(S+1)^{2}}$ is the identity of dimension $(S+1)^{2}$ and $\Phi_{\infty}$ is an infinite dimensional matrix with only zero entries. In this section, we give a proof of \textbf{Conjecture 1} (\ref{conj2}) in the limiting case as $L\rightarrow\infty$. Then we shall verify the structure of the density matrix (\ref{denl}) explicitly.

We first realize from (\ref{lamb}) that as $L\rightarrow\infty$, $\lambda^{L-1}(l,S)\rightarrow \delta_{l,0}$. Therefore only the first term with $l=0$ is left in (\ref{sum}) and contributes to the final result. So that we need only to calculate $K^{\dagger}_{0}(\hat{\Omega})$:
\begin{eqnarray}
	K^{\dagger}_{0}(\hat{\Omega})&=&\int \mathrm{d}\hat{\Omega}_{1} \mathrm{d}\hat{\Omega}_{L} 
	\left(u_{1}a^{\dagger}_{1}+v_{1}b^{\dagger}_{1}\right)^{S} \left(uu^{\ast}_{1}+vv^{\ast}_{1}\right)^{J}    \nonumber \\ &&\cdot
\left(u^{\ast}_{1}v^{\ast}_{L}-v^{\ast}_{1}u^{\ast}_{L}\right)^{S-J}
\left(uu^{\ast}_{L}+vv^{\ast}_{L}\right)^{J}
\left(u_{L}a^{\dagger}_{L}+v_{L}b^{\dagger}_{L}\right)^{S}. \label{k0}
\end{eqnarray} 

It is useful to know transformation properties of the integrand in (\ref{k0}) under $SU(2)$.
The pair of variables $(u, v)$ defined in (\ref{spin}) and bosonic annihilation operators $(a, b)$ in the Schwinger representation both transform as spinors under $SU(2)$. That is to say, if we take an arbitrary element $\mathbf{D}\in SU(2)$ (a $2\times 2$ unitary matrix with determinant $1$), then $(u, v)$, \textit{etc.} transform according to
\begin{eqnarray}
\left( \begin{array}{c}
	u \\ v 
\end{array}\right)
\rightarrow \mathbf{D}
\left( \begin{array}{c}
	u \\ v
\end{array} \right). \label{tran}
\end{eqnarray}
On the other hand, $(u^{\ast}, v^{\ast})$, $(-v, u)$, $(a^{\dagger}, b^{\dagger})$ and $(-b, a)$ transform conjugately to $(u, v)$. That is to say $(u^{\ast}, v^{\ast})$, \textit{etc.} transform according to
\begin{eqnarray}
	\left( \begin{array}{c}
	u^{\ast} \\ v^{\ast} 
\end{array}\right)
\rightarrow \mathbf{D}^{\ast}
\left( \begin{array}{c}
	u^{\ast} \\ v^{\ast}
\end{array}\right). \label{ctra}
\end{eqnarray}
The combinations appeared in $K^{\dagger}_{0}(\hat{\Omega})$ (\ref{k0})
\begin{eqnarray}
	u_{1}a^{\dagger}_{1}+v_{1}b^{\dagger}_{1},\ \ uu^{\ast}_{1}+vv^{\ast}_{1},\ \ u^{\ast}_{1}v^{\ast}_{L}-v^{\ast}_{1}u^{\ast}_{L},\ \ uu^{\ast}_{L}+vv^{\ast}_{L},\ \
	u_{L}a^{\dagger}_{L}+v_{L}b^{\dagger}_{L} \label{comb}
\end{eqnarray}
as well as $A^{\dagger}_{J}$ in (\ref{aope}), boundary operator $B^{\dagger}$ in (\ref{bope1}), \textit{etc.} all transform covariantly under $SU(2)$, \textit{i.e.} those expressions keep their form in the new (transformed) coordinates. 

These transformation properties (\ref{tran}), (\ref{ctra}) can be used to simplify the $K^{\dagger}_{0}(\hat{\Omega})$ integral. We first make a $SU(2)$ transform
\begin{eqnarray}
	\mathbf{D}_{u_{L}}=
\left( \begin{array}{cc}
	u^{\ast}_{L} & v^{\ast}_{L} \\
	-v_{L} & u_{L}
\end{array} \right), \qquad
\mathbf{D}_{u_{L}}
\left( \begin{array}{c}
	u_{L} \\ v_{L}
\end{array} \right)=\left( \begin{array}{c}
	1 \\ 0
\end{array} \right), \label{utra}
\end{eqnarray}
under the part of the integral (\ref{k0}) over $\hat{\Omega}_{1}$. Then this part of integral  becomes
\begin{eqnarray}
	\int \mathrm{d}\hat{\Omega}_{1} 
\left(u_{1}a^{\dagger}_{1}+v_{1}b^{\dagger}_{1}\right)^{S} \left(uu^{\ast}_{1}+vv^{\ast}_{1}\right)^{J}  \left(-v^{\ast}_{1}\right)^{S-J}.
\end{eqnarray}
This can be calculated using binomial expansion and the result is
\begin{eqnarray}
	\frac{4\pi}{S+1}\left(ua^{\dagger}_{1}+vb^{\dagger}_{1}\right)^{J}  \left(-b^{\dagger}_{1}\right)^{S-J}. \label{part1}
\end{eqnarray}
Then we make an inverse transform in (\ref{part1}) using $\mathbf{D}^{-1}_{u_{L}}=\mathbf{D}^{\dagger}_{u_{L}}$, consequently (\ref{k0}) is put in a form with a single integral over $\hat{\Omega}_{L}$ remaining:
\begin{eqnarray}
K^{\dagger}_{0}(\hat{\Omega})&=&\frac{4\pi}{S+1}\left(ua^{\dagger}_{1}+vb^{\dagger}_{1}\right)^{J} \label{rema} \\
&&\cdot\int \mathrm{d}\hat{\Omega}_{L} 
\left(a^{\dagger}_{1}v^{\ast}_{L}-b^{\dagger}_{1}u^{\ast}_{L}\right)^{S-J}
\left(uu^{\ast}_{L}+vv^{\ast}_{L}\right)^{J}   \left(u_{L}a^{\dagger}_{L}+v_{L}b^{\dagger}_{L}\right)^{S}. \nonumber
\end{eqnarray}
Now we make another $SU(2)$ transform using
\begin{eqnarray}
	\mathbf{D}_{u}=
\left( \begin{array}{cc}
	u^{\ast} & v^{\ast} \\
	-v & u
\end{array} \right), \qquad
\mathbf{D}_{u}
\left( \begin{array}{c}
	u \\ v
\end{array} \right)=\left( \begin{array}{c}
	1 \\ 0
\end{array} \right),
\end{eqnarray}
then the remaining integral over $\hat{\Omega}_{L}$ in (\ref{rema}) becomes
\begin{eqnarray}
\int \mathrm{d}\hat{\Omega}_{L} 
\left(a^{\dagger}_{1}v^{\ast}_{L}-b^{\dagger}_{1}u^{\ast}_{L}\right)^{S-J}
\left(u^{\ast}_{L}\right)^{J}   \left(u_{L}a^{\dagger}_{L}+v_{L}b^{\dagger}_{L}\right)^{S}. \label{reml}
\end{eqnarray} 
Using again binomial expansion, this integral (\ref{reml}) yields
\begin{eqnarray}
\frac{4\pi}{S+1}
\left(a^{\dagger}_{1}b^{\dagger}_{L}-b^{\dagger}_{1}a^{\dagger}_{L}\right)^{S-J}
\left(a^{\dagger}_{L}\right)^{J}. \label{part2}
\end{eqnarray} 
At last we make an inverse transform in (\ref{part2}) using $\mathbf{D}^{-1}_{u}=\mathbf{D}^{\dagger}_{u}$ and plug the result into (\ref{rema}), the final form is
\begin{eqnarray}
K^{\dagger}_{0}(\hat{\Omega})
=\left(\frac{4\pi}{S+1}\right)^{2}A^{\dagger}_{J}. \label{fina}
\end{eqnarray}
This expression is consistent with \textbf{Conjecture 1} (\ref{conj2}), which also proves that $\{|\mathrm{G}; J, \hat{\Omega}\rangle\}$ is a set of eigenvectors of the density matrix as $L\rightarrow\infty$. Let us denote the density matrix in the limit by $\boldsymbol{\rho}_{\infty}$. Then (\ref{fina}) leads to the result (see (\ref{appl}))
\begin{eqnarray}
	\boldsymbol{\rho}_{\infty}|\mathrm{G}; J, \hat{\Omega}\rangle=
	\frac{1}{(S+1)^{2}}\,|\mathrm{G}; J, \hat{\Omega}\rangle. \label{appll}
\end{eqnarray}

We find from (\ref{appll}) that the limiting eigenvalue
\begin{eqnarray}
	\Lambda_{\infty}=\frac{1}{(S+1)^{2}}, \qquad L\to\infty
\end{eqnarray}
is independent of $J$. Any vector of the $(S+1)^2$-dimensional subspace spanned by the set $\{|\mathrm{G}; J, \hat{\Omega}\rangle\}$ is an eigenvector of $\boldsymbol{\rho}_{\infty}$ with the same eigenvalue $1/(S+1)^{2}$. Therefore $\boldsymbol{\rho}_{\infty}$ acts on this subspace as (proportional to) the identity $I_{(S+1)^{2}}$. So that we have proved explicitly that the density matrix takes the form (\ref{denl}) in the large block limit. The limiting density matrix is proportional to a projector $\mathbf{P}_{(S+1)^{2}}$ on the degenerate ground states of the block Hamiltonian $H_{B}$ (\ref{degels})
\begin{eqnarray}
	\boldsymbol{\rho}_{L}\to\boldsymbol{\rho}_{\infty}=\frac{1}{(S+1)^{2}}\,\mathbf{P}_{(S+1)^{2}} , \qquad L\to\infty. \label{largedens}
\end{eqnarray}
In addition, we also derive from the eigenvalues that the von Neumann entropy $S_{\mathrm{v\ N}}=-\sum^{S}_{J=0}(2J+1)\Lambda_{\infty}\ln\Lambda_{\infty}$ coincides with the R\'enyi entropy $S_{\mathrm{R}}(\alpha)=\frac{1}{1-\alpha}\ln\left\{\sum^{S}_{J=0}(2J+1)\Lambda^{\alpha}_{\infty}\right\}$ and is equal to the saturated value 
\begin{eqnarray}
	S_{\mathrm{v\ N}}=S_{\mathrm{R}}(\alpha)=\ln(S+1)^{2} , \qquad L\to\infty.
\end{eqnarray}

\subsection{The Density Matrix and Correlation Functions}
\label{sec:denandcorr}

The relation between the density matrix and correlation functions was studied in \cite{AAH,JK,KHH,XKHK}.
It was shown in \cite{JK} that the density matrix contains information of all correlation functions in the ground state. The original proof was for spin $S=1/2$. In this section we generalize the result to generic spin-$S$ as in \cite{XKHK} and the proof is written in a form applicable but not restricted to the VBS state. 

The Hilbert space associated with a spin-$S$ is $(2S+1)$-dimensional. Therefore we could choose a basis of $(2S+1)^{2}$ linearly independent matrices such that an arbitrary operator defined in the Hilbert space can be written as a superposition over the basis. Let us denote the basis by $\{A_{ab}\ | \ a,b=1,\ldots,2S+1\}$, in which each matrix $A_{ab}$ is labeled by a pair of indices $a$ and $b$ with totally $(2S+1)^{2}$ possible combinations. The matrix element is defined as
\begin{eqnarray}
	(A_{ab})_{kl}=\delta_{ak}\delta_{bl}, \qquad k,l=1,\ldots,2S+1. \label{aab}
\end{eqnarray}
In addition to $\{A_{ab}\}$, we introduce an equivalent \textit{conjugate basis} $\{\bar{A}_{ab}\}$ such that
\begin{eqnarray}
	(\bar{A}_{ab})_{kl}=\delta_{al}\delta_{bk}, \qquad a,b,k,l=1,\ldots,2S+1. \label{abar}
\end{eqnarray}
These matrices (\ref{aab}) and (\ref{abar}) are actually matrix representation of operators $\{|S,m\rangle\langle S,m^{\prime}| \ |\ m,m^{\prime}=-S,\ldots,S\}$. They are normalized such that
\begin{eqnarray}
	\mathrm{tr}(\bar{A}_{ab}A_{cd})=\sum_{k,l}(\bar{A}_{ab})_{kl}(A_{cd})_{lk}=\sum_{k,l}\delta_{al}\delta_{bk}\delta_{cl}\delta_{dk}=\delta_{ac}\delta_{bd}. \label{traceaa}
\end{eqnarray}
Here `$\mathrm{tr}$' takes trace at one and the same site. Because of the completeness of $\{A_{ab}\}$ at each site, the density matrix of the block ($B$) can be written as (see (\ref{den1}))
\begin{eqnarray}
	\boldsymbol{\rho}_{B}=\mathrm{tr}_{E}|\mathrm{G}\rangle\langle \mathrm{G}|=\sum_{\{a_{j}b_{j}\}}\left(\bigotimes_{j\in B}A_{a_{j}b_{j}}\right)\mathrm{coeff}\{a_{j}b_{j}\}, \label{rhoblock}
\end{eqnarray}
where $|\mathrm{G}\rangle$ denotes the unique ground state (\textit{e.g.} the VBS state), $\mathrm{tr}_{E}$ takes traces of sites outside the block (\textit{i.e.} sites in the environment $E$) and $\mathrm{coeff}\{a_{j}b_{j}\}$ denotes the coefficient. Using the normalization property (\ref{traceaa}), the coefficient $\mathrm{coeff}\{a_{j}b_{j}\}$ with label $j$ taking values within the block can be expressed as
\begin{eqnarray}
	\mathrm{coeff}\{a_{j}b_{j}\}&=&\sum_{\{c_{j}d_{j}\}}\prod_{j\in B}\mathrm{tr}(\bar{A}_{a_{j}b_{j}}A_{c_{j}d_{j}})\mathrm{coeff}\{c_{j}d_{j}\} \nonumber \\
	&=&\mathrm{tr}_{B}\left[\left(\bigotimes_{j\in B}\bar{A}_{a_{j}b_{j}}\right)\boldsymbol{\rho}_{B}\right] \nonumber \\
&=&\mathrm{tr}\left[\left(\bigotimes_{j\in B}\bar{A}_{a_{j}b_{j}}\right)|\mathrm{G}\rangle\langle \mathrm{G}|\right]\nonumber \\
&=&\langle \mathrm{G}|\left(\bigotimes_{j\in B}\bar{A}_{a_{j}b_{j}}\right)|\mathrm{G}\rangle. \label{coeajbj}
\end{eqnarray}
Here $\mathrm{tr}_{B}$ takes traces of sites within the block and $\mathrm{tr}$ takes traces of all lattice sites. Combing (\ref{rhoblock}) with (\ref{coeajbj}), we have the final form
\begin{eqnarray}
	\boldsymbol{\rho}_{B}=\sum_{\{a_{j}b_{j}\}}\left(\bigotimes_{j\in B}A_{a_{j}b_{j}}\right)\langle \mathrm{G}|\left(\bigotimes_{j\in B}\bar{A}_{a_{j}b_{j}}\right)|\mathrm{G}\rangle. \label{rhocorr}
\end{eqnarray}
This is the expression of the density matrix with entries related to multi-point correlation functions $\langle \mathrm{G}|\left(\bigotimes_{j\in B}\bar{A}_{a_{j}b_{j}}\right)|\mathrm{G}\rangle$ in the ground state. All possible combinations $\{a_{j}b_{j}\}$ are involved in the summation. Therefore, we have prove for generic spin-$S$ that the density matrix contains information of all correlation functions. The matrix elements are all multi-point correlators.

\section{The One--dimensional Inhomogeneous Model}
\label{sec:1dinhom}

The most general $1$-dimensional model is the inhomogeneous model in which spins at different lattice site can take different values. As a special case of the generalized model defined in \S$\,\ref{sec:generalized}\,$, we associate a positive integer number (called multiplicity numbers, see \S$\,\ref{sec:generalhamil}\,$) to each bond of the lattice and denote by $M_{ij}$ ($M_{ij}=M_{ji}$) the multiplicity number between sites $i$ and $j$. They are related to bulk spins by the following relation which ensures the existence of a unique ground state
\begin{eqnarray}
	2S_{j}=M_{j-1,j}+M_{j,j+1}, \qquad \forall\ j \label{smrela00}
\end{eqnarray}
with $2S_{0}=M_{01}$ and $2S_{N+1}=M_{N, N+1}$ for ending spins. (Equation (\ref{smrela00}) is a special case of the more general relation (\ref{condition1}).) The condition for solvability of relation (\ref{smrela00}) is
\begin{eqnarray}
	\sum^{N+1}_{j=0}(-1)^{j}S_{j}=0. \label{condition00}
\end{eqnarray}
Solution to relation (\ref{smrela00}) under condition (\ref{condition00}) is 
\begin{eqnarray}
	M_{j,j+1}=2\sum^{j}_{l=0}(-1)^{j-l}S_{l}\geq 1. \label{solution00}
\end{eqnarray}
(More details can be found in \cite{KK}.)
Now we defined the Hamiltonian of the inhomogeneous AKLT model according to (\ref{generalizedh}) as
\begin{eqnarray}
	H=\sum^{N}_{j=0}\ \sum^{S_{j}+S_{j+1}}_{J=S_{j}+S_{j+1}-M_{j,j+1}+1} C_{J}(j,j+1)\,\pi_{J}(j, j+1). \label{hami00}
\end{eqnarray}
Here the projector $\pi_{J}(j, j+1)$ describes interactions between neighboring spins $j$ and $j+1$, which projects the bond spin $\boldsymbol{J}_{j, j+1}\equiv\boldsymbol{S}_{j}+\boldsymbol{S}_{j+1}$ onto the subspace with total spin $J$ ($J=S_{j}+S_{j+1}-M_{j,j+1}+1, \ldots, S_{j}+S_{j+1}$). An explicit expression of $\pi_{J}(j,j+1)$ is given in \S$\,\ref{sec:proj}\,$ and \cite{KK}. The coefficient $C_{J}(j,j+1)$ can take an arbitrary positive value. This Hamiltonian (\ref{hami00}) has a unique ground state (VBS state, see \S$\,\ref{sec:unique}\,$).

Following \cite{XKHK2}, we study the entanglement of the unique VBS ground state of the inhomogeneous model (\ref{hami00}) in this section.

\subsection{The VBS Ground State}
\label{sec:inhomvbs}

The unique VBS ground state of the Hamiltonian (\ref{hami00}) is given in the Schwinger representation by \cite{AAH,KK}
\begin{eqnarray}
	|\mathrm{VBS}\rangle \equiv
 \prod^{N}_{j=0}
\left(a^{\dagger}_{j}b^{\dagger}_{j+1}-b^{\dagger}_{j}a^{\dagger}_{j+1}\right)^{M_{j,j+1}}|\mathrm{vac}\rangle, \label{inhomvbs}
\end{eqnarray}
where $a^{\dagger}$, $b^{\dagger}$ are bosonic creation operators defined in exactly the same way as in \S$\,\ref{sec:vbssch}\,$, the constraint on the total boson occupation number is now $(a^{\dagger}_{j}a_{j}+b^{\dagger}_{j}b_{j})/2=S_{j}$. The pure state density matrix of the VBS ground state (\ref{inhomvbs}) is
\begin{eqnarray}
	\boldsymbol{\rho}=\frac{|\mathrm{VBS}\rangle\langle \mathrm{VBS}|}{\langle \mathrm{VBS}|\mathrm{VBS}\rangle}. \label{pure00}
\end{eqnarray}
Normalization of the VBS state is (calculation similar to those in \S$\,\ref{sec:norm}\,$)
\begin{eqnarray}
	\langle \mathrm{VBS}|\mathrm{VBS}\rangle=\frac{\displaystyle\prod^{N+1}_{j=0}(2S_{j}+1)!}{\displaystyle\prod^{N}_{j=0}(M_{j,j+1}+1)}. \label{normvbs00}
\end{eqnarray}
(See \cite{XKHK2} for more details.)

\subsection{The Block Density Matrix}
\label{sec:deninhom}

We take a block of $L$ contiguous bulk spins as a subsystem, which starts from site $k$ and ends at site $k+L-1$. Using the coherent state basis (\ref{cohebas}) and completeness relation (\ref{comp}), tracing out degrees of freedom outside the block, $\boldsymbol{\rho}_{L}$ can be written as \cite{KHH,XKHK2}
\begin{eqnarray}
	&&\boldsymbol{\rho}_{L}=
	\label{roug00} \\
	&&\frac{
\displaystyle\int\left[\prod^{k-1}_{j=0}\prod^{N+1}_{j=k+L}\mathrm{d}\hat{\Omega}_{j}\right]\prod^{k-2}_{j=0}\prod^{N}_{j=k+L}\left[\frac{1}{2}(1-\hat{\Omega}_{j}\cdot\hat{\Omega}_{j+1})\right]^{M_{j,j+1}}
	B^{\dagger}|\mathrm{VBS}_{L}\rangle\langle \mathrm{VBS}_{L}|B}
	{\displaystyle\left[\prod^{k+L-1}_{j=k}\frac{(2S_{j}+1)!}{4\pi}\right]
\int\left[\prod^{N+1}_{j=0}\mathrm{d}\hat{\Omega}_{j}\right]\prod^{N}_{j=0}\left[\frac{1}{2}(1-\hat{\Omega}_{j}\cdot\hat{\Omega}_{j+1})\right]^{M_{j,j+1}}}. \nonumber 
\end{eqnarray}
Here the boundary operator $B$ and block VBS state $\left|\mathrm{VBS}_{L}\right\rangle$ are defined as
\begin{eqnarray}
	&&B\equiv \left(u_{k-1}b_{k}-v_{k-1}a_{k}\right)^{M_{k-1,k}}\left(a_{k+L-1}v_{k+L}-b_{k+L-1}u_{k+L}\right)^{M_{k+L-1,k+L}}, \nonumber \\ \label{bope00} \\
	&&|\mathrm{VBS}_{L}\rangle \equiv
 \prod^{k+L-2}_{j=k}
\left(a^{\dagger}_{j}b^{\dagger}_{j+1}-b^{\dagger}_{j}a^{\dagger}_{j+1}\right)^{M_{j,j+1}}|\mathrm{vac}\rangle, \label{vbsl00}
\end{eqnarray}
respectively. After performing integrals over $\hat{\Omega}_{j}$ ($j=0, 1, \ldots, k-2, k+L+1, \ldots, N, N+1$) in the numerator and all integrals in the denominator, the density matrix $\boldsymbol{\rho}_{L}$ turns out to be independent of spins outside the block. This property has been proved for the homogeneous AKLT model in \S$\,\ref{sec:1dssden}\,$ (see also \cite{FKR,KHH,XKHK}). Therefore, we can re-label spins within the block for notational convenience. Let $k=1$ and the density matrix takes the form
\begin{eqnarray}
	\boldsymbol{\rho}_{L}=\frac{\displaystyle\prod^{L}_{j=0}(M_{j,j+1}+1)}{\displaystyle\prod^{L}_{j=1}(2S_{j}+1)!}\frac{1}{(4\pi)^{2}}
	\int \mathrm{d}\hat{\Omega}_{0}\mathrm{d}\hat{\Omega}_{L+1}\,B^{\dagger}|\mathrm{VBS}_{L}\rangle\langle \mathrm{VBS}_{L}|B \label{matr00}
\end{eqnarray}
with
\begin{eqnarray}
&&B^{\dagger}=\left(u^{\ast}_{0}b^{\dagger}_{1}-v^{\ast}_{0}a^{\dagger}_{1}\right)^{M_{0,1}}\left(a^{\dagger}_{L}v^{\ast}_{L+1}-b^{\dagger}_{L}u^{\ast}_{L+1}\right)^{M_{L,L+1}}, \label{bope100} \\
	&&|\mathrm{VBS}_{L}\rangle=\prod^{L-1}_{j=1}
\left(a^{\dagger}_{j}b^{\dagger}_{j+1}-b^{\dagger}_{j}a^{\dagger}_{j+1}\right)^{M_{j,j+1}}|\mathrm{vac}\rangle. \label{vbsl100}	
\end{eqnarray}
Again, the remaining two integrals in (\ref{matr00}) are kept in the present form for later use.

\subsection{Ground States of the Block Hamiltonian}
\label{sec:gsbhinhom}

The block Hamiltonian with the re-labeling $k=1$ reads
\begin{eqnarray}
	H_{B}=\sum^{L-1}_{j=1}\ \sum^{S_{j}+S_{j+1}}_{J=S_{j}+S_{j+1}-M_{j,j+1}+1} C_{J}(j,j+1)P_{J}(j, j+1). \label{blockhamire00}
\end{eqnarray}
Now the degenerate ground states are constructed in a similar way as in \S$\,\ref{sec:gsbhs}\,$. The new $A^{\dagger}_{J}$ operator is defined as:
\begin{eqnarray}
 A^{\dagger}_{J}\equiv
\left(ua^{\dagger}_{1}+vb^{\dagger}_{1}\right)^{J_{-}+J}  \left(a^{\dagger}_{1}b^{\dagger}_{L}-b^{\dagger}_{1}a^{\dagger}_{L}\right)^{J_{+}-J}
\left(ua^{\dagger}_{L}+vb^{\dagger}_{L}\right)^{-J_{-}+J}, \label{aope00}
\end{eqnarray}
where $J_{-}\equiv(M_{01}-M_{L,L+1})/2$, $J_{+}\equiv(M_{01}+M_{L,L+1})/2$ and $|J_{-}|\leq J\leq J_{+}$. Then the set of ground states of the block Hamiltonian (\ref{blockhamire00}) is
\begin{eqnarray}
	|\mathrm{G}; J, \hat{\Omega}\rangle \equiv
A^{\dagger}_{J}|\mathrm{VBS}_{L}\rangle, \qquad J=|J_{-}|, \ldots, J_{+}. \label{eigeinhom}
\end{eqnarray}
To prove that any state $|\mathrm{G}; J, \hat{\Omega}\rangle$ is a zero-energy ground state of (\ref{blockhamire00}), we essentially repeat the arguments as in \S$\,\ref{sec:gsbhs}\,$ for any site $j$ and bond $(j, j+1)$:
\begin{enumerate}
	\item The total power of $a^{\dagger}_{j}$ and $b^{\dagger}_{j}$ is $2S_{j}$, so that we have spin-$S_{j}$ at site $j$;
	\item $-\frac{1}{2}(M_{j-1,j}+M_{j+1,j+2})\leq J^{z}_{j,j+1}\equiv S^{z}_{j}+S^{z}_{j+1}\leq \frac{1}{2}(M_{j-1,j}+M_{j+1,j+2})$ by a binomial expansion, so that the maximum value of the bond spin $J_{j,j+1}$ is $\frac{1}{2}(M_{j-1,j}+M_{j+1,j+2})=S_{j}+S_{j+1}-M_{j,j+1}$ (from $SU(2)$ invariance, see \cite{AAH}).
\end{enumerate}
Therefore, the state $|\mathrm{G}; J, \hat{\Omega}\rangle$ defined in (\ref{eigeinhom}) has spin-$S_{j}$ at site $j$ and no projection onto the $J_{j, j+1}>S_{j}+S_{j+1}-M_{j,j+1}$ subspace for any bond.

Parallelly, we also introduce an orthogonal basis in description of the degenerate zero-energy ground states of $H_{B}$ (\ref{blockhamire00}), \textit{i.e.} the degenerate VBS states. Using the same notations as in \S$\,\ref{sec:gsbhs}\,$, the operator $A^{\dagger}_{J}$ defined in (\ref{aope00}) can be expanded as (see \cite{Ha,XKHK2})
\begin{eqnarray}
	A^{\dagger}_{J}&=&\sqrt{\frac{(J_{+}+J+1)!(J_{-}+J)!(J_{+}-J)!(-J_{-}+J)!}{2J+1}} 
	\sum^{J}_{M=-J}X_{JM} \label{aexp00} \\
	&&\cdot\sum^{m_{1}+m_{L}=M}_{m_{1}, m_{L}}(\frac{1}{2}M_{01}, m_{1}; \frac{1}{2}M_{L,L+1}, m_{L}|J, M)\ \psi^{\dagger}_{\frac{1}{2}M_{01},m_{1}}\otimes\psi^{\dagger}_{\frac{1}{2}M_{L,L+1},m_{L}}, \nonumber
\end{eqnarray}
where $(\frac{1}{2}M_{01}, m_{1}; \frac{1}{2}M_{L,L+1}, m_{L}|J, M)$ is the Clebsch-Gordan coefficient. Again, the particular form of the sum over $m_{1}$ and $m_{L}$ in (\ref{aexp00}) is identified as a single spin state creation operator
\begin{eqnarray}
	\Psi^{\dagger}_{JM}\equiv\sum^{m_{1}+m_{L}=M}_{m_{1}, m_{L}}(\frac{1}{2}M_{01}, m_{1}; \frac{1}{2}M_{L,L+1}, m_{L}|J, M)\ \psi^{\dagger}_{\frac{1}{2}M_{01},m_{1}}\otimes\psi^{\dagger}_{\frac{1}{2}M_{L,L+1},m_{L}}.\nonumber \\ \label{crea200}
\end{eqnarray}
So that the set of \textit{degenerate VBS states} $\{|\mathrm{VBS}_L(J,M)\rangle\}$ is defined as
\begin{eqnarray}
	|\mathrm{VBS}_L(J,M)\rangle\equiv \Psi^{\dagger}_{JM}|\mathrm{VBS}_L\rangle, \quad J=|J_{-}|,...,J_{+}, \quad M=-J, ...,J. \label{devb00}
\end{eqnarray}
Then these $(M_{01}+1)(M_{L,L+1}+1)$ states (\ref{devb00}) are mutually orthogonal (the proof is exactly the same as in \S$\,\ref{sec:gsbhs}\,$). Also, the state $|\mathrm{G}; J, \hat{\Omega}\rangle$ written as a linear superposition over these degenerate VBS states reads
\begin{eqnarray}
	|\mathrm{G};J,\hat{\Omega}\rangle&=&\sqrt{\frac{(J_{+}+J+1)!(J_{-}+J)!(J_{+}-J)!(-J_{-}+J)}{2J+1}}
	\nonumber \\
	&&\cdot\sum^{J}_{M=-J}X_{JM}|\mathrm{VBS}_{L}(J, M)\rangle.
	\label{line00}
\end{eqnarray}

Therefore, as seen from (\ref{line00}), the rank of set of states $\{|\mathrm{G}; J, \hat{\Omega}\rangle\}$ with the same $J$ value is $2J+1$ and the total number of linearly independent states of the set $\{|\mathrm{G}; J, \hat{\Omega}\rangle\}$ is $\sum^{J_{+}}_{J=|J_{-}|}(2J+1)=(M_{01}+1)(M_{L,L+1}+1)$, which is exactly the degeneracy of the ground states of (\ref{blockhamire00}). So that $\{|\mathrm{G}; J, \hat{\Omega}\rangle\}$ forms a complete set of zero-energy ground states. The orthogonal set $\{|\mathrm{VBS}_L(J,M)\rangle\}$ also forms a complete set of zero-energy ground states, which differs from $\{|\mathrm{G}; J, \hat{\Omega}\rangle\}$ by a change of basis.

\subsection{Diagonalization of the Density Matrix}
\label{sec:diadeninhom}

The density matrix is diagonalized in \S$\,\ref{sec:eigvecs}\,$ and \S$\,\ref{sec:alterproof}\,$ for the homogeneous AKLT model. The analysis can be made in parallel for the inhomogeneous model.

The statement of \textbf{Theorem} \ref{theorem2} is still valid here. \textit{i.e.} Eigenvectors of the density matrix $\boldsymbol{\rho}_{L}$ (\ref{matr00}) with non-zero eigenvalues are given by the set $\{|\mbox{G}; J, \hat{\Omega}\rangle\}$ (\ref{eigeinhom}) or $\{|\mathrm{VBS}_L(J,M)\rangle\}$ (\ref{devb00}). This explicit construction of eigenvectors yields a direct diagonalization of the density matrix.

Again, we prove the theorem by re-writing the density matrix $\boldsymbol{\rho}_{L}$ (\ref{matr00}) as a projector in diagonal form onto the orthogonal degenerate VBS states $\{|\mathrm{VBS}_{L}(J, M)\rangle\}$ introduced in (\ref{devb00}).

Take expression (\ref{matr00}) and integrate over $\hat{\Omega}_{0}$ and $\hat{\Omega}_{L+1}$ using binomial expansions and 
\begin{eqnarray}
	\int^{1}_{-1}\mathrm{d}x (1+x)^{m}(1-x)^{n}=\frac{m!n!}{(m+n+1)!}2^{m+n+1}.
\end{eqnarray}
Then we have
\begin{eqnarray}
	\boldsymbol{\rho}_{L}&=&	\frac{\displaystyle\prod^{L-1}_{j=1}(M_{j,j+1}+1)}{\displaystyle\prod^{L}_{j=1}(2S_{j}+1)!}\sum^{M_{01}}_{p=0}\sum^{M_{L,L+1}}_{q=0}\left(\begin{array}{c}M_{01}\\p\end{array}\right)\left(\begin{array}{c}M_{L,L+1}\\q\end{array}\right)
\nonumber \\
&&(b^{\dagger}_{1})^{p}(a^{\dagger}_{1})^{M_{01}-p}(a^{\dagger}_{L})^{q}(b^{\dagger}_{L})^{M_{L,L+1}-q}|\mathrm{VBS}_{L}\rangle
\nonumber \\
&&\langle\mathrm{VBS}_{L}|(b_{L})^{M_{L,L+1}-q}(a_{L})^{q}(a_{1})^{M_{01}-p}(b_{1})^{p}. \label{treat100}
\end{eqnarray}
The particular combinations of bosonic operators appeared in (\ref{treat100}) are recognized up to a constant as spin creation operators $\psi^{\dagger}_{\frac{1}{2}M_{01},\frac{1}{2}M_{01}-p}$ and \\ $\psi^{\dagger}_{\frac{1}{2}M_{L,L+1},q-\frac{1}{2}M_{L,L+1}}$ at site $1$ and site $L$, respectively. They commute with all bond operators $\left(a^{\dagger}_{j}b^{\dagger}_{j+1}-b^{\dagger}_{j}a^{\dagger}_{j+1}\right)^{M_{j,j+1}}$, so that we could simplify the right hand side of (\ref{treat100}) using definition (\ref{crea200}) and the following identity:
\begin{eqnarray}
&&\sum^{M_{01}}_{p=0}\sum^{M_{L,L+1}}_{q=0}\psi^{\dagger}_{\frac{1}{2}M_{01},\frac{1}{2}M_{01}-p}\otimes\psi^{\dagger}_{\frac{1}{2}M_{L,L+1},q-\frac{1}{2}M_{L,L+1}}|\mathrm{vac}\rangle_{1,L}
\nonumber \\
&&{}_{1,L}\langle\mathrm{vac}|\psi_{\frac{1}{2}M_{01},\frac{1}{2}M_{01}-p}\otimes\psi_{\frac{1}{2}M_{L,L+1},q-\frac{1}{2}M_{L,L+1}} \nonumber \\
&=&\sum^{M_{01}}_{p=0}\sum^{M_{L,L+1}}_{q=0}|\frac{1}{2}M_{01},\frac{1}{2}M_{01}-p\rangle_{1}\langle\frac{1}{2}M_{01},\frac{1}{2}M_{01}-p| \nonumber \\
&&\otimes|\frac{1}{2}M_{L,L+1},q-\frac{1}{2}M_{L,L+1}\rangle_{L}\langle\frac{1}{2}M_{L,L+1},q-\frac{1}{2}M_{L,L+1}| \nonumber \\
&=&\sum^{J_{+}}_{J=|J_{-}|}\sum^{J}_{M=-J}|J,M\rangle_{1,L}\langle J,M| \nonumber \\
&=&\sum^{J_{+}}_{J=|J_{-}|}\sum^{J}_{M=-J}\Psi^{\dagger}_{JM}|\mathrm{vac}\rangle_{1,L}\langle \mathrm{vac}|\Psi_{JM}.
\end{eqnarray}
The resultant final form of density matrix $\boldsymbol{\rho}_{L}$ is then
\begin{eqnarray}
	&&\boldsymbol{\rho}_{L}=\frac{\displaystyle\prod^{L-1}_{j=1}(M_{j,j+1}+1)}{\displaystyle\prod^{L}_{j=1}(2S_{j}+1)!}M_{01}!M_{L,L+1}!\sum^{J_{+}}_{J=|J_{-}|}\sum^{J}_{M=-J}\Psi^{\dagger}_{JM}|\mathrm{VBS}_{L}\rangle\langle \mathrm{VBS}_{L}|\Psi_{JM} \label{fipr00} \nonumber \\
&&=\frac{\displaystyle\prod^{L-1}_{j=1}(M_{j,j+1}+1)}{\displaystyle\prod^{L}_{j=1}(2S_{j}+1)!}M_{01}!M_{L,L+1}!\sum^{J_{+}}_{J=|J_{-}|}\sum^{J}_{M=-J}|\mathrm{VBS}_{L}(J, M)\rangle\langle \mathrm{VBS}_{L}(J, M)|. \nonumber \\ \label{finalrho00}
\end{eqnarray}

The set of degenerate VBS states $\{|\mathrm{VBS}_{L}(J, M)\rangle\}$ with $J=|J_{-}|,\ldots,J_{+}$ and $M=-J,\ldots,J$ forms an orthogonal basis. These $(M_{01}+1)(M_{L,L+1}+1)$ states also forms a complete set of zero-energy ground states of the block Hamiltonian (\ref{blockhamire00}). So that in expression (\ref{finalrho00}) we have re-written the density matrix as a projector in diagonal form over an orthogonal basis. Each degenerate VBS state $|\mathrm{VBS}_{L}(J, M)\rangle$ is an eigenvector of the density matrix, so as any of the state $|\mbox{G}; J, \hat{\Omega}\rangle$ (because of the degeneracy of corresponding eigenvalues of the density matrix, see \S$\,\ref{sec:eigvalinhom}\,$). Thus we have generalized \textbf{Theorem} \ref{theorem2} to the inhomogeneous case.

\subsection{Eigenvalues of the Density Matrix}
\label{sec:eigvalinhom}

Given the diagonalized form (\ref{finalrho00}), again eigenvalues of the density matrix $\boldsymbol{\rho}_{L}$ are derived from normalization of degenerate VBS states with an explicit expression in terms of Wigner $3j$-symbols. 

Similarly, we first calculate the integral of the norm square of $|\mbox{G};J,{\hat\Omega}\rangle$over the unit vector $\hat{\Omega}$
\begin{eqnarray}
 &&\frac{1}{4\pi}\int \mathrm{d}{\hat \Omega}\,\langle\mathrm{G};J,\hat{\Omega}|\mathrm{G};J,\hat{\Omega}\rangle  \label{novb00} \\
&=& \frac{(J_{+}+J+1)!(J_{-}+J)!(J_{+}-J)!(-J_{-}+J)!}{(2J+1)!}\,\langle \mathrm{VBS}_L(J,M)|\mathrm{VBS}_L(J,M)\rangle.\nonumber
\end{eqnarray}
This expression (\ref{novb00}) also states that normalization of the degenerate VBS state is independent of $\hat{\Omega}$ and/or $M$.

Let us consider the integral involved in (\ref{novb00}). Using coherent state basis (\ref{cohebas}) and completeness relation (\ref{comp}) as before, we obtain
\begin{eqnarray}
&& \frac{1}{4\pi}\int \mathrm{d}{\hat\Omega}\,\langle\mathrm{G};J,{\hat\Omega}|\mathrm{G};J,{\hat\Omega}\rangle
\label{inne00} \\
&=& \frac{1}{4\pi}\left[\prod^{L}_{j=1}\frac{(2S_{j}+1)!}{4\pi}\right] \int \mathrm{d}{\hat\Omega} \int \left[\prod^{L}_{j=1} \mathrm{d}{\hat\Omega}_{j}\right] \prod^{L-1}_{j=1}\left[\frac{1}{2}(1-\hat{\Omega}_{j}\cdot\hat{\Omega}_{j+1})\right]^{M_{j,j+1}} \nonumber \\
&&\cdot\left[\frac{1}{2}(1-\hat{\Omega}_{1}\cdot\hat{\Omega}_{L})\right]^{J_{+}-J}
\left[\frac{1}{2}(1+\hat{\Omega}_{1} \cdot\hat{\Omega})\right]^{J_{-}+J}
\left[\frac{1}{2}(1+\hat{\Omega} \cdot\hat{\Omega}_{L})\right]^{-J_{-}+J}. \nonumber
\end{eqnarray}
Now we expand $\left[\frac{1}{2}(1-\hat{\Omega}_i \cdot\hat{\Omega}_j)\right]^{M_{ij}}$ in terms of spherical harmonics
\begin{eqnarray}
\left[\frac{1}{2}(1-\hat{\Omega}_i \cdot\hat{\Omega}_j)\right]^{M_{ij}}
=\frac{4\pi}{M_{ij}+1}\sum^{M_{ij}}_{l=0}\lambda(l,M_{ij})\sum^{l}_{m=-l} Y_{lm}(\hat{\Omega}_{i})
Y^{\ast}_{lm}(\hat{\Omega}_{j}) \label{expansion00}
\end{eqnarray}
with
\begin{equation}
\lambda(l,M_{ij})=\frac{(-1)^l M_{ij}!(M_{ij}+1)!}{(M_{ij}-l)!(M_{ij}+l+1)!}. \label{lambdalm00}
\end{equation}
Then integrate over ${\hat\Omega}$ and from ${\hat\Omega}_2$ to ${\hat\Omega}_{L-1}$, the right hand side of (\ref{inne00}) is equal to
\begin{eqnarray}
&& \frac{4\pi \displaystyle\prod^{L}_{j=1}(2S_{j}+1)!}{\left[\displaystyle\prod^{L-1}_{j=1}(M_{j,j+1}+1)\right](J_{-}+J+1)(J_{+}-J+1)(-J_{-}+J+1)} \nonumber \\
&&\sum^{M_{<}}_{l} \sum^{J_{+}-J}_{l_{\alpha}=0} \sum^{J_{<}}_{l_{\beta}=0}
\sum^{l}_{m=-l} \sum^{l_{\alpha}}_{m_{\alpha}=-l_{\alpha}} \sum^{l_{\beta}}_{m_{\beta}=-l_{\beta}}
\left[\prod^{L-1}_{j=1}\lambda(l,M_{j,j+1})\right] \nonumber \\
&&\cdot\lambda(l_{\alpha},J_{+}-J) \lambda(l_{\beta},J_{-}+J) \lambda(l_{\beta},-J_{-}+J)
\int \mathrm{d}{\hat\Omega}_{1} \int \mathrm{d}{\hat\Omega}_{L}
\nonumber \\
&&\cdot Y_{l,m}({\hat\Omega}_{1})Y_{l_{\alpha},m_{\alpha}}({\hat\Omega}_{1})Y_{l_{\beta},m_{\beta}}({\hat\Omega}_1)
Y^{\ast}_{l,m}({\hat\Omega}_{L})Y^{\ast}_{l_{\alpha},m_{\alpha}}({\hat\Omega}_{L})Y^{\ast}_{l_{\beta},m_{\beta}}({\hat\Omega}_{L}).\nonumber\\
\label{pre100}
\end{eqnarray}
Where we have $M_{<}\equiv \mathrm{min}\{M_{j,j+1},j=1,\ldots,L-1\}$ and $J_{<}\equiv \mathrm{min}\{J_{-}+J, -J_{-}+J\}$, both being the minimum of the corresponding set. Now we carry out remaining integrals in (\ref{pre100}) using
\begin{eqnarray}
&&\int \mathrm{d}{\hat\Omega}\, Y_{l,m}({\hat\Omega})Y_{l_{\alpha},m_{\alpha}}({\hat\Omega})Y_{l_{\beta},m_{\beta}}({\hat\Omega})
\nonumber \\
&=&\sqrt{\frac{(2l+1)(2l_{\alpha}+1)(2l_{\beta}+1)}{4\pi}}
\left(\begin{array}{c c c}
l & l_{\alpha} & l_{\beta} \\
0   &  0  & 0 
\end{array}\right)
\left(\begin{array}{c c c}
l & l_{\alpha} & l_{\beta} \\
m & m_{\alpha} & m_{\beta} 
\end{array}\right). \nonumber\\ \label{wign00}
\end{eqnarray}
The result after integration can be further simplified by applying the following orthogonality relation
\begin{eqnarray}
	\sum_{m,m_{\alpha}}(2l_{\beta}+1)
	\left(\begin{array}{c c c}
	l & l_{\alpha} & l_{\beta} \\
	m & m_{\alpha} & m_{\beta} 
	\end{array}\right)
	\left(\begin{array}{c c c}
	l & l_{\alpha} & l^{\prime}_{\beta} \\
	m & m_{\alpha} & m^{\prime}_{\beta} 
	\end{array}\right)=\delta_{l_{\beta}l^{\prime}_{\beta}} \delta_{m_{\beta}m^{\prime}_{\beta}}, \label{3jor00}
\end{eqnarray}
where $\left(\begin{array}{c c c}
l & l_{\alpha} & l_{\beta} \\
m & m_{\alpha} & m_{\beta} 
\end{array}\right)$, \textit{etc.} are the Wigner $3j$-symbols.

So that finally expression (\ref{pre100}) is equal to
\begin{eqnarray}
&&\frac{\displaystyle\prod^{L}_{j=1}(2S_{j}+1)!}{\left[\displaystyle\prod^{L-1}_{j=1}(M_{j,j+1}+1)\right](J_{-}+J+1)(J_{+}-J+1)(-J_{-}+J+1)} \nonumber \\
&&\sum^{M_{<}}_{l} \sum^{J_{+}-J}_{l_{\alpha}=0} \sum^{J_{<}}_{l_{\beta}=0}
\left[\prod^{L-1}_{j=1}\lambda(l,M_{j,j+1})\right]
\lambda(l_{\alpha},J_{+}-J) \lambda(l_{\beta},J_{-}+J) \lambda(l_{\beta},-J_{-}+J)
\nonumber \\
&&\cdot(2l+1)(2l_{\alpha}+1)(2l_{\beta}+1)
\left(\begin{array}{c c c}
l & l_{\alpha} & l_{\beta} \\
0   &  0  & 0 
\end{array}\right)^{2}. \label{renovb00}
\end{eqnarray}
The explicit value of $\left(\begin{array}{c c c}
l & l_{\alpha} & l_{\beta} \\
0   &  0  & 0 
\end{array} \right)$ is given by
\begin{eqnarray}
&&\left(\begin{array}{c c c}
l & l_{\alpha} & l_{\beta} \\
0   &  0  & 0 
\end{array} \right) \label{3jva00} \\
&=&(-1)^{g} \sqrt{\frac{(2g-2l)!(2g-2l_{\alpha})!(2g-2l_{\beta})!}{(2g+1)!}}\frac{g!}{(g-l)!(g-l_{\alpha})!(g-l_{\beta})!},
\nonumber
\end{eqnarray}
if $l+l_{\alpha}+l_{\beta}=2g$ ($g \in \mathbf{N}$), otherwise zero. 

Combining results of (\ref{fipr00}), (\ref{novb00}) and (\ref{renovb00}), we arrive at the following result for eigenvalues:
Eigenvalues $\Lambda(J)$ $(J=|J_{-}|,\ldots,J_{+})$ of the density matrix are independent of $\hat{\Omega}$ and/or $M$ in defining eigenvectors (see (\ref{eigeinhom}) and (\ref{devb00})). An explicit expression is given by the following triple sum
	\begin{eqnarray}
&&\Lambda(J)		=\frac{\displaystyle\prod^{L-1}_{j=1}(M_{j,j+1}+1)}{\displaystyle\prod^{L}_{j=1}(2S_{j}+1)!}{M_{01}!M_{L,L+1}!}\,\langle\mathrm{VBS}_{L}(J,M)|\mathrm{VBS}_{L}(J,M)\rangle  \nonumber\\
&&=\frac{(2J+1)!M_{01}!M_{L,L+1}!}{(J_{+}+J+1)!(J_{-}+J+1)!(J_{+}-J+1)!(-J_{-}+J+1)!} \nonumber \\
&&\sum^{M_{<}}_{l} \sum^{J_{+}-J}_{l_{\alpha}=0} \sum^{J_{<}}_{l_{\beta}=0}
\left[\prod^{L-1}_{j=1}\lambda(l,M_{j,j+1})\right]
\lambda(l_{\alpha},J_{+}-J) \lambda(l_{\beta},J_{-}+J) \lambda(l_{\beta},-J_{-}+J)
\nonumber \\
&&\cdot(2l+1)(2l_{\alpha}+1)(2l_{\beta}+1)
\left(\begin{array}{c c c}
l & l_{\alpha} & l_{\beta} \\
0   &  0  & 0 
\end{array}\right)^{2}. \label{eivaex00}
	\end{eqnarray}

\subsection{The Large Block Limit}
\label{sec:largeinhom}

In this section, we generalize the characteristic properties (\S$\,\ref{sec:larges}\,$) of the limiting density matrix to the inhomogeneous model.

Let us apply the density matrix $\boldsymbol{\rho}_{L}$ (\ref{matr00}) to the state $|\mathrm{G}; J, \hat{\Omega}\rangle$ (\ref{eigeinhom}) and get
\begin{eqnarray}
	&&\boldsymbol{\rho}_{L}|\mathrm{G}; J, \hat{\Omega}\rangle \label{appl100} \\
	&=&\frac{\displaystyle\prod^{L}_{j=0}(M_{j,j+1}+1)}{\displaystyle\prod^{L}_{j=1}(2S_{j}+1)!}\frac{1}{(4\pi)^{2}}
	\int \mathrm{d}\hat{\Omega}_{0}\mathrm{d}\hat{\Omega}_{L+1}\,B^{\dagger}|\mathrm{VBS}_{L}\rangle\langle \mathrm{VBS}_{L}|BA^{\dagger}_{J}|\mathrm{VBS}_{L}\rangle. \nonumber
\end{eqnarray}
Using the coherent state basis (\ref{cohebas}) and completeness relation (\ref{comp}), the factor $\langle \mathrm{VBS}_{L}|BA^{\dagger}_{J}|\mathrm{VBS}_{L}\rangle$
in (\ref{appl1}) can be re-written as
\begin{eqnarray}
	&&\langle \mathrm{VBS}_{L}|BA^{\dagger}_{J}|\mathrm{VBS}_{L}\rangle \label{appl200} \\
	&=&\displaystyle\left[\prod^{L}_{j=1}\frac{(2S_{j}+1)!}{4\pi}\right]\int \left(\prod^{L}_{j=1}\mathrm{d}\hat{\Omega}_{j}\right)\prod^{L-1}_{j=1}\left[\frac{1}{2}(1-\hat{\Omega}_{j}\cdot\hat{\Omega}_{j+1})\right]^{M_{j,j+1}}
	\nonumber \\
	&&\cdot\left(u_{0}v_{1}-v_{0}u_{1}\right)^{M_{01}} \left(uu^{\ast}_{1}+vv^{\ast}_{1}\right)^{J_{-}+J}\left(u^{\ast}_{1}v^{\ast}_{L}-v^{\ast}_{1}u^{\ast}_{L}\right)^{J_{+}-J}
\nonumber \\	&&\cdot\left(uu^{\ast}_{L}+vv^{\ast}_{L}\right)^{-J_{-}+J}\left(u_{L}v_{L+1}-v_{L}u_{L+1}\right)^{M_{L,L+1}}. \nonumber
\end{eqnarray}
We plug the expression (\ref{appl200}) into (\ref{appl100}). Using transformation properties under $SU(2)$ and a binomial expansion, the integral over $\hat{\Omega}_{0}$ yields that
\begin{eqnarray}
	\int \mathrm{d}\hat{\Omega}_{0}\left(u^{\ast}_{0}b^{\dagger}_{1}-v^{\ast}_{0}a^{\dagger}_{1}\right)^{M_{01}}\left(u_{0}v_{1}-v_{0}u_{1}\right)^{M_{01}}
	=\frac{4\pi}{M_{01}+1}\left(u_{1}a^{\dagger}_{1}+v_{1}b^{\dagger}_{1}\right)^{M_{01}}.\nonumber\\ \label{int000}
\end{eqnarray}
Similarly we can perform the integral over $\hat{\Omega}_{L+1}$. Then using expansion (\ref{expansion00}) and orthogonality of spherical harmonics, other integrals over $\hat{\Omega}_{j}$ with $j=2, \ldots, L-1$ in (\ref{appl200}) can be performed. As a result, the following expression is obtained from (\ref{appl100}):
\begin{eqnarray}
\boldsymbol{\rho}_{L}|\mathrm{G}; J, \hat{\Omega}\rangle =  \frac{1}{(4\pi)^{2}}\sum^{M_{<}}_{l=0}(2l+1)\left[\prod^{L-1}_{j=1}\lambda(l,M_{j,j+1})\right]
K^{\dagger}_{l}(\hat{\Omega})
\left|\mathrm{VBS}_{L}\right\rangle. \label{sum00}
\end{eqnarray}
The operator $K^{\dagger}_{l}(\hat{\Omega})$ involved in (\ref{sum00}) is defined as
\begin{eqnarray}
	K^{\dagger}_{l}(\hat{\Omega})&\equiv&\int \mathrm{d}\hat{\Omega}_{1} \mathrm{d}\hat{\Omega}_{L}\, 
	P_{l}(\hat{\Omega}_{1} \cdot \hat{\Omega}_{L})
	\left(u_{1}a^{\dagger}_{1}+v_{1}b^{\dagger}_{1}\right)^{M_{01}} \left(uu^{\ast}_{1}+vv^{\ast}_{1}\right)^{J_{-}+J}
\nonumber \\ 
&&\cdot \left(u^{\ast}_{1}v^{\ast}_{L}-v^{\ast}_{1}u^{\ast}_{L}\right)^{J_{+}-J}
\left(uu^{\ast}_{L}+vv^{\ast}_{L}\right)^{-J_{-}+J}
\left(u_{L}a^{\dagger}_{L}+v_{L}b^{\dagger}_{L}\right)^{M_{L,L+1}}. \nonumber \\ \label{inte00}
\end{eqnarray} 
It is expressed as an integral depending on the order $l$ of the Legendre polynomial $P_{l}(\hat{\Omega}_{1} \cdot \hat{\Omega}_{L})$.

There was no ambiguity in defining the large block limit in the homogeneous AKLT model (see \S$\,\ref{sec:larges}\,$). However, in the inhomogeneous model we must specify the behavior of ending spins in the large block limit. So we define the large block limit as when $L\rightarrow\infty$, the two ending spins approach definite values, namely, $M_{01}\rightarrow S_{-}$ and $M_{L,L+1}\rightarrow S_{+}$. Then we realize from (\ref{lambdalm00}) that as $L\rightarrow\infty$, $\prod^{L-1}_{j=1}\lambda(l,M_{j,j+1})\rightarrow \delta_{l,0}$. Therefore only the first term with $l=0$ is left in (\ref{sum00}). So that we need only to calculate the limiting $K^{\dagger}_{0}(\hat{\Omega})$:
\begin{eqnarray}
	K^{\dagger}_{0}(\hat{\Omega})&\stackrel{L\rightarrow\infty}{\longrightarrow}&\int \mathrm{d}\hat{\Omega}_{1} \mathrm{d}\hat{\Omega}_{L} 
	\left(u_{1}a^{\dagger}_{1}+v_{1}b^{\dagger}_{1}\right)^{S_{-}} \left(uu^{\ast}_{1}+vv^{\ast}_{1}\right)^{J_{-}+J}    \nonumber \\ &&\cdot
\left(u^{\ast}_{1}v^{\ast}_{L}-v^{\ast}_{1}u^{\ast}_{L}\right)^{J_{+}-J}
\left(uu^{\ast}_{L}+vv^{\ast}_{L}\right)^{-J_{-}+J}
\left(u_{L}a^{\dagger}_{L}+v_{L}b^{\dagger}_{L}\right)^{S_{+}}. \nonumber\\ \label{k000}
\end{eqnarray} 
Here both $J_{-}$ and $J_{+}$ take the limiting values $\frac{1}{2}(S_{-}-S_{+})$ and $\frac{1}{2}(S_{-}+S_{+})$, respectively.

Using transformation properties of the integrand in (\ref{k000}) under $SU(2)$, the $K^{\dagger}_{0}(\hat{\Omega})$ integral is simplified and carried out as
\begin{eqnarray}
K^{\dagger}_{0}(\hat{\Omega})
=\frac{(4\pi)^{2}}{(S_{-}+1)(S_{+}+1)}\,A^{\dagger}_{J}. \label{fina00}
\end{eqnarray}
This expression states that $\{|\mathrm{G}; J, \hat{\Omega}\rangle\}$ is a set of eigenvectors of the density matrix as $L\rightarrow\infty$. Let us denote the density matrix in the limit by $\boldsymbol{\rho}_{\infty}$. Then (\ref{fina00}) leads to the result (see (\ref{sum00}))
\begin{eqnarray}
	\boldsymbol{\rho}_{\infty}|\mathrm{G}; J, \hat{\Omega}\rangle=
	\frac{1}{(S_{-}+1)(S_{+}+1)}\,|\mathrm{G}; J, \hat{\Omega}\rangle. \label{appll00}
\end{eqnarray}

We find from (\ref{appll00}) that the limiting eigenvalue
\begin{eqnarray}
	\Lambda_{\infty}=\frac{1}{(S_{-}+1)(S_{+}+1)}, \qquad L\to\infty
\end{eqnarray}
is independent of $J$. Any vector of the $(S_{-}+1)(S_{+}+1)$-dimensional subspace spanned by the set $\{|\mathrm{G}; J, \hat{\Omega}\rangle\}$ is an eigenvector of $\boldsymbol{\rho}_{\infty}$ with the same eigenvalue $\frac{1}{(S_{-}+1)(S_{+}+1)}$. Therefore $\boldsymbol{\rho}_{\infty}$ is proportional to a projector $\mathbf{P}_{(S_{-}+1)(S_{+}+1)}$:
\begin{eqnarray}
	\boldsymbol{\rho}_{L}\to\boldsymbol{\rho}_{\infty}=\frac{1}{(S_{-}+1)(S_{+}+1)}\,\mathbf{P}_{(S_{-}+1)(S_{+}+1)}, \label{limitrho00}
\end{eqnarray}
which is a generalization of (\ref{largedens}) to the inhomogeneous model. In the expression (\ref{limitrho00}) $S_{-}$ and $S_{+}$ denote the limiting spin values at the left and right boundary sites of the block, respectively. In addition, we also derive from the eigenvalues that the von Neumann entropy $S_{\mathrm{v\ N}}$ coincides with the R\'enyi entropy $S_{\mathrm{R}}$ and is equal to the saturated value $\ln\left[(S_{-}+1)(S_{+}+1)\right]$.

\section{The One--dimensional SU(n) Model}
\label{sec:su(n)}

In previous sections, we have discussed VBS states with spins in different representations of $SU(2)$. Our discussion essentially exhausted all possible variations of $SU(2)$ VBS states in $1$-dimension with open boundary conditions. The AKLT model and the VBS ground state can be generalized to the $SU(n)$ case as in \cite{GR,GRS,RSSTG,RTFSG}. In this section we study the entanglement of the VBS state in the $SU(n)$ version with open boundary conditions. Our treatment follows \cite{KHK}.

\subsection{The Hamiltonian and the SU(n) VBS State}
\label{sec:su(n)hamil}

Let us first define the model. Consider a $1$-dimensional lattice with spins sitting on each site. What we mean by a `spin' in our model is an adjoint representation of $SU(n)$. Our spin chain consists of $N$ adjoint representations of $SU(n)$ in the bulk and fundamental and conjugate representations of $SU(n)$ on the two boundaries. This construction corresponds to the spin-$1$ model in \S$\,\ref{sec:1dspin1}\,$ where we have the adjoint representation of $SU(2)$ (\textit{i.e.} spin-$1$) in the bulk and the fundamental representation of $SU(2)$ (\textit{i.e.} spin-$1/2$) at two ends (the fundamental representation $\square$ and its conjugate representation $\bar{\square}$ are equivalent for $SU(2)$). Unlike previous sections, in this section we shall reverse the order of discussion for better understanding. That is, we construct the $SU(n)$ VBS state before writing down the Hamiltonian. The Hamiltonian will be constructed in such a way that the VBS state shall be the ground state.

\begin{figure}
	\centering
		\includegraphics[width=5in]{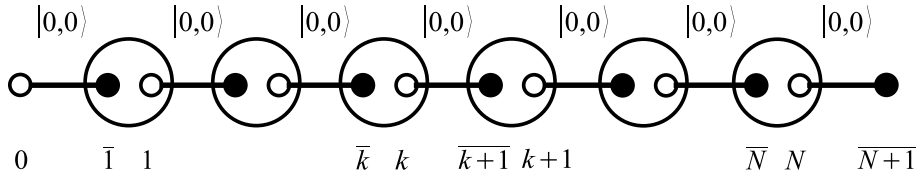}
	\caption{Construction of the $SU(n)$ VBS state. White and black dots represent the $SU(n)$ fundamental and its conjugate representations, respectively. A solid line connecting two dots corresponds to the singlet state $|0,0\rangle$ and a large circle denotes the projection onto the adjoint representation.}
	\label{construction}
\end{figure}

\begin{figure}
	\centering
		\includegraphics[width=5in]{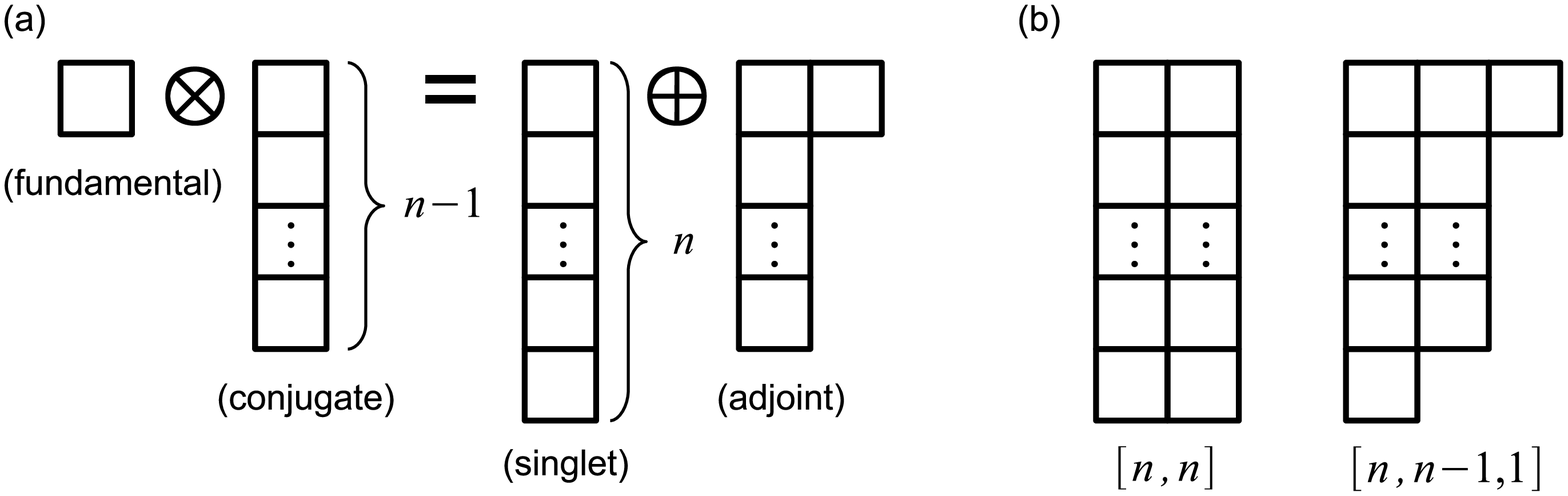}
	\caption{(a)The decomposition rule for the tensor product of $\square \otimes {\bar \square}$. (b)The Young tableaux corresponding to [$n,n$] and [$n,n-1,1$].}
	\label{construction2}
\end{figure}

Now let us first construct an $SU(n)$ VBS state which consists of $N$ adjoint representations of $SU(n)$ in the bulk and fundamental and conjugate representations of $SU(n)$ on the boundary. First, we prepare sites $k$ ($k=0, 1, ..., N$) and ${\bar k}$ ($k=1, 2, ..., N+1$) and arrange $SU(n)$ singlets consisting of a fundamental ($\square$) and its conjugate ($\bar {\square}$) representations as shown in Figure \ref{construction} (see also \cite{NO}).
We assign $|j\rangle$ $(j=0,1,2, ..., n-1)$ to the fundamental representation, while $|{\bar{j}}\rangle$ $(j=0,1,2, ..., n-1)$ to the conjugate representation. $|{\bar j}\rangle$ can be represented by the tensor product of $(n-1)$ $|j\rangle$'s as
\begin{equation}
|\bar{j}\rangle \equiv \frac{1}{\sqrt{(n-1)!}}\sum_{\alpha_2, ..., \alpha_n} \epsilon^{j \alpha_2,...,\alpha_n}|\alpha_2,...,\alpha_n \rangle
\label{conjugate}
\end{equation}
where $\epsilon^{j \alpha_2,...,\alpha_n}$ is a totally antisymmetric tensor of rank $n$. Using $|j\rangle$ and $|{\bar j}\rangle$, an $SU(n)$ singlet state $|0,0\rangle$ can be represented as a maximally entangled state:
\begin{equation}
|0,0\rangle = \frac{1}{\sqrt{N}}\sum_{j=0}^{N-1}|j\rangle |\bar{j}\rangle.
\end{equation}
The above relation can be easily confirmed by inserting the resolution of the identity $I=\sum_{j=0}^{n-1}|j\rangle \langle j|$ and substituting (\ref{conjugate}). Next, we prepare the adjoint representation of $SU(n)$ by projecting the tensor product $\square \otimes {\bar \square}$ onto an $(n^2-1)$-dimensional subspace (the dimension of the adjoint representation of $SU(n)$ is equal to the number of generators). This procedure corresponds to large circles in Figure \ref{construction}. In Figure \ref{construction2}(a) we visualize the decomposition rule $\square \otimes \bar{\square}=(\mathrm{singlet}) \oplus (\mathrm{adjoint})$. Then we have obtained the $SU(n)$ adjoint representation at each composite site ($k,{\bar k}$). Henceforth we shall call this composite site $k$. 
Finally, we can represent the $SU(n)$ generalized VBS state as
\begin{equation}
|\mathrm{VBS}\rangle = \left(\bigotimes_{k=1}^N \mathbf{P}_{k{\bar k}}\right)|0,0\rangle_{0{\bar 1}} |0,0\rangle_{1{\bar 2}} \dots |0,0\rangle_{N\overline{N+1}},
\label{gs}
\end{equation}
where $\mathbf{P}_{k{\bar k}}$ is a projection operator onto an adjoint representation of $SU(n)$. 

After construction of the VBS state (\ref{gs}), we now write down the Hamiltonian along the same line as the $SU(2)$ AKLT model:
\begin{eqnarray}
	&&H = \sum_{k=1}^{N-1} H(k,k+1)+H(0,1)+H(N,N+1), \nonumber\\
	&&H(k,k+1) = \sum_{Y}C_{Y}(k,k+1)\,\pi_{Y}(k,k+1),
\label{Ham}
\end{eqnarray}
where $Y$ is a Young tableau which is neither [$n, n$] nor [$n$, $n-1$, 1]. Here we have assigned [$\kappa_1, ..., \kappa_{\lambda_1}$] to the Young tableau $Y$, where $\kappa_j$ is the number of boxes in the $j^{\mathrm{th}}$ column and $\lambda_1$ is the number of boxes in the first row. $\pi_{Y}(k,k+1)$ is a projection operator which projects $(\mathrm{adjoint})\otimes(\mathrm{adjoint})$ onto a representation characterized by $Y$ and the coefficient $C_Y(k,k+1)$ can be an arbitrary positive number. The reason why [$n, n$] and [$n$, $n-1$, 1] are excluded from the sum is the following: Since $\square$ at site $k$ and $\bar \square$ at site $k+1$ have already formed a singlet in the ground state (\ref{gs}), the possible representations obtained from the decomposition of $(\mathrm{adjoint})\otimes(\mathrm{adjoint})$ are restricted to [$n, n$] and [$n$, $n-1$, 1] (as graphically shown in Figure \ref{construction2}(b)). 
$H(0,1)$ and $H(N,N+1)$ are boundary terms which assure the uniqueness of the ground state of this Hamiltonian. $H(0,1)$ and $H(N,N+1)$ can be written in terms of the projection operators acting on the tensor products $(\mathrm{fundamental})\otimes(\mathrm{adjoint})$ and $(\mathrm{conjugate})\otimes(\mathrm{adjoint})$, respectively. 
By construction, the $SU(n)$ VBS state (\ref{gs}) is a zero-energy ground state of this Hamiltonian (\ref{Ham}). We note here that another construction of the Hamiltonian by M. Greiter and S. Rachel \cite{GR} is similar but slightly different from ours.

\subsection{The Block Density Matrix}
\label{sec:su(n)denmatr}

Next, we consider the density matrix of a block subsystem of the VBS ground state $|\mathrm{VBS}\rangle$ (\ref{gs}). To calculate the density matrix, it is more convenient to recast the chain of singlets $|0,0\rangle_{0{\bar 1}} |0,0\rangle_{1{\bar 2}} \dots |0,0\rangle_{N\overline{N+1}}$ in (\ref{gs}) in a different form. Let us first look at a chain of two singlets $|0,0\rangle_{0{\bar 1}} |0,0\rangle_{1{\bar 2}}$. We can rewrite this product state as
\begin{equation}
|0,0\rangle_{0{\bar 1}} |0,0\rangle_{1{\bar 2}}=
\frac{1}{n}\sum_{l=0}^{n-1}\sum_{m=0}^{n-1}|l,m\rangle_{0{\bar 2}} 
|l,-m\rangle_{{\bar 1}1},
\label{important_relation}
\end{equation}
where $|m,n\rangle$ is a basis of the maximally entangled state defined by
\begin{equation}
|l,m\rangle = (U_{l,m}\otimes I)|0,0\rangle.
\end{equation}
Here $I$ is an $n$-dimensional identity matrix and $U_{l,m}=X^lZ^m$ ($m,n=0, 1, ..., n-1$) are generalized Pauli matrices, where the unitary operators $X$ and $Z$ act on $|j\rangle$ as $X|j\rangle=|j+1 ({\rm mod}.n)\rangle$ and $Z|j\rangle=\omega^j|j\rangle$ with $\omega=\mathrm{e}^{2\pi \mathrm{i}/n}$, respectively. 
One can easily show the relation (\ref{important_relation}) by using the fact that $|0,0\rangle$ is invariant under the action of $(U_{l,m}\otimes U_{l,-m})$, see \cite{F}. This procedure can be regarded as a multi-dimensional generalization of entanglement swapping. (The maximally entangled state basis (\ref{alph}) and a similar relation (\ref{maxi}) have been used in \S$\,\ref{sec:1dspin1VBS}\,$ for the $SU(2)$ spin-$1$ VBS state.) By repeatedly using the relation (\ref{important_relation}), we can generalize in a straightforward way to a chain of singlet states:
\begin{eqnarray}
&&|0,0\rangle_{0{\bar 1}} |0,0\rangle_{1{\bar 2}} \dots |0,0\rangle_{N\overline{N+1}}\nonumber\\
&=&\frac{1}{n^N}
\sum_{(l_1,m_1)}\cdots \sum_{(l_N,m_N)}|l_1,-m_1\rangle_{{\bar 1}1}\cdots |l_N,-m_N\rangle_{{\bar N}N}\nonumber\\
&&\cdot (U_{l_1,m_1}\cdots U_{l_N,m_N}\otimes I)|0,0\rangle_{0\overline{N+1}},
\label{singlet_product}
\end{eqnarray}
where $(m_k,n_k)$ ($k=1, 2, ..., N$) runs from (0,0) to $(n-1,n-1)$. To obtain the ground state $|\mathrm{VBS}\rangle$ from (\ref{singlet_product}), we still have to make a projection onto the subspace of adjoint representation at each site $k$. Since the decomposition rule $\square \otimes \bar{\square}=(\mathrm{singlet}) \oplus (\mathrm{adjoint})$ and the fact that $|0,0\rangle$ is an $SU(n)$ singlet, the vector space of the adjoint representation is spanned by $|l,-m\rangle$ ($(l,m)\ne (0,0)$). Then the only thing to do is to omit the summation over $(l_k,m_k)=(0,0)$ in (\ref{singlet_product}). As a result, the $SU(n)$ generalized VBS state can be re-written as:
\begin{eqnarray}
|\mathrm{VBS}\rangle &=&\frac{1}{(n^2-1)^{N/2}}\sum_{\stackrel{(l_1,m_1)}{\ne(0,0)}}\cdots\sum_{\stackrel{(l_N,m_N)}{\ne(0,0)}}|l_1,-m_1\rangle_{{\bar 1}1}\cdots |l_N,-m_N\rangle_{{\bar N}N}\nonumber\\
&&\cdot (U_{l_1,m_1}\cdots U_{l_N,m_N}\otimes I)|0,0\rangle_{0\overline{N+1}},
\end{eqnarray}
where we have already normalized $|\mathrm{VBS}\rangle$ by the factor $1/(n^2-1)^{N/2}$. 

Now we calculate the density matrix of a block of contiguous spins of length $L$. Assume that the block starts from site $k$ and stretches up to $k+L-1$, where $k\ge 1$ and $k+L-1 \le N$. The reduced density matrix is obtained by taking the trace over the sites $j=0,1, ..., k-1$ and $j=k+L, ..., N, \overline{N+1}$ outside the block as
\begin{eqnarray}
\boldsymbol{\rho}_{L} &=& \mathrm{tr}_{1,...,k-1,k+L,...,N,0,\overline{N+1}}\left[\,|\mathrm{VBS}\rangle \langle \mathrm{VBS}|\,\right] 
\nonumber \\
&=&\frac{1}{(n^2-1)^N} \sum_{\stackrel{(l_1,m_1)}{\ne(0,0)}}\cdots\sum_{\stackrel{(l_{k-1},m_{k-1})}{\ne(0,0)}}\sum_{\stackrel{(l_{k+L},m_{k+L})}{\ne(0,0)}}\nonumber\\
&&\cdots\sum_{\stackrel{(l_N,m_N)}{\ne(0,0)}}
\sum_{\stackrel{(l_k,m_k)}{\ne(0,0)}}\sum_{\stackrel{(l_k',m_k')}{\ne(0,0)}}\cdots \sum_{\stackrel{(l_{L+k-1},m_{L+k-1})}{\ne(0,0)}}\sum_{\stackrel{(l_{L+k-1}',m_{L+k-1}')}{\ne(0,0)}}
\nonumber \\
&& \cdot|l_k,-m_k\rangle_{{\bar k}k}\langle l_k',-m_k'|\cdots|l_{L+k-1},-m_{L+k-1}\rangle_{\overline{L+k-1}L+k-1}\langle l_{L+k-1}',-m_{L+k-1}'| \nonumber \\
&& \cdot\mathrm{tr}_{0,\overline{N+1}}\left[(U_1 V U_2 \otimes I)|0,0\rangle_{0\overline{N+1}}\langle 0,0|(U_1 V' U_2 \otimes I)^\dagger\right],
\label{RDM1}
\end{eqnarray}
where $U_1=U_{l_1,m_1}\cdots U_{l_{k-1},m_{k-1}}$, $U_2=U_{l_{L+k},m_{L+k}}\cdots U_{l_N,m_N}$,\\ $V=U_{l_k,m_k}\cdots U_{l_{L+k-1},m_{L+k-1}}$ and $V'=U_{l_k',m_k'}\cdots U_{l_{L+k-1}',m_{L+k-1}'}$.
To rewrite the complicated expression (\ref{RDM1}), we use the following property of $|0,0\rangle$:
\begin{equation}
(S \otimes T)|0,0\rangle =(ST^t\otimes I)|0,0\rangle=(I\otimes TS^t)|0,0\rangle,
\end{equation}
where $S$ and $T$ are $n$-dimensional unitary operations acting on $|j\rangle$ and $|{\bar j}\rangle$, respectively, and the superscript $t$ denotes the transposition. Using this property and the cyclic property of the trace, we can simplify the last part of (\ref{RDM1}) as
\begin{eqnarray}
&&\mathrm{tr}_{0,\overline{N+1}}\left[(U_1 V U_2 \otimes I)|0,0\rangle_{0\overline{N+1}}\langle 0,0|(U_1 V' U_2 \otimes I)^\dagger\right] \nonumber \\
&=&\mathrm{tr}_{0,\overline{N+1}}\left[(U_1\otimes I)(V \otimes I)(U_2 \otimes I)|0,0\rangle_{0\overline{N+1}}\langle 0,0|(U_2 \otimes I)^\dagger(V' \otimes I)^\dagger(U_1 \otimes I)^\dagger\right] \nonumber \\
&=&\mathrm{tr}_{0,\overline{N+1}}\left[(V \otimes I)(I \otimes U_2^t)|0,0\rangle_{0\overline{N+1}}\langle 0,0|(I \otimes U_2^t)^\dagger(V' \otimes I)^\dagger\right] \nonumber \\
&=&\mathrm{tr}_{0,\overline{N+1}}\left[(I \otimes U_2^t)(V \otimes I)|0,0\rangle_{0\overline{N+1}}\langle 0,0|(V' \otimes I)^\dagger(I \otimes U_2^t)^\dagger\right] \nonumber \\
&=&\mathrm{tr}_{0,\overline{N+1}}\left[(V \otimes I)|0,0\rangle_{0\overline{N+1}}\langle 0,0|(V' \otimes I)^\dagger\right].
\label{trace_relation}
\end{eqnarray}
Since (\ref{trace_relation}) does not depend on $(l_1,m_1), \cdots, (l_{k-1},m_{k-1})$ and $(l_{k+L},m_{k+L}), \cdots,\\ (l_N,k_N)$, we can rewrite the density matrix (\ref{RDM1}) as 
\begin{eqnarray}
\boldsymbol{\rho}_L
&=&\frac{1}{(n^2-1)^L}
\sum_{\stackrel{(l_k,m_k)}{\ne(0,0)}}\sum_{\stackrel{(l_k',m_k')}{\ne(0,0)}}\cdots \sum_{\stackrel{(l_{L+k-1},m_{L+k-1})}{\ne(0,0)}}\sum_{\stackrel{(l_{L+k-1}',m_{L+k-1}')}{\ne(0,0)}}
|l_k,-m_k\rangle_{{\bar k}k}\langle l_k',-m_k'|\nonumber\\
&&\cdots|l_{L+k-1},-m_{L+k-1}\rangle_{\overline{L+k-1}L+k-1}\langle l_{L+k-1}',-m_{L+k-1}'|\nonumber\\ 
&&\cdot\mathrm{tr}_{0,\overline{N+1}}\left[(V \otimes I)|0,0\rangle_{0\overline{N+1}}\langle 0,0|(V' \otimes I)^\dagger\right].
\label{RDM2}
\end{eqnarray}
From the form of the density matrix in (\ref{RDM2}), we immediately notice that the density matrix does not depend on both the starting site $k$ and the total length of the chain $N$. The same property for $SU(2)$ VBS states has been discussed in \S$\,\ref{sec:1ds1den}\,$, \S$\,\ref{sec:1dssden}\,$ and \S$\,\ref{sec:deninhom}\,$ (this property was proved in \cite{FKR} for spin-$1$, in \cite{XKHK} for Spin-$S$, and in \cite{XKHK2} for the inhomogeneous $SU(2)$ models, respectively). We can regard the above result as an $SU(n)$ generalization of those results.

\subsection{Spectrum of the Density Matrix and Entropies}
\label{sec:su(n)spec}

Since the density matrix $\boldsymbol{\rho}_{L}$ (\ref{RDM2}) is independent of both $k$ and $N$, we can set $N=L$ without loss of generality. We can further reduce the original problem to that of the reduced density matrix of two ending spins ($\square$ and $\bar \square$) using the Schmidt decomposition of a bipartite pure state (see the introduction \S$\,\ref{sec:intro1}\,$ for a brief discussion). Suppose that $|B\cup E\rangle$ is a bipartite pure state of a total system $B\cup E$. Then there exists orthonormal states $|B_j\rangle$ for the subsystem $B$, and orthonormal states $|E_j\rangle$ for $E$ such that 
\begin{equation}
|B\cup E\rangle=\sum_j \sqrt{\lambda_j}\ |B_j\rangle\otimes |E_j\rangle,
\label{Schmidt}
\end{equation}
where $\lambda_j(>0)$ satisfy $\sum_j \lambda_j=1$. The proof of the above theorem using the singular value decomposition can be found in \S$\,2.1.10\,$ and \S$\,2.5\,$ of \cite{NC}. From (\ref{Schmidt}), one can immediately notice that 
the set of non-vanishing eigenvalues of $\boldsymbol{\rho}_B=\mathrm{tr}_E\left[\,|B\cup E\rangle\langle B\cup E|\,\right]$ coincides with that of $\boldsymbol{\rho}_E=\mathrm{tr}_B\left[\,|B\cup E\rangle\langle B\cup E|\,\right]$. 

Now we can reduce the eigenvalue-problem of $\boldsymbol{\rho}_L$ to that of the density matrix for the two ending spins $\boldsymbol{\rho}_{\hat L}$. This density matrix $\boldsymbol{\rho}_{\hat L}$ takes the following form:
\begin{equation}
\boldsymbol{\rho}_{\hat L}=\frac{1}{(n^2-1)^L}\sum_{\stackrel{(l_1,m_1)}{\ne(0,0)}}\cdots \sum_{\stackrel{(l_L,m_L)}{\ne(0,0)}} (U \otimes I)|0,0\rangle_{0,\overline{L+1}}\langle 0,0|(U \otimes I)^\dagger,
\label{edge_RDM}
\end{equation}
where $U=U_{l_1,m_1}\cdots U_{l_L,m_L}$. To evaluate the eigenvalues of $\boldsymbol{\rho}_{\hat L}$, it is convenient to formulate the action of $(U_{l,m}\otimes I)$ as a transfer matrix. Let us first see the action of $(U_{l',m'}\otimes I)$ on a state $|l,m\rangle$:
\begin{equation}
(U_{l',m'}\otimes I)|l,m\rangle
=(X^{l'}Z^{m'}X^{l}Z^{m}\otimes I)|0,0\rangle
=\omega^{m'l}|l+l',m+m'\rangle,
\end{equation}
where we have used the relation $ZX=\omega XZ$ with $\omega=\mathrm{e}^{2\pi\mathrm{i}/n}$.
Using the above relation, we can prove that 
\begin{equation}
(U_{l',m'}\otimes I)|l,m\rangle\langle l,m|(U_{l',m'}\otimes I)^\dagger =|l+l',m+m'\rangle\langle l+l',m+m'|.
\end{equation}
Next, we assign the vector 
$(0, ..., 0,1((l,m)^{\mathrm{th}}$ entry), 0, ..., 0)$^t$ to the state $|l,m\rangle \langle l,m|$. This one to one correspondence plays an essential role in our analysis. 
From this bijection, the operation\\ $\sum_{(l',m')\ne(0,0)}(U_{l',m'}\otimes I)|l,m\rangle\langle l,m|(U_{l',m'}\otimes I)^\dagger$ can be written in terms of $(n^2 \times n^2)$-dimensional matrix as
\begin{equation}
T \equiv
\begin{array}{c}
\longleftarrow \hspace{2mm} n^2 \hspace{2mm} \longrightarrow \\
\left(\begin{array}{ccccc}
0 & 1 & 1 & \cdots & 1 \\
1 & 0 & 1 & \cdots & 1 \\
1 & 1 & 0 & \cdots & 1 \\
\vdots & \vdots & \vdots & \ddots & \vdots \\
1 & 1 & 1 & \cdots & 0 \\
\end{array}\right).
\end{array}
\end{equation}
This transfer matrix can be diagonalized by the following unitary matrix:
\begin{equation}
U_c=\frac{1}{n}\left(
\begin{array}{ccccc}
1 & 1 & 1 & \cdots & 1 \\
1 & \zeta & \zeta^2 & \cdots   & \zeta^{n^2-1} \\
1 & \zeta ^2 & \zeta^4 & \cdots & \zeta^{2(n^2-1)} \\
\vdots & \vdots & \vdots & \ddots & \vdots \\
1 & \zeta^{n^2-1} & \zeta^{2(n^2-1)} & \cdots & \zeta^{(n^2-1)^2}\\
\end{array}\right),
\end{equation}
where $\zeta={\rm exp}(2\pi \mathrm{i}/n^2)$. Then we can obtain the explicit form of the reduced density matrix $\boldsymbol{\rho}_{\hat L}$ as
\begin{eqnarray}
\boldsymbol{\rho}_{\hat L}&=&\frac{1}{(n^2-1)^L}T^L (1, 0, ..., 0)^t\nonumber\\
&=&\frac{1}{(n^2-1)^L} U_c [{\rm diag}(n^2-1,-1, ..., -1)]^L U_c^\dagger (1, 0, ..., 0)^t \nonumber \\
&=& \frac{1}{n^2}(1+(n^2-1)p_n(L))|0,0\rangle_{0\overline{N+1}}\langle 0,0|\nonumber\\
 &&+ \frac{1}{n^2}\sum_{(l,m)\ne(0,0)}(1-p_n(L))|l,m\rangle_{0\overline{N+1}}\langle l,m|,
\label{RDM3}
\end{eqnarray}
where we have used the relation, $1+\zeta^k+\zeta^{2k}+\cdots +\zeta^{(n^2-1)k}=0$ ($1\le k \le n^2-1$) and $p_n(L)=(\frac{-1}{n^2-1})^L$.
Substituting $n=2$ into (\ref{RDM3}), one can reproduce the result of the $SU(2)$ Spin-$1$ VBS state obtained in \S$\,\ref{sec:1dspin1}\,$ (see also \cite{FKR}). 

Let us now start the evaluation of the von Neumann and the R\'enyi entropies of a block of $L$ contiguous spins. 
First, we shall examine the von Neumann entropy of the block. 
From the Schmidt decomposition and the definition of the von Neumann entropy $S_{\mathrm{v\ N}}\left[\boldsymbol{\rho}_{L}\right]=S_{\mathrm{v\ N}}\left[\boldsymbol{\rho}_{\hat L}\right]=-\mathrm{tr}_{1,2,...,L}(\boldsymbol{\rho}_{\hat L}\ln \boldsymbol{\rho}_{\hat L})$, we obtain
\begin{eqnarray}
S_{\mathrm{v\ N}}&=& \ln n^{2} -\frac{1+(n^2-1)p_n(L)}{n^2}\ln (1+(n^2-1)p_n(L)) \nonumber\\
&&-(n^2-1)\frac{1-p_n(L)}{n^2}\ln(1-p_n(L))
\end{eqnarray}
with $p_n(L)=(\frac{-1}{n^2-1})^L$.
Similarly to the $SU(2)$ VBS states \cite{FKR,KHH,XKHK,XKHK2} and the $XY$ spin chains in the gapped regime \cite{IJK,IJK2,FIJK0,FIJK}, 
$S_{\mathrm{v\ N}}[\boldsymbol{\rho}_{L}]$ is bounded by $2 \ln n$ in the limit of large block sizes $L \to \infty$ and approaches to this value exponentially fast in $L$. This is a partial proof of the conjecture proposed in \cite{VLRK}, that the von Neumann entropy of a large block of spins in gapped spin chains shows saturation. 
Next we shall examine the R\'enyi entropy of our system. From the definition of the R\'enyi entropy $S_{\mathrm{R}}(\alpha)=\frac{1}{1-\alpha}\ln \left[\mathrm{tr}(\boldsymbol{\rho}_L^{\alpha})\right]$ ($\alpha \ne 1$ and $\alpha >0$),  
\begin{equation}
S_{\mathrm{R}}(\alpha)=\frac{1}{1-\alpha}\ln(\lambda_{0,0}(L)^\alpha + (n^2-1)\lambda_{l,m \ne 0,0}(L)^\alpha),
\end{equation}
where 
\begin{equation}
\lambda_{l,m}(L) =\left\{
\begin{array}{cc}
\frac{1}{n^2}(1+(n^2-1)p_n(L)), & (l,m)=(0,0)\\ \\
\frac{1}{n^2}(1-p_n(L)) & (l,m)\ne(0,0).
\end{array}\right.
\label{lambda}
\end{equation}

\subsection{The Density Matrix as a Projector}
\label{projectorsun}

We have obtained the spectrum and derived the entropies of the density matrix in \S$\,\ref{sec:su(n)spec}\,$. Our treatment avoided explicit construction of the eigenvectors by reducing the problem to the density matrix of the two ending spins. In this section, we shall show that \textbf{Theorem} \ref{theorem1} in \S$\,\ref{sec:den0spec}\,$ is also valid for our $SU(n)$ version VBS state. \textit{i.e.} The eigenvectors of $\boldsymbol{\rho}_{L}$ with non-zero eigenvalues are degenerate ground states of the block Hamiltonian (The block Hamiltonian is defined similarly as in the $SU(2)$ model, see (\ref{blockhsun}) below). Indeed, the proof of \textbf{Theorem} \ref{theorem1} is completely applicable in our $SU(n)$ case because it only relies on the definition of the density matrix and the fact that the VBS state has no projection on any of the subspaces specified in the Hamiltonian. 
 
\begin{figure}
	\centering
		\includegraphics[width=5in]{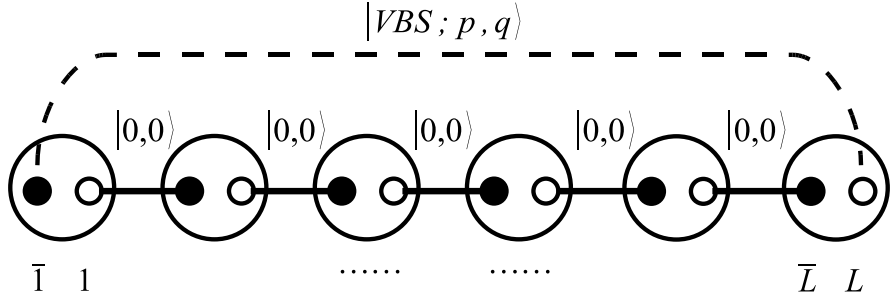}
	\caption{Construction of the degenerate VBS states $|\mathrm{VBS}; p,q\rangle$ for the block. A white (black) dot represents the $SU(n)$ fundamental (conjugate) representation. A large circle denotes the projection onto the adjoint representation. The dashed line corresponds to the state $|\mathrm{VBS}; p,q\rangle$.}
	\label{construction_open}
\end{figure}

The $SU(n)$ block Hamiltonian, defined along the same line as in the $SU(2)$ model, is the sum of interaction terms within the block:
\begin{eqnarray}
H_{B}=\sum_{k=1}^{L-1} H(k,k+1). \label{blockhsun}
\end{eqnarray}
The number of degrees of freedom in the block subsystem can be counted from the von Neumann entropy as $n^2$. This number coincides with the number of degenerate ground states of the block Hamiltonian $H_{B}$. \textit{i.e.}
\begin{eqnarray}
	D=deg=n^{2},
\end{eqnarray}
where $D$ is the number of non-zero eigenvalues (dimension of the \textit{support}) of the density matrix and $deg$ is the number of degenerate ground states of the block Hamiltonian. A basis of the degenerate ground states can be constructed as follows:
\begin{eqnarray}
|\mathrm{VBS}; p,q\rangle &\equiv& C_{p,q}\sum_{\stackrel{(l_1,m_1)}{\ne(0,0)}}\cdots \sum_{\stackrel{(l_L,m_L)}{\ne(0,0)}} |l_1,-m_1\rangle_{\bar 11}\cdots|l_{L-1},-m_{L-1}\rangle_{\overline{L-1}L-1} \nonumber \\
&&\cdot \mathbf{P}_{L \bar L}\left((U_{p,q}U_{l_1,m_1}...U_{l_{L-1},m_{L-1}} \otimes I)|0,0\rangle_{L \bar L}\right),
\label{Edgestate}
\end{eqnarray}
where $C_{p,q}$ is a normalization factor and $\mathbf{P}_{L \bar L}$ is a projector onto the adjoint representation of $SU(n)$. This set of states (\ref{Edgestate}) can be called the \textit{degenerate VBS states}. Any linear combination of (\ref{Edgestate}) is apparently a ground state of $H_{B}$ (\ref{blockhsun}). 
The graphical representation of the construction of this state is shown in Figure \ref{construction_open}. 
The following orthogonality relation of these generate VBS states (\ref{Edgestate}) holds:
\begin{equation}
	\langle \mathrm{VBS}; p,q| \mathrm{VBS}; r,s \rangle =C_{p,q}^2 (n^2-1)^{L}\delta_{p,r}\delta_{q,s}\lambda_{-p,-q}(L) 
\end{equation}
where the subscripts $-p$ and $-q$ are modulo $n$. This can be shown as follows:
\begin{eqnarray}
&&\langle \mathrm{VBS}; p,q| \mathrm{VBS}; r,s \rangle\nonumber\\ 
&=& C_{p,q} C_{r,s} \sum_{\stackrel{(l_1,m_1)}{\ne(0,0)}}\cdots\sum_{\stackrel{(l_{L-1},m_{L-1})}{\ne(0,0)}} \nonumber \\
&& {}_{L \bar L}\langle 0,0| (U_{p,q}U_{l_1,m_1}...U_{l_{L-1},m_{L-1}})^\dagger \mathbf{P}_{L\bar L}(U_{r,s}U_{l_1,m_1}...U_{l_{L-1},m_{L-1}} \otimes I)|0,0\rangle_{L \bar L} 
\nonumber \\
&=& C_{p,q} C_{r,s}\sum_{\stackrel{(l,m)}{\ne(0,0)}} \sum_{\stackrel{(l_1,m_1)}{\ne(0,0)}} \cdots \sum_{\stackrel{(l_{L-1},m_{L-1})}{\ne (0,0)}}\nonumber\\
&& {}_{L \bar L}\langle 0,0| (U_{l_1,m_1}...U_{l_{L-1},m_{L-1}}\otimes I)^\dagger (U_{p,q}\otimes I)^\dagger |l,m\rangle_{L\bar L}
\nonumber \\
&&\cdot {}_{L \bar L}\langle l,m|
(U_{r,s}\otimes I)(U_{l_1,m_1}...U_{l_{L-1},m_{L-1}} \otimes I)|0,0\rangle_{L \bar L} 
\nonumber \\
&=& C_{p,q}C_{r,s}(n^2-1)^{L-1}\nonumber\\
&&\cdot\sum_{(l',m')} \lambda_{l',m'}(L-1) {}_{L \bar L}\langle p+l',q+m'|(1-|0,0\rangle_{L \bar L}\langle 0,0|)|r+l',s+m'\rangle_{L \bar L}
\nonumber \\
&=& C_{p,q}^2 (n^2-1)^{L}\delta_{p,r}\delta_{q,s}\lambda_{-p,-q}(L).
\end{eqnarray}
Here we have recalled (\ref{edge_RDM}) and used the relation
\begin{eqnarray}
&& \sum_{\stackrel{(l_1,m_1)}{\ne(0,0)}} \cdots \sum_{\stackrel{(l_{L-1},m_{L-1})}{\ne (0,0)}}\nonumber\\
&&(U_{l_1,m_1}...U_{l_{L-1},m_{L-1}} \otimes I)|0,0\rangle_{L \bar L}\langle 0,0|  (U_{l_1,m_1}...U_{l_{L-1},m_{L-1}}\otimes I)^{\dagger}\nonumber \\
&=&(n^2-1)^{L-1}\boldsymbol{\rho}_{\widehat{L-1}}=(n^2-1)^{L-1}\sum_{(l,m)}\lambda_{l,m}(L-1)|l,m\rangle_{L\bar L}\langle l,m|.
\end{eqnarray}
The explicit form of the normalization factors $C_{p,q}$ are given by
$C_{p,q}= 1/\sqrt{(n^2-1)^{L}\lambda_{-p,-q}(L)}$.
Now, we could write $\boldsymbol{\rho}_L$ in terms of this basis of degenerate VBS states. 
By the original definition, 
\begin{eqnarray} 
\boldsymbol{\rho}_L &=& \mathrm{tr}_{0,\overline{L+1}}\left[\,|\mathrm{VBS}\rangle \langle \mathrm{VBS}|\,\right] \nonumber \\
&=& \frac{1}{(n^2-1)^L}\sum_{(p,q)}\sum_{\stackrel{(l_1,m_1)}{\ne(0,0)}}\sum_{\stackrel{(l'_1,m'_1)}{\ne(0,0)}}\cdots \sum_{\stackrel{(l_{L-1},m_{L-1})}{\ne (0,0)}}\sum_{\stackrel{(l'_{L-1},m'_{L-1})}{\ne (0,0)}}
\nonumber \\
&& |l_1,-m_1\rangle \langle l'_1,-m'_1|
\cdots |l_{L-1},-m_{L-1}\rangle \langle l'_{L-1},-m'_{L-1}|
\nonumber \\
&&\cdot \mathbf{P}_{L\bar L}(U_{p,q}U_{l_1,m_1}...U_{l_{L-1},m_{L-1}} \otimes I)|0,0\rangle_{L \bar L} \langle 0,0|(U_{p,q}U_{l'_1,m'_1}...U_{l'_{L-1},m'_{L-1}} \otimes I)^\dagger \mathbf{P}_{L \bar L}.\nonumber\\
\end{eqnarray}
Then by comparing with (\ref{Edgestate}), we obtain 
\begin{equation}
\boldsymbol{\rho}_L=\sum_{(p,q)} \lambda_{-p,-q}(L) |\mathrm{VBS}; p,q\rangle \langle \mathrm{VBS}; p,q|,
\end{equation}
where $\lambda_{-p,-q}(=\lambda_{n-p,n-q})$  was defined in (\ref{lambda}). 
Therefore we conclude that the density matrix $\boldsymbol{\rho}_{L}$ of a block of $L$ contiguous spins in the $SU(n)$ VBS state is completely characterized by the degenerate ground states $\{\,|\mathrm{VBS}; p,q \rangle\,\}$ of the block Hamiltonian $H_{B}$ (\ref{blockhsun}).

\subsection{The Large Block Limit}
\label{sec:limitsun}

Now we consider the large block limit, \textit{i.e.} $L \to \infty$. In this case, $p_{n}(L)\to 0$ and $\lambda_{l,m}$ (\ref{lambda}) become degenerate. So that great simplification occurs in the expressions of entropies as $L\to\infty$:
\begin{eqnarray}
S_{\mathrm{v\ N}}&=& \ln n^{2}, \nonumber\\
S_{\mathrm{R}}(\alpha)&=&\frac{1}{1-\alpha}\ln \left( \left(\frac{1}{n^2}\right)^\alpha+(n^2-1)\left(\frac{1}{n^2}\right)^\alpha \right) \nonumber \\
&=& \frac{1}{1-\alpha}\ln(n^2)^{1-\alpha} \nonumber \\
&=& \ln n^{2}. \label{renyisun}
\end{eqnarray}
We notice that the R\'enyi entropy is independent of $\alpha$ and coincides with the von Neumann entropy in the large block limit. The same saturation behavior was observed in all our $SU(2)$ cases in \S$\,\ref{sec:1ds1limit}\,$, \S$\,\ref{sec:larges}\,$, \S$\,\ref{sec:largeinhom}\,$. This means that the density matrix of a large block is proportional to a $n^2$-dimensional identity matrix. 
In other words, a sufficiently large block of neighboring spins in our $SU(n)$ VBS ground state is maximally entangled with the rest of the chain.

In the limit of large block sizes, \textit{i.e.} $L \to \infty$, $\boldsymbol{\rho}_{L}$ can be written as
\begin{equation}
\boldsymbol{\rho}_{L}\to\boldsymbol{\rho}_{\infty} = \frac{1}{n^2}\sum_{(p,q)}|\mathrm{VBS}; p,q\rangle \langle \mathrm{VBS}; p,q|=\frac{1}{n^{2}}\,\mathbf{P}_{n^{2}}, \qquad L\to\infty. \label{limitrhosun}
\end{equation}
The limiting density matrix is proportional to a projector $\mathbf{P}_{n^{2}}$ which projects on the $n^{2}$-dimensional subspace spanned by the degenerate ground states of the block Hamiltonian. This structure of the limiting density matrix is also a generalization of the corresponding results for $SU(2)$ density matrices in \S$\,\ref{sec:1ds1limit}\,$, \S$\,\ref{sec:larges}\,$, \S$\,\ref{sec:largeinhom}\,$.

\section{Summary}
\label{sec:sum1}

We have studied the entanglement in the VBS ground state of the AKLT model in this review. The AKLT model is formulated on an arbitrary connected graph or a lattice. The Hamiltonian (\ref{basich}), (\ref{generalizedh}) is a sum of projectors which describe interactions between nearest neighbors. The condition of uniqueness of the ground state relates the spin value at each vertex (site) with multiplicity numbers associated with edges incident to the vertex (bonds connected to the site), see (\ref{basiccond}), (\ref{condition1}), (\ref{condition2}). The unique ground state is known as the Valence-Bond-Solid state (\ref{basicvbs}), (\ref{generalizedvbs}).

To study the entanglement, the graph (lattice) is divided into two parts: the \textit{block} and the \textit{environment}. We investigate the density matrix $\boldsymbol{\rho}_{B}$ of the block and show that it has many zero eigenvalues. We describe the subspace (called the \textit{support}) of eigenvectors of $\boldsymbol{\rho}_{B}$ with non-zero eigenvalues. It has been proved (see \textbf{Theorem} \ref{theorem1} in \S$\,\ref{sec:den0spec}\,$) that this subspace is the degenerate \textit{ground space} of some Hamiltonian which is called the \textit{block Hamiltonian} $H_{B}$ (\ref{subsystemh}). The block Hamiltonian is a part of the original AKLT Hamiltonian describing interactions of spins inside of the block.

The entanglement can be measured by the von Neumann entropy or the R\'enyi entropy of the density matrix $\boldsymbol{\rho}_{B}$. Most eigenvalues of $\boldsymbol{\rho}_{B}$ vanish and have no contribution to the entanglement entropies. The density matrix takes the form of a projector on the \textit{ground space} of $H_{B}$ multiplied by another matrix (see also \cite{XK}).

\begin{figure}
	\centering
		\includegraphics[width=3in]{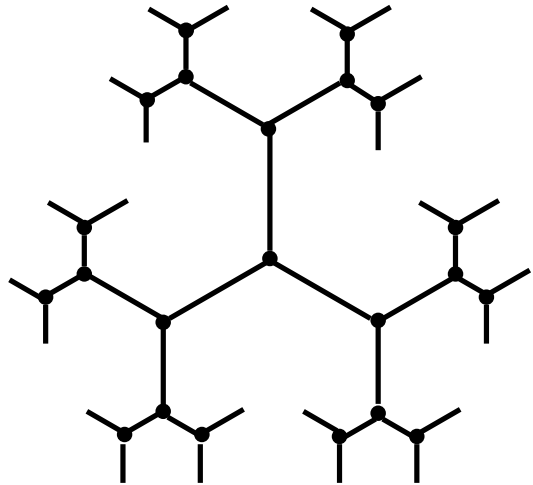}
	\caption{A $2$-dimensional Cayley tree. Each dot represents a spin-$3S/2$ in the bulk and a spin-$S/2$ on the boundary. Each solid line represents the bond connecting a pair of interacting spins. This tree structure has no loop.}
	\label{fig:Cayley}
\end{figure}

A complete analysis of the density matrix for a variety of $1$-dimensional AKLT models (including the $SU(n)$ generalization) has been presented. The block density matrix $\boldsymbol{\rho}_{L}$ for a subsystem of $L$ contiguous bulk spins has been diagonalized with non-zero eigenvalues calculated (see also \cite{FKR,KHH,KHK,XKHK,XKHK2}). (The general notation of the density matrix $\boldsymbol{\rho}_{B}$ is changed to $\boldsymbol{\rho}_{L}$ for these $1$-dimensional models as emphasizing the dependence on the size $L$ of the block.) We find that in all these cases the \textit{support} coincides with the \textit{ground space}, so their dimensions are equal $D=deg$. In the large block limit $L \to \infty$\footnote{As $L\to\infty$, the size of the whole spin chain also goes to infinity.}, all non-zero eigenvalues become the same and the density matrix is proportional to a projector (\ref{1dsirholimit}), (\ref{largedens}), (\ref{limitrho00}), (\ref{limitrhosun}). The von Neumann entropy equals the R\'enyi entropy and both take the saturated value $S_{\mathrm{v\ N}}=S_{\mathrm{R}}=\ln D=\ln (deg)$.

Moreover, it turns out that the block Hamiltonian $H_{B}$ defines the density matrix $\boldsymbol{\rho}_{L}$ completely in the large block limit $L\to\infty$. The zero-energy ground states of the block Hamiltonian $H_{B}$ span the subspace that the density matrix $\boldsymbol{\rho}_{L}$ projects onto. So that $\boldsymbol{\rho}_{L}$ can be represented as the zero-temperature limit of the canonical ensemble density matrix defined by $H_{B}$:
\begin{eqnarray}
	\boldsymbol{\rho}_{L}=\lim_{\beta\to +\infty}\left(\frac{\mathrm{e}^{-\beta H_{B}}}{\mathrm{tr}\left[\ \mathrm{e}^{-\beta H_{B}}\ \right]}\right), \qquad L\to\infty. \label{enlim}
\end{eqnarray}
In the zero-temperature limit, contributions from excited states of $H_{B}$ all vanish and the right hand side of (\ref{enlim}) turns into a projector onto the ground states of the block Hamiltonian.

For more complicated graphs or lattices, non-zero eigenvalues of the density matrix are still unknown. One open problem is to calculate those eigenvalues. One may start with the Cayley tree (also known as the Bethe tree), see Figure \ref{fig:Cayley}.
The picture shows a Cayley tree with each bulk vertex connected to three edges. The uniqueness condition (\ref{condition1}) requires that we shall have spin-$3S/2$ in the bulk and spin-$S/2$ on the boundary. A symmetric block subsystem consists of spins enclosed by a circle centered at the center of the tree. The degeneracy of ground states of the block Hamiltonian is $(S+1)^{N_{\partial B}}$, where $N_{\partial B}$ is the number of sites on the boundary of the block. An exact explicit expression for the non-zero eigenvalues (finite block) is expected because there is no loop. It is also interesting to study the large block limit. In all known examples \cite{FKR,KHK,XKHK,XKHK2} where the density matrix of a large block has been calculated, all non-zero eigenvalues approach the same value $1/D=1/deg$. So that the entanglement entropies are saturated, \textit{i.e.} $S_{\mathrm{v\ N}}=S_{\mathrm{R}}=\ln{D}=\ln{(deg)}$. Therefore the density matrix of a large block is proportional to a projector on the ground space of the block Hamiltonian, \textit{i.e.} $\boldsymbol{\rho}_{B}=\frac{1}{D}\,\mathbf{P}_{D}=\frac{1}{deg}\,\mathbf{P}_{deg}$. However, this might not be the case for the Cayley tree. According to the area law, we expect that in the large block limit (a circular block centered at the center of the tree), the entropy be proportional to the length of the boundary with some coefficient $\alpha$, \textit{i.e.} $S_{\mathrm{v\ N}}=\alpha N_{\partial B}$. It is interesting to calculate $\alpha$ and we expect that it will be smaller than $\ln(S+1)$. Another open problem is the generalization of the VBS state to other Lie algebras beyond $SU(n)$ and the study of the entanglement.

It is also important to calculate non-zero eigenvalues of $\boldsymbol{\rho}_{B}$ for graphs with loops. 
\begin{figure}
	\centering
		\includegraphics[width=3.6in]{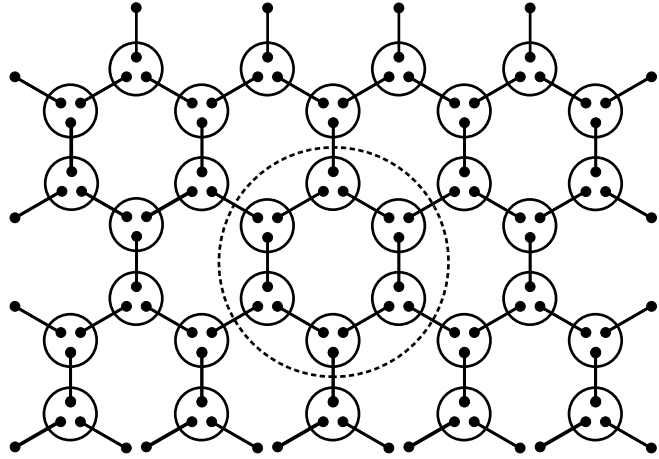}
	\caption{The basic model on a $2$-dimensional hexagonal lattice. Each spin-$3/2$ in the bulk is represented by three small dots representing spin-$1/2$'s enclosed by a solid circle (a symmetrization). The solid lines antisymmetrize the connected two spin-$1/2$ states. The block is given by a dashed large circle. Each line cut by the dashed circle results in a free spin-$1/2$ on the boundary of the block.}
	\label{fig:Hexagonal}
\end{figure}
For example, consider a basic AKLT model defined on the $2$-dimensional hexagonal lattice, see Figure \ref{fig:Hexagonal}.
The basic model has spin-$3/2$ in the bulk and the block is a large circle. Note that each spin on the boundary of the block contributes effectively a free spin-$1/2$. Then according to Kastura's formula (\ref{kastura}), the degeneracy of the ground states of the block Hamiltonian is $2^{N_{\partial B}}$, where $N_{\partial B}$ denotes the number of site (spins) on the boundary of the block. There is at most $2^{N_{\partial B}}$ number of non-zero eigenvalues of the density matrix. The entropy takes a saturated value if all these eigenvalues are equal in the large block limit. Similar to the case of the Cayley tree, in the large block limit the entropy should be proportional to the size (area) of the boundary, \textit{i.e.} $S_{\mathrm{v\ N}}=\alpha N_{\partial B}$. We expect that the coefficient $\alpha$ be greater than $0$ but smaller than $\ln2$. The value $N_{\partial B}\ln2$ is an upper bound of the von Neumann entropy because this is the logarithm of the dimension of the Hilbert space (number of non-zero eigenvalues of the density matrix).




\section*{Acknowledgements}

We want  to thank Dr. Hosho Katsura for fruitful discussions. One of the authors (VK) has productive discussions during his visit to Erwin Schrodinger Institute in Vienna.  The project was supported by NSF Grant DMS-0503712.



\begin{thebibliography}{00}



\bibitem{AKLT1}
A. Affleck, T. Kennedy, E. H. Lieb and H. Tasaki, Rigorous Results on Valence-Bond Ground States in Antiferromagnets, \textit{Phys. Rev. Lett.} \textbf{59}, 799-802 (1987).

\bibitem{AKLT2}
A. Affleck, T. Kennedy, E. H. Lieb and H. Tasaki, Valence Bond Ground States in Isotropic Quantum Antiferromagnets, \textit{Commun. Math. Phys.} \textbf{115}, 477-528 (1988).

\bibitem{AFOV}
L. Amico, R. Fazio, A. Osterloh and V. Vedral, Entanglement in many-body systems, \textit{Rev. Mod. Phys.} \textbf{80}, 517-576 (2008). \textit{Preprint} arXiv:quant-ph/0703044v3.

\bibitem{ABV}
M. C. Arnesen, S. Bose and V. Vedral, Natural Thermal and Magnetic Entanglement in $1$D Heisenberg Model, \textit{Phys. Rev. Lett.} \textbf{87}, 017901 (2001). \textit{Preprint} arXiv:quant-ph/0009060v2.

\bibitem{AAH}
D. P. Arovas, A. Auerbach and F. D. M. Haldane, Extended Heisenberg models of antiferromagnetism: Analogies to the fractional quantum Hall effect, \textit{Phys. Rev. Lett.} \textbf{60}, 531-534 (1988).

\bibitem{A}
A. Auerbach, \textit{Interacting Electrons and Quantum Magnetism} (Springer,New York, 1998).

\bibitem{BBPS}
C. H. Bennett, H. J. Bernstein, S. Popescu and B. Schumacher, Concentrating Partial Entanglement by Local Operations, \textit{Phys. Rev.} A\,\textbf{53}, 2046-2052 (1996). \textit{Preprint} arXiv:quant-ph/9511030.

\bibitem{BD}
C. H. Bennett and D. P. DiVincenzo, Quantum information and computation, \textit{Nature} \textbf{404}, 247-255 (2000). 

\bibitem{BH}
S. Bose and D. Home, Dualism in Entanglement and Testing Quantum to Classical Transition of Identicity (2005), \textit{Preprint} arXiv:quant-ph/0505217v1.

\bibitem{BM}
G. K. Brennen and A. Miyake, Measurement-based quantum computer in the gapped ground state of a two-body Hamiltonian, \textit{Phys. Rev. Lett.} \textbf{101}, 010502 (2008). \textit{Preprint} arXiv:0803.1478.

\bibitem{CBO}
A. T. Costa Jr., S. Bose and Y. Omar, Entanglement of Two Impurities through Electron Scattering, \textit{Phys. Rev. Lett.} \textbf{96}, 230501 (2006). \textit{Preprint} arXiv:quant-ph/0503183v1.

\bibitem{CEP}
M. Cramer, J. Eisert and M. B. Plenio, Statistical Dependence of Entanglement-Area Laws, \textit{Phys. Rev. Lett.} \textbf{98}, 220603 (2007). \textit{Preprint} arXiv:quant-ph/0611264.

\bibitem{CEPD}
M. Cramer, J. Eisert, M. B. Plenio and J. Dreisig, On an entanglement-area-law for general bosonic harmonic lattice systems, \textit{Phys. Rev.} A\,\textbf{73}, 012309 (2006). \textit{Preprint} arXiv:quant-ph/0505092.

\bibitem{D}
P. A. M. Dirac, Note on Exchange Phenomena in the Thomas Atom, \textit{Proc. Cambr. Phil. Soc.} \textbf{26}, 376-385 (1930). 

\bibitem{EC}
J. Eisert and M. Cramer, Single-copy entanglement in critical spin chains, \textit{Phys. Rev.} A\,\textbf{72}, 042112 (2005).  \textit{Preprint} arXiv:quant-ph/0506250v3.

\bibitem{F}
H. Fan, Distinguishability and indistinguishability by LOCC, \textit{Phys. Rev. Lett.} \textbf{92}, 177905 (2004). \textit{Preprint} arXiv:quant-ph/0311026v1.

\bibitem{FK}
H. Fan and V. E. Korepin, Quantum Entanglement of the Higher Spin Matrix Product States (2008) (in preparation).

\bibitem{FKR}
H. Fan, V. E. Korepin and V. Roychowdhury, Entanglement in a Valence-Bond-Solid State, \textit{Phys. Rev. Lett.} \textbf{93}, 227203 (2004). \textit{Preprint} arXiv:quant-ph/0406067.

\bibitem{FKR2}
H. Fan, V. E. Korepin and V. Roychowdhury, Valence-Bond-Solid state entanglement in a $2$-D Cayley tree (2005), \textit{Preprint} arXiv:quant-ph/0511150.

\bibitem{FKRHB}
H. Fan, V. E. Korepin, V. Roychowdhury, C. Hadley and S. Bose, Boundary effects to the entanglement entropy and two-site entanglement of the spin-$1$ valence-bond solid, \textit{Phys. Rev.} B\,\textbf{76}, 014428 (2007). \textit{Preprint} arXiv:quant-ph/0605133.

\bibitem{FS}
H. Fan and L. Lloyd, Entanglement of eta-pairing state with off-diagonal long-range order, \textit{J. Phys.} A\,\textbf{38}, 5285 (2005). \textit{Preprint} arXiv:quant-ph/0405130.






\bibitem{FIJK0}
F. Franchini, A. R. Its, B.-Q. Jin and V. E. Korepin, Analysis of entropy of XY Spin Chain, \textit{Proceedings of the Third Feynman Workshop} (2006). \textit{Preprint} arXiv:quant-ph/0606240v1.

\bibitem{FIJK}
F. Franchini, A. R. Its, B.-Q. Jin and V. E. Korepin, Ellipses of Constant Entropy in the $XY$ Spin Chain, \textit{J. Phys.} A\,\textbf{40}, 8467 (2007). \textit{Preprint} arXiv:quant-ph/0609098v5.

\bibitem{FIK}
F. Franchini, A. R. Its and V. E. Korepin, R\'enyi Entropy of the $XY$ Spin Chain, \textit{J. Phys.} A\,\textbf{41}, 025302 (2008). \textit{Preprint} arXiv:0707.2534.

\bibitem{FM}
W.-D. Freitag and E. Muller-Hartmann, Complete analysis of two spin correlations of valence bond solid chains for all integer spins, \textit{Z. Phys.} B\,\textbf{83}, 381 (1991).

\bibitem{GMC}
J. J. Garcia-Ripoll, M. A. Martin-Delgado and J. I. Cirac, Implementation of Spin Hamiltonians in Optical Lattices, \textit{Phys. Rev. Lett.} \textbf{93}, 250405 (2004). \textit{Preprint} arXiv:cond-mat/0404566.

\bibitem{GRAC}
S. Ghosh, T. F. Rosenbaum, G. Aeppli and S. N. Coppersmith, Entangled quantum state of magnetic dipoles, \textit{Nature} \textbf{425}, 48-51 (2003); V. Vedral, Entanglement hits the big time, \textit{Nature} \textbf{425}, 28-29 (2003).

\bibitem{GR}
M. Greiter and S. Rachel, Valence bond solids for $SU(n)$ spin chains: exact models, spinon confinement, and the Haldane gap, \textit{Phys. Rev.} B\,\textbf{75}, 184441 (2007). \textit{Preprint} arXiv:cond-mat/0702443v2.

\bibitem{GRS}
M. Greiter, S. Rachel and D. Schuricht, Exact results for $SU(3)$ spin chains: trimer states, valence bond solids, and their parent Hamiltonians, \textit{Phys. Rev.} B\,\textbf{75}, 060401(R) (2007). \textit{Preprint} arXiv:cond-mat/0701354v1.

\bibitem{GESG}
D. Gross, J. Eisert, N. Schuch and D. Perez-Garcia, Measurement-based quantum computation beyond the one-way model, \textit{Phys. Rev.} A\,\textbf{76}, 052315 (2007), \textit{Preprint} arXiv:0706.3401v1.

\bibitem{GDLL}
S.-J. Gu, S.-S. Deng, Y.-Q. Li and H.-Q. Lin, Entanglement and Quantum Phase Transition in the Extended Hubbard Model, \textit{Phys. Rev. Lett.} \textbf{93}, 086402 (2004). \textit{Preprint} arXiv:quant-ph/0405067v1.

\bibitem{GT}
O. G\"uhne and G. T\'oth, Entanglement detection, \textit{Phys. Rep.} \textbf{474}, 1 (2009). \textit{Preprint} arXiv:0811.2803v3.

\bibitem{Had}
C. Hadley, Single copy entanglement in a gapped quantum spin chain, \textit{Phys. Rev. Lett.} \textbf{100}, 170001 (2008). \textit{Preprint} arXiv:0801.3681.

\bibitem{HKAHB}
M. Hagiwara, K. Katsumata, I. Affleck, B. I. Halperin and J. P. Renard,
Observation of S=$1/2$ degrees of freedom in an S=$1$ linear-chain Heisenberg antiferromagnet, 
\textit{Phys. Rev. Lett.} \textbf{65}, 3181 (1990).

\bibitem{H}
F. D. M. Haldane, Continuum dynamics of the 1-D Heisenberg antiferromagnet: Identification with the $O(3)$ nonlinear sigma model, \textit{Phys. Lett.} \textbf{93A}, 464-468 (1983); Nonlinear Field Theory of Large-Spin Heisenberg Antiferromagnets: Semiclassically Quantized Solitons of the One-Dimensional Easy-Axis N\'eel State, \textit{Phys. Rev. Lett.} \textbf{50}, 1153-1156 (1983).

\bibitem{HR0}
F. D. M. Haldane and E. H. Rezayi, Finite-Size Studies of the Incompressible State of the Fractionally Quantized Hall Effect and its Excitations, \textit{Phys. Rev. Lett.} \textbf{54}, 237 (1985).

\bibitem{Ha}
M. Hamermesh, \textit{Group Theory and Its Application to Physical Problems} (Dover Publications, New York, 1989).

\bibitem{HZS}
M. Haque, O. Zozulya and K. Schoutens, Entanglement Entropy in Fermionic Laughlin States, \textit{Phys. Rev. Lett.} \textbf{98}, 060401 (2007). \textit{Preprint}	arXiv:cond-mat/0609263v2.

\bibitem{Har}
F. Harary, \textit{Graph Theory} (Addison-Weslay Publ. Comp. Reading, Massachusets, 1969).

\bibitem{Has}
M. B. Hastings, An Area Law for One Dimensional Quantum Systems, \textit{J. Stat. Mech.} P08024 (2007).	\textit{Preprint} arXiv:0705.2024v3.

\bibitem{HJPW}
P. Hayden, R. Jozsa, D. Petz and A. Winter, Structure of states which satisfy strong subadditivity of quantum entropy with equality, \textit{Commun. Math. Phys.} \textbf{246}(2), 359-374 (2004). \textit{Preprint} arXiv:quant-ph/0304007v2.

\bibitem{HLW}
C. Holzhey, F. Larsen and F. Wilczek, Geometric and Renormalized Entropy in Conformal Field Theory, \textit{Nucl. Phys.} B\,\textbf{424}, 443 (1994). \textit{Preprint} arXiv:hep-th/9403108v1.

\bibitem{ILO}
S. Iblisdir, J. I. Latorre and R. Orus, Entropy and Exact Matrix Product Representation of the Laughlin Wave Function, \textit{Phys.Rev.Lett.} \textbf{98}, 060402 (2007). \textit{Preprint} cond-mat/0609088.

\bibitem{IJK}
A. R. Its, B.-Q. Jin and V. E. Korepin, Entanglement in $XY$ Spin Chain, \textit{J. Phys.} A\,\textbf{38}, 2975 (2005). \textit{Preprint} arXiv:quant-ph/0409027.

\bibitem{IJK2}
A. R. Its, B.-Q. Jin and V. E. Korepin, Entropy of $XY$ Spin Chain and Block Toeplitz Determinants, \textit{Fields Institute Communications, Universality and Renormalization} [editors I. Bender and D. Kreimer] \textbf{50}, 151 (2007). \textit{Preprint} arXiv:quant-ph/0606178v3.

\bibitem{JK}
B.-Q. Jin and V. E. Korepin, Quantum Spin Chain, Toeplitz Determinants and Fisher-Hartwig Conjecture, \textit{J. Stat. Phys.} \textbf{116}, 79 (2004). \textit{Preprint} arXiv:quant-ph/0304108.

\bibitem{JKLPPSW}
R. Jozsa, M. Koashi, N. Linden, S. Popescu, S. Presnell, D. Shepherd and A. Winter, Entanglement Cost of Generalised Measurements (2003), \textit{Preprint} arXiv:quant-ph/0303167v1.

\bibitem{KHH}
H. Katsura, T. Hirano and Y. Hatsugai, Exact Analysis of Entanglement in Gapped Quantum Spin Chains, \textit{Phys. Rev.} B\,\textbf{76}, 012401 (2007). \textit{Preprint} arXiv:cond-mat/0702196.

\bibitem{KHK}
H. Katsura, T. Hirano and V. E. Korepin, Entanglement in an $SU(n)$ Valence-Bond-Solid State, \textit{J. Phys.} A\,\textbf{41}, 135304 (2008). \textit{Preprint} arXiv:0711.3882v2.

\bibitem{KM}
J. P. Keating and F. Mezzadri, Random Matrix Theory and Entanglement in Quantum Spin Chains, \textit{Commun. Math. Phys.} \textbf{252}, 543-579 (2004). \textit{Preprint} arXiv:quant-ph/0407047.

\bibitem{KLT}
T. Kennedy, E. H. Lieb and H. Tasaki, A two-dimensional isotropic quantum antiferromagnet with unique disordered ground state, \textit{J. Stat. Phys.} \textbf{53}, 383-415 (1988).

\bibitem{KK}
A. N. Kirillov and V. E. Korepin, Correlation Functions in Valence Bond Solid Ground State, \textit{Sankt Petersburg Mathematical Journal} \textbf{1}, 47 (1990); \textit{Algebra and Analysis} \textbf{1}, 47 (1989); \textit{Preprint}  http://insti.physics.sunysb.edu/\~korepin/kir.pdf \footnote{If your browser displays an error message for this link, you can get
the PDF file using Google search for `The valence bond solid in quasicrystals'.}

\bibitem{KSZ1}
A. Kl\"umper, A. Schadschneider and J. Zittartz, Equivalence and Solution of Anisotropic Spin-$1$ Models and Generalized $t$-$J$ Fermion Models in one Dimension, \textit{J. Phys.} A\,\textbf{24}, L955-L959 (1991).

\bibitem{KSZ}
A. Kl\"umper, A. Schadschneider and J. Zittartz, Groundstate Properties of a Generalized VBS-Model, \textit{Z. Phys.} B\,\textbf{87}, 281-287 (1992).

\bibitem{K}
V. E. Korepin, Universality of Entropy Scaling in One Dimensional Gapless Models, \textit{Phys. Rev. Lett.} \textbf{92}, 096402 (2004). \textit{Preprint} arXiv:cond-mat/0311056.

\bibitem{LORV}
J. I. Latorre, R. Orus, E. Rico and J. Vidal, Entanglement entropy in the Lipkin-Meshkov-Glick model, \textit{Phys. Rev.} A\,\textbf{71}, 064101 (2005). \textit{Preprint} cond-mat/0409611.

\bibitem{LRV}
J. I. Latorre, E. Rico and G. Vidal, Ground state entanglement in quantum spin chains, \textit{Quant. Inf. Comput.} \textbf{4}, 48-92 (2004). \textit{Preprint} arXiv:quant-ph/0304098v4.

\bibitem{L}
S. Lloyd, A potentially realizable quantum computer, \textit{Science} \textbf{261}, 1569-1671 (1993); Envisioning a Quantum Supercomputer, \textit{ibid} \textbf{263}, 695 (1994).


\bibitem{NC}
M. A. Nielsen and I. L. Chuang, \textit{Quantum Computation and Quantum Information} (Cambridge Univ. Press, Cambridge, 2000).

\bibitem{NO}
Z. Nussinov and G. Ortiz, A symmetry principle for Topological Quantum Order (2007), \textit{Preprint} arXiv:cond-mat/0702377v5.

\bibitem{O}
R. Orus, Geometric Entanglement in a One-Dimensional Valence Bond Solid State, \textit{Phys. Rev.} A\,\textbf{78}, 062332 (2008). \textit{Preprint} arXiv:0808.0938v2.

\bibitem{OL}
R. Orus and J. I. Latorre, Universality of entanglement and quantum-computation complexity, \textit{Phys. Rev.} A\,\textbf{69}, 052308 (2004). \textit{Preprint} arXiv:quant-ph/0311017v3.

\bibitem{OAFF}
A. Osterloh, L. Amico, G. Falci and R. Fazio, Scaling of entanglement close to quantum phase transitions, \textit{Nature} \textbf{416}, 608 (2002). \textit{Preprint} arXiv:quant-ph/0202029v2.

\bibitem{ON}
T. J. Osborne and M. A. Nielsen, Entanglement in a simple quantum phase transition, \textit{Phys. Rev.} A\,\textbf{66}, 032110 (2002). \textit{Preprint} arXiv:quant-ph/0202162v1.

\bibitem{P}
J. K. Pachos, Three-spin interactions and entanglement in optical lattices, \textit{International Journal of Quantum Information} \textbf{4}, 541 (2006). \textit{Preprint} arXiv:quant-ph/0505225v1.

\bibitem{PP}
J. K. Pachos and M. B. Plenio, Three-Spin Interactions in Optical Lattices and Criticality in Cluster Hamiltonians, \textit{Phys. Rev. Lett.} \textbf{93}, 056402 (2004). \textit{Preprint}	arXiv:quant-ph/0401106v2.

\bibitem{PS}
V. Popkov and M. Salerno, Logarithmic divergence of the block entanglement entropy for the ferromagnetic Heisenberg model, \textit{Phys. Rev.} A\,\textbf{71}, 012301 (2005). \textit{Preprint} arXiv:quant-ph/0404026.

\bibitem{RSSTG}
S. Rachel, D. Schuricht, B. Scharfenberger, R. Thomale and M. Greiter, Spontaneous Parity Violation in a Quantum Spin Chain, \textit{J. Phys.: Conf. Ser.} \textbf{200}, 022049 (2010). \textit{Preprint} arXiv:0905.4895v1.

\bibitem{RTFSG}
S. Rachel, R. Thomale, M. Fuehringer, P. Schmitteckert and M. Greiter, Spinon confinement and the Haldane gap in $SU(n)$ spin chains, \textit{Phys. Rev.} B\,\textbf{80}, 180420(R) (2009). \textit{Preprint} arXiv:0904.3882v1.

\bibitem{RB1}
R. Raussendorf and H. Briegel, A one-way quantum computer, \textit{Phys. Rev. Lett.} \textbf{86}, 5188 (2001). \textit{Preprint} arXiv:quant-ph/0010033v1; \textit{Method for quantum computing}, U.S. Patent 7277872 (2007).

\bibitem{RB}
E. Rico and H. J. Briegel, $2$D Multipartite Valence Bond States in Quantum Antiferromagnets, \textit{Annals. Phys.} \textbf{323}, 2115-2131 (2008). \textit{Preprint} arXiv:0710.2349v2.

\bibitem{S}
E. Schr\"odinger, Die gegenw\"artige Situation in der Quantenmechanik, \textit{Naturwissenschaften} \textbf{23}, 807-812; 823-828; 844-849 (1935); The Present Situation in Quantum Mechanics (translated by John D. Trimmer), \textit{Proceedings of the American Philosophical Society} \textbf{124}, 323-38. \textit{Preprint} http://www.tu-harburg.de/rzt/rzt/it/QM/cat.html

\bibitem{VC}
F. Verstraete and J. I. Cirac, Valence-bond states for quantum computation, \textit{Phys. Rev.} A\,\textbf{70}, 060302(R) (2004). \textit{Preprint} arXiv:quant-ph/0311130v1.

\bibitem{VMC}
F. Verstraete, M. A. Mart\'in-Delgado and J. I. Cirac, Diverging Entanglement Length in Gapped Quantum Spin Systems, \textit{Phys. Rev. Lett.} \textbf{92}, 087201 (2004). \textit{Preprint} arXiv:quant-ph/0311087v2.

\bibitem{VLRK}
G. Vidal, J. I. Latorre, E. Rico and A. Kitaev, Entanglement in quantum critical phenomena, \textit{Phys. Rev.Lett.} \textbf{90}, 227902 (2003). \textit{Preprint} arXiv:quant-ph/0211074v1.

\bibitem{XKHK}
Y. Xu, H. Katsura, T. Hirano and V. E. Korepin, Entanglement and Density Matrix of a Block of Spins in AKLT Model, \textit{J. Stat. Phys.} \textbf{133}, 347-377 (2008). \textit{Preprint} arXiv:0802.3221.

\bibitem{XKHK2}
Y. Xu, H. Katsura, T. Hirano and V. E. Korepin, Block Spin Density Matrix of the Inhomogeneous AKLT Model, \textit{Quantum Inf. Process.} \textbf{7}, 153-174 (2008). \textit{Preprint} arXiv:0804.1741.

\bibitem{XK}
Y. Xu and V. E. Korepin, Entanglement of the Valence-Bond-Solid State on an Arbitrary Graph, \textit{J. Phys.} A\,\textbf{41}, 505302 (2008). \textit{Preprint} arXiv:0805.3542.

\bibitem{ZR}
P. Zanardi and M. Rasetti, Holonomic Quantum Computation, \textit{Phys. Lett.} A\,\textbf{264}, 94-99 (1999). \textit{Preprint}	arXiv:quant-ph/9904011v3; A. Marzuoli and M. Rasetti, Spin network quantum simulator, \textit{Phys. Lett.} A\,\textbf{306}, 79-87 (2002). \textit{Preprint} arXiv:quant-ph/0209016v1; M. Rasetti, A consistent Lie algebraic representation of quantum phase and number operators (2002), \textit{Preprint} arXiv:cond-mat/0211081.


\end{thebibliography}
\end{document}